\documentclass[11pt]{article}
\usepackage{color,soul}
\usepackage{comment}
\usepackage{amsmath}
\usepackage{graphicx,psfrag,epsf}
\usepackage{enumerate}
\usepackage{xurl} 
\usepackage{amssymb, amsfonts, enumitem, xcolor, booktabs}
\usepackage{authblk} 
\usepackage{amsthm} 
\usepackage{bm} 
\usepackage{subcaption}
\usepackage{caption}
\usepackage{xr-hyper}
\usepackage[colorlinks=true,linkcolor=blue,citecolor=blue]{hyperref}
\usepackage{multirow}
\usepackage{setspace}
\def\spacingset#1{\renewcommand{\baselinestretch}%
{#1}\small\normalsize} \spacingset{1}
\usepackage{natbib}

\newcommand{\matern}{Mat\'{e}rn }

\newcommand{\R}{\mathbb{R}}

\newcommand{\vt}{\mathbf}
\newcommand{\mc}{\mathcal}

\newtheorem{proposition}{Proposition}
\newtheorem{theorem}{Theorem}

\newcommand{\articletitle}{\bf Bag of DAGs: Inferring Directional Dependence in Spatiotemporal Processes}

\graphicspath{ {./Figure/} }

\usepackage[hmargin=1.0in,vmargin=1.0in]{geometry}

\graphicspath{ {./Figure/} }

\begin{document}

\title{\bf \articletitle}
\author[1,*]{Bora Jin}
\affil[1]{Department of Biostatistics, Johns Hopkins University}
\affil[*]{email: bjin9@jh.edu} 
\author[2]{Michele Peruzzi}
\affil[2]{Department of Biostatistics, University of Michigan-Ann Arbor}
\author[3]{David Dunson}
\affil[3]{Department of Statistical Science, Duke University}
\date{}
\vspace{-6em}
\maketitle

\spacingset{1.45}
\begin{abstract}
We propose a class of nonstationary processes to characterize space- and time-varying directional associations in point-referenced data. We are motivated by spatiotemporal modeling of air pollutants in which local wind patterns are key determinants of the pollutant spread, but information regarding prevailing wind directions may be missing or unreliable. We propose to map a discrete set of wind directions to edges in a sparse directed acyclic graph (DAG), accounting for uncertainty in directional correlation patterns across a domain. The resulting Bag of DAGs processes (BAGs) lead to interpretable nonstationarity and scalability for large data due to sparsity of DAGs in the bag. We outline Bayesian hierarchical models using BAGs and illustrate inferential and performance gains of our methods compared to other state-of-the-art alternatives. We analyze fine particulate matter using high-resolution data from low-cost air quality sensors in California during the 2020 wildfire season. An \texttt{R} package is available on GitHub. 
\end{abstract}


\section{Introduction} 
\label{sec:intro}

\begin{figure}
\centering
\includegraphics[width=0.9\textwidth]{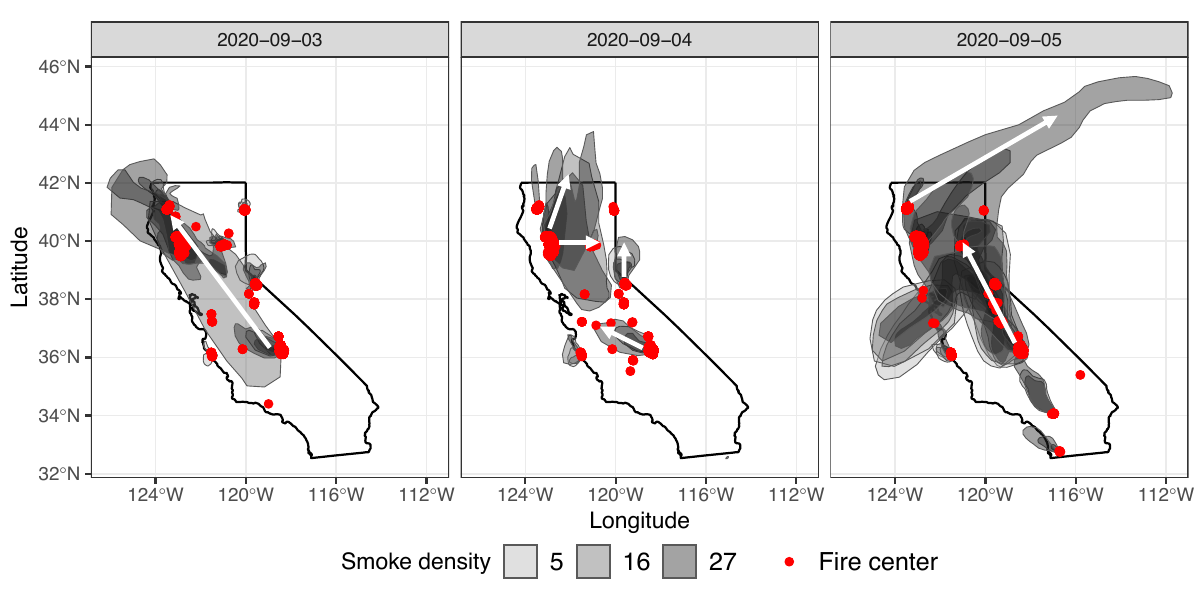}
\caption{California with wildfire centers and heavy smoke areas on September 3 -- 5, 2020. White arrows of arbitrary lengths visually assist to indicate directions of smoke spread.}
\label{mot:CAfire}
\end{figure}

In the spread of aqueous or air pollutants, local dynamics of currents or winds influence the strength of spatial and temporal correlations. Figure \ref{mot:CAfire} illustrates daily incidents in California in which smoke from wildfires exhibits regional and temporal patterns of spread in different directions. Realistic models of pollutant spread should allow correlations between locations $\bm{l}$ and $\bm{l}'$ to peak if one is downwind of the other after a certain temporal lag. In principle, one can construct rich models for point-referenced data via Gaussian processes (GPs). Let $w(\cdot) \sim \mbox{GP}(0, \vt{C}(\bm{\theta}))$ denote a zero-mean GP across a $d$-dimensional domain $\mc{D}$. A covariance function $\vt{C}(\bm{\theta}): \mc{D}\times \mc{D}\rightarrow \R$ specifies associations across locations, indexed by parameters $\bm{\theta}$. A realization of $w(\cdot)$ at an arbitrary set of locations $\mc{L} = \{\bm{l}_1,\dots,\bm{l}_n\}$ follows a multivariate Gaussian distribution with mean zero and a covariance matrix $\vt{C}_\mc{L}(\bm{\theta})$ whose $(i,j)$ element is $\vt{C}(\bm{l}_i, \bm{l}_j\mid \bm{\theta}) = \mbox{cov}(w(\bm{l}_i), w(\bm{l}_j))$.

We are particularly interested in settings in which covariance depends on relative positions of pairs of locations. A nonstationary covariance function in a GP can be well-suited in such settings; refer to the constructions of nonstationary covariance functions via deformation, convolutions of kernels, basis functions, and kernel-based methods in \cite{sampson_constructions_2010}. However, these constructions increase model complexity and may lead to difficulties in interpretation, uncertainty quantification and computation. 

Several attempts have been made to handle such difficulties. To improve interpretability, researchers have explicitly involved spatially- and/or temporally-varying covariates in nonstationary covariance functions. For instance, \cite{calder_dynamic_2008} and \cite{neto_accounting_2014} let covariance parameters depend on wind direction to characterize nonstationarity in air pollutants. Other examples include \cite{schmidt_considering_2011}, \cite{risser_regression-based_2015}, and \cite{ingebrigtsen_estimation_2015}. However, obtaining reliable covariate data can be troublesome especially at a sufficiently high spatial and temporal resolution. To enhance computational efficiency in nonstationary models, sparsity is often exploited. For instance, \cite{krock_nonstationary_2021} encourage sparsity in a precision matrix of stochastic coefficients in a basis expansion of a latent spatial process, and \cite{kidd_bayesian_2022} take advantage of a sparse Cholesky factor of a precision matrix. However, these approaches do not characterize uncertainty in the estimated precision matrix or specify a valid process, leading to an inability to perform model-based posterior predictions at new locations. 

Mixture-based methods have also been proposed to characterize nonstationarity and to relax Gaussian assumptions by introducing spatial dependence in atoms and/or in mixing weights \citep{gelfand_bayesian_2005, duan_generalized_2007, rodriguez_latent_2010, fuentes_multivariate_2013}. These methods are flexible and may lead to more accurate predictions. However, they are unable to easily characterize directional dependence in data, which is our primary objective. Overcoming the aforementioned challenges in current nonstationary modeling, we aim to construct valid and scalable nonstationary processes that yield interpretable directional dependence and enable uncertainty quantification, without requiring nonstationarity-informing covariates.

We propose a novel class of nonstationary processes by placing a prior over directional edges within sparse directed acyclic graphs (DAGs) to characterize varying directional dependence structures in space and time. We build a DAG with nodes corresponding to spatial blocks. Directed edges encode directional dependence among nodes. When a directed edge exists between two nodes, variability in the child node is partly explained by variability of its parent. Since we are uncertain about the true directional dependence, we treat the edges, corresponding to prevailing winds or currents, as unknown. We call the resulting model a Bag of directed Acyclic Graphs process (BAG). The construction is appealing in terms of interpretability, flexibility, and uncertainty quantification.

As we limit spatiotemporal dependence via conditional independence assumptions based on DAGs, BAGs are related to the literature on scalable GPs using sparse DAGs \citep{vecchia_estimation_1988, datta_hierarchical_2016, katzfuss_general_2021, peruzzi_highly_2022}. For instance, one can construct a nearest neighbor GP (NNGP; \citealt{datta_hierarchical_2016}) by restricting parent sets to include a few closest neighbors with equal weights or with spatially varying weights \citep{zheng_nearest-neighbor_2022} or develop a meshed GP \citep{peruzzi_highly_2022} by fixing a patterned DAG over domain partitions. Due to the shared construction based on sparse DAGs, BAGs retain computational gains and nice properties of the aforementioned methods. However, methods based on fixed sparse DAGs may perform poorly when directional dependence is important. Hence, we extend these methods by allowing data to determine a sparse DAG. In doing so, we also widen the perspective towards DAGs as a tool to infer the interpretable dependence structure they can give rise to, rather than merely for computational scalability. 

We formally introduce BAGs and Gaussian BAGs (G-BAGs), along with Bayesian hierarchical models, and investigate the resulting nonstationarity. We demonstrate inferential benefits of our approach via simulation studies and applications to air quality data. A Markov chain Monte Carlo (MCMC) sampler is implemented in an \texttt{R} package \texttt{bags}, and the code to reproduce all analyses in this paper is provided on GitHub. Exact links are provided in Supplementary Material S1.

\section{Spatiotemporal Process Modeling Using BAGs} 
\label{sec:allforBAG}

\subsection{DAG Specifications}

Consider a process, $\{w(\bm{l})\in \R\mid\bm{l} \in \mc{D}\}$, defined over $\mc{D} \subseteq \R^d$ for spatial data or $\mc{D} \subseteq \R^{d+1}$ for spatiotemporal data. Let $\mc{S} = \{\vt{s}_1,\dots,\vt{s}_k\}$ denote a fixed reference set of locations in $\mc{D}$, which may or may not coincide with locations of the data points $\mc{T} = \{\vt{t}_1,\dots,\vt{t}_n\}$. We first divide $\mc{D}$ into $M$ disjoint regions, $\mc{D} = \cup_{i=1}^M\mc{D}_i$ with $\mc{D}_i \cap \mc{D}_j=\emptyset$ for $i\neq j$. This partitioning similarly divides $\mc{S}$ into $M$ disjoint reference subsets where $\mc{S} = \cup_{i=1}^M\mc{S}_i$ with $\mc{S}_i = \mc{S} \cap \mc{D}_i$. Elements in $\mc{S}_i$ are enumerated as $\{\vt{s}_{i_1},\dots,\vt{s}_{i_{k_i}}\}$ with $\{i_1,\dots,i_{k_i}\} \subseteq \{1,\dots,k\}$ and $k = \sum_{i=1}^Mk_i$. 

\begin{figure}[htbp]
\centering
\includegraphics[width=0.8\textwidth]{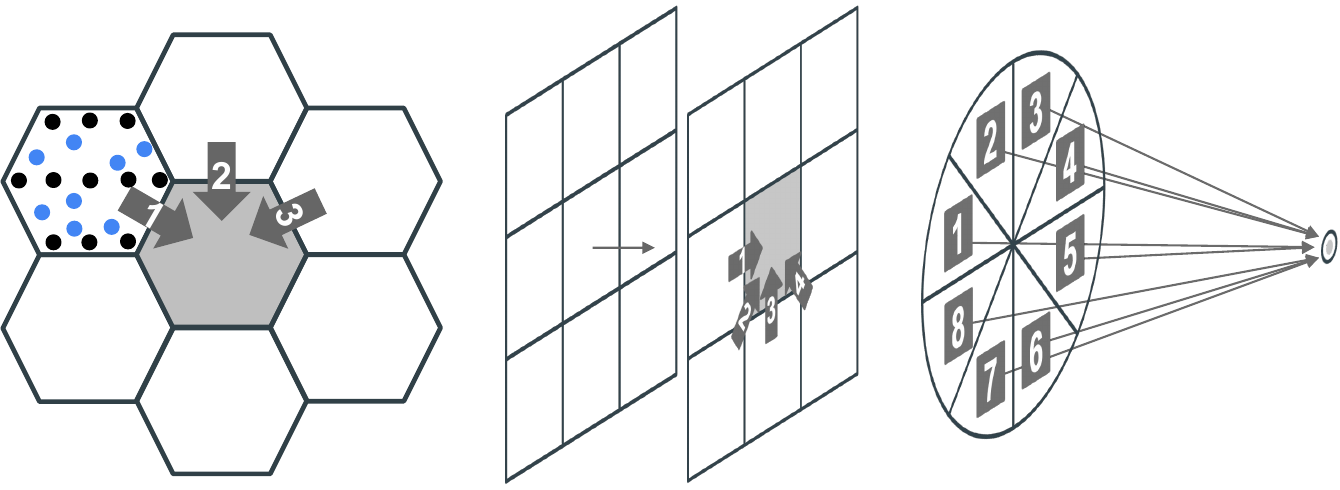}
\caption{Examples of partitions and possible directed edges for spatial (left) and spatiotemporal (middle and right) domains. Each dot is a location, and the locations are omitted for visual simplicity except on the first partition block of the left panel. Arrows show a pool of potential directed edges to choose for shaded regions.}
\label{method:ex}
\end{figure}

Figure \ref{method:ex} illustrates examples of domain partitions where partitioned blocks consist of individual or groups of locations. The left and the middle panels show respective tessellations via hexagons and rectangles. Blue dots are reference locations in $\mc{S}$, while black dots are other locations. Same-colored dots in a block collectively become a node of a DAG, rendering two nodes in each block. Overlaying directed edges between nodes completes specifications of the DAG. 

Consider a realization of the process $w(\cdot)$ over $\mc{S}$, $\vt{w}_{\mc{S}} = (\vt{w}_1^T,\dots,\vt{w}_M^T)^T$ where $\vt{w}_i = (w(\vt{s}_{i_1}), \dots, w(\vt{s}_{i_{k_i}}))^T$ is a realized vector over $\mc{S}_i$ for $i=1,\dots, M$. After choosing an arbitrary order of $M$ partitioned regions, the joint density $p(\vt{w}_{\mc{S}}) = p(\vt{w}_1, \dots, \vt{w}_M)$ is rewritten as a product of conditional densities:
\begin{align}
    p(\vt{w}_{\mc{S}}) =p(\vt{w}_1) \prod_{i=2}^M p(\vt{w}_i\mid\vt{w}_1,\dots,\vt{w}_{i-1}). 
    \label{eq:jointdensity}
\end{align}
We can represent equation \eqref{eq:jointdensity} as a DAG $\mc{G}_{\mc{S}} = (A, \Gamma)$ with nodes $A = \{ a_1, \dots, a_M \}$ and directed edges $\Gamma = \{([a_i] \rightarrow a_i) \mid  i = 1, \dots, M \}$. The random vector $\vt{w}_i$ of length $k_i$ is collectively mapped to a single node $a_i$, and $[a_i]$ represents a set of parents of $a_i$. Specifically, $[a_1]=\emptyset$ and $[a_i]=\{a_1, \dots, a_{i-1}\}$ for $i=2,\dots,M$ in equation \eqref{eq:jointdensity}.

Models with fixed sparse DAGs drop some edges in $\Gamma$ and build a new DAG $\mc{G}^*_{\mc{S}}$ leading to a joint density $p^*$ as a product of conditional densities with reduced conditioning sets:
\begin{align}
    p^*(\vt{w}_{\mc{S}}) = \prod_{i=1}^M p(\vt{w}_i\mid\vt{w}_{[i]}). \label{pstar}
\end{align}
The parent set $[a_i]$ in the new DAG $\mc{G}^*_{\mc{S}}$ is now a subset of $\{a_1,\dots,a_{i-1}\}$, and $\vt{w}_{[i]}$ collects realized values in $[a_i]$, so that $\vt{w}_{[i]} = \{\vt{w}_j\mid a_j \in [a_i]\}$. Since $\mc{G}^*_{\mc{S}}$ is a DAG, $p^*(\vt{w}_{\mc{S}})$ is a proper joint density \citep{lauritzen_graphical_1996} with some base density $p$.

A potential pitfall of fixed-DAG models is the use of an inappropriate sparse DAG for inference when directional dependence is important. If each of the true $\vt{w}_i$ depends on parent nodes from a certain direction, fixed-DAG models that include locations from other directions in place of locations from the correct direction may be sub-optimal in terms of Kullback-Leibler divergence to the true model as shown in the following proposition. 

\begin{proposition}
Suppose data are generated by equation \eqref{pstar} with some base density $p$, yielding a density $p^*_0$ corresponding to a sparse DAG $\mc{G}^*_0 = \{V, \Gamma_0\}$ with nodes $V = \{v_1,\dots,v_M\}$ and directed edges $\Gamma_0 = \{([v_i]_0 \rightarrow v_i) \mid i=1,\dots,M\}$. Each $v_i$ is located downwind of $[v_i]_0$. Let $p^*_1$ be a density with a DAG $\mc{G}^*_1 = \{V, \Gamma_1\}$ where $\Gamma_1 = \{([v_i]_1 \rightarrow v_i)\mid i=1,\dots,M\}$ and $[v_i]_1 \subseteq [v_i]_0$ for all $i=1,\dots,M$. Each $[v_i]_1$ collects locations only from the correct directions. Define another density $p^*_2$ with a DAG $\mc{G}^*_2 = \{V, \Gamma_2\}$. The edge set $\Gamma_2 = \{([v_i]_2 \rightarrow v_i)\mid i=1,\dots,M\}$ has $[v_i]_2 = [v_i]_1$ for all $i$ but some $j \in \{1,\dots,M\}$, for which $[v_j]_2 \not\subset [v_j]_0$. The set $[v_j]_2$ includes locations from wrong directions. Then $p^*_2$ has larger Kullback-Leibler divergence to $p^*_0$ than $p^*_1$ does: $\mbox{KL}(p^*_0\|p^*_2) > \mbox{KL}(p^*_0\|p^*_1)$. 
\label{prop:KL}
\end{proposition} 

Proposition \ref{prop:KL} implies that edges should be placed to match directional dependence in the data. Because one cannot establish such dependence a priori, models using fixed sparse DAGs will poorly explain changing local dynamics of the dependence. We provide the proof and illustrate the intuition behind Proposition \ref{prop:KL} in Supplementary Material S2.1.

\subsection{Unknown DAG}
\label{sec:unknownDAG}

We take a different perspective and allow a DAG to be inferred from data. We first choose a realistic set, referred to as a ``bag'', of possible directions based on prior knowledge about a domain. The possible directions in the bag are enumerated as $h=1,\ldots,K$. Then we introduce a latent variable $z_i$ storing the direction along which dependence to node $a_i$ is allowed to flow. The event $z_i=h$ assigns $a_i$ a parent from the $h$th direction. We denote the resulting parent node as $[a_i\mid z_i=h]$. 

Conditional on the latent vector of assigned directions $Z=(z_1,\dots,z_M)$, we define a joint density $\tilde{p}$, modifying equation \eqref{pstar} as follows:
\begin{align}
    \tilde{p}(\vt{w}_{\mc{S}}\mid Z) = \prod_{i=1}^M p(\vt{w}_i\mid\vt{w}_{[i\mid z_i]}), \label{ptilde_cond}
\end{align}
with $\vt{w}_{[i\mid z_i]} = \{\vt{w}_j \mid a_j \in [a_i\mid z_i] \subset \{a_1,\dots,a_{i-1}\}\}$. The directions $z_1,\dots,z_M$ are a priori independent with the joint prior probability $q(Z) = \prod_{i=1}^M q_i(z_i)$ and $\sum_{z_i=1}^K q_i(z_i) = 1$ for $i \in \{1,\dots,M\}$, and marginalizing them out yields $\tilde{p}(\vt{w}_{\mc{S}}) = \sum_{Z\in \text{bag of DAGs}}\tilde{p}(\vt{w}_{\mc{S}}\mid Z) q(Z)$ $= \sum_{Z}\left\{\prod_{i=1}^M p(\vt{w}_i\mid \vt{w}_{[i\mid z_i]})q_i(z_i)\right\}.$ The summation $\sum_{Z\in \text{bag of DAGs}}$ is over $K^M$ possible $Z$'s and is hereafter denoted by $\sum_{Z}$ for notational brevity. Then Proposition \ref{prop:jointdensity} holds, and its proof is provided in Supplementary Material S2.2.

\begin{proposition}
$\tilde{p}(\vt{w}_{\mc{S}}) = \sum_{Z}\left\{\prod_{i=1}^M p(\vt{w}_i\mid \vt{w}_{[i\mid z_i]})q_i(z_i)\right\}$ is a proper joint density.
\label{prop:jointdensity}
\end{proposition}

In Figure \ref{method:ex}, the numbered arrows represent competing assumptions on the dependence structure relevant for the shaded regions. Panels from left to right correspond to $K = 3, 4, 8$, respectively, considering a subset of all directions to ensure acyclicity of inferred directed graphs. Prior knowledge about the domain can be utilized to choose one direction over the other on each axis. If no background knowledge is available, one may allow partition blocks from any directions in the past to be parents as illustrated on the right panel of Figure \ref{method:ex}. This construction of ensuring acyclicity guarantees an inferred DAG to be unique although a DAG is identifiable only up to its Markov equivalence class from observational data alone \citep{castelletti_bayesian_2021}. Proposition \ref{prop:uniqueDAG} shows that the Markov equivalence class of an inferred DAG in the proposed framework is of size 1, implying the DAG is identifiable. The proof is given in Supplementary Material S2.2.

\begin{proposition}
Let $\mc{G}$ be an inferred DAG starting with a bag of directions chosen to ensure acyclicity. Then the Markov equivalence class of $\mc{G}$ is of size 1. 
\label{prop:uniqueDAG}
\end{proposition}

Our current implementation of BAGs is based on tessellations and a simplifying assumption that each node selects at most one parent from neighboring blocks in space and takes the block covering the same spatial region at the previous time point as the temporal parent. The middle DAG specification of Figure \ref{method:ex} is an example. This modeling choice is solely for the implementation and made to increase interpretability, clarity in exposition and computational tractability. Directed edges in tessellations lead to intuitive interpretations; for example, the directed edge 1 on the middle panel of Figure \ref{method:ex} is interpreted as the direction from west to east. Moreover, tessellations enable stochastic search of directed edges to proceed analogously for all partition blocks due to the same neighbor structure and thus the same bag of directions. Theoretical conclusions or properties of BAGs to appear later in this paper do not rely on the simplifying assumption or tessellations. We show robustness of BAGs to the simplifying assumption in simulation studies and robustness to domain tessellations and directed edges in real applications.

\subsection{Defining a Coherent Process} 
\label{sec:BAG}

The above discussion focuses on finite dimensional $\vt{w}_\mc{S}$. We extend the finite density to a valid process over all locations in the domain. We label the set of non-reference locations as $\mc{U} = \mc{D} \setminus \mc{S}$, grouped into disjoint sets $\mc{U}_1, \dots, \mc{U}_M$ such that $\cup_{i=1}^M \mc{U}_i = \mc{U}$ by the same partitioning in $\mc{D}$. We extend the DAG over $\mc{S}$ to a larger DAG $\tilde{\mc{G}}$. Nodes of $\tilde{\mc{G}}$ are the union of $A = \{ a_1, \dots, a_M \}$ and $B = \{b_1, \dots, b_M\}$ where $\mc{U}_i$ is mapped to a node $b_i$ for $i=1,\dots,M$. The construction of $\tilde{\mc{G}}$ is completed by assigning directed edges from nodes in $A$ to nodes in $B$, ensuring acyclicity of $\tilde{\mc{G}}$. There are several possible ways to place these directed edges. Assuming that $\mc{S}_i \neq \emptyset$ for all $i$, one can fix the edges as $a_i \to b_i$ with $[b_i] = a_i$, implying that local reference locations become the parent set for the non-reference locations in the same partitioned region. Kolmogorov consistency conditions are then easily verified following a similar approach as in Appendix A of \cite{peruzzi_highly_2022}, proving the validity of the resulting process. However, a better alternative lets $[b_i\mid z_i] = a_i \cup [a_i\mid z_i]$, which allows modeling at any non-reference locations to depend not only on the local reference locations but also on the reference locations' parents learned from data. This way, we can perform predictions based on inferred directions. 

Proposition \ref{prop:Kolmogorov} shows that our proposed stochastic process, which entails randomness in choosing directed edges among $A$ and $B$, satisfies the Kolmogorov consistency conditions; the proof is in Supplementary Material S2.2. Assume conditional independence of non-reference locations given the local reference locations and their parents so that 
\begin{align}
    \tilde{p}(\vt{w}_{\mc{U}}\mid \vt{w}_{\mc{S}},Z) =  \prod_{i=1}^M \prod_{\vt{u}\in\mc{U}_i}p(w(\vt{u})\mid \vt{w}_{i},\vt{w}_{[i\mid z_i]}).\label{ptildeU_cond}
\end{align} 
Equations $\eqref{ptilde_cond}$ and $\eqref{ptildeU_cond}$ suffice to describe the joint density over any finite subset of locations $\mc{L}\subset \mc{D}$. With $\mc{U}_{\mc{L}} = \mc{L}\setminus\mc{S}$,
\begin{align}
    \tilde{p}(\vt{w}_{\mc{L}}) = \int \left\{ \sum_{Z}\tilde{p}(\vt{w}_{\mc{U}_{\mc{L}}}\mid \vt{w}_{\mc{S}},Z)\tilde{p}(\vt{w}_{\mc{S}}\mid Z)q(Z)\right\}\prod_{\vt{s}_i\in\mc{S}\setminus\mc{L}}d\{w(\vt{s}_i)\}.
    \label{ptildeL}
\end{align}

\begin{proposition}
The collection of finite dimensional densities in equation \eqref{ptildeL} satisfies the Kolmogorov conditions. \label{prop:Kolmogorov}
\end{proposition}

Proposition \ref{prop:Kolmogorov} implies that there exists a stochastic process associated with the collection of finite dimensional densities in equation \eqref{ptildeL}, which we call a BAG. In summary, via domain partitioning and mixtures of DAGs, our approach generates a valid spatiotemporal process, leading to inferential advantages by incorporating parameter estimation and prediction at arbitrary locations into a coherent framework. BAGs can also be embedded seamlessly within Bayesian hierarchies as a prior process. 

When modeling spatiotemporal variation, we are interested in understanding covariances between process realizations at different locations. The covariance of BAGs between arbitrary locations $\bm{l}_1, \bm{l}_2\in\mc{D}$ has the following form: $\mbox{cov}_{\tilde{p}}(w(\bm{l}_1), w(\bm{l}_2)) =  E_{Z}\{\mbox{cov}_{\tilde{p}}(w(\bm{l}_1), w(\bm{l}_2) \mid  Z)\} +  \mbox{cov}_{Z}\{E_{\tilde{p}}(w(\bm{l}_1) \mid Z), E_{\tilde{p}}(w(\bm{l}_2) \mid Z)\}$ where $E_{\tilde{p}}$ and $E_{Z}$ indicate expectation with respect to $\tilde{p}$ and $Z$, respectively. If $E_{\tilde{p}}(w(\bm{l}) \mid Z) = 0$ for any $\bm{l}\in \mc{D}$, which can be easily the case, $\mbox{cov}_{\tilde{p}}(w(\bm{l}_1), w(\bm{l}_2)) =  \sum_{Z}\mbox{cov}_{\tilde{p}}(w(\bm{l}_1), w(\bm{l}_2) \mid Z)q(Z)$. This suggests that pairwise covariances between process realizations can be studied starting from the conditional covariances given $Z$. We further study BAG-induced nonstationary covariance under the Gaussian assumption. 

\subsection{Gaussian BAGs} 
\label{sec:GBAG}

We now discuss BAGs under the typical distributional assumption that the base process is a zero-centered $\mbox{GP}(0, \vt{C}(\bm{\theta}))$ with a valid base covariance function $\vt{C}$ parametrized by $\bm{\theta}$. Then, with a Gaussian base density, equation \eqref{ptilde_cond} becomes 
\begin{align}
    \tilde{p}(\vt{w}_{\mc{S}} \mid Z) = \prod_{i=1}^MN(\vt{w}_i; H_{i\mid z_i}\vt{w}_{[i\mid z_i]}, R_{i\mid z_i}) = N(\vt{w}_{\mc{S}}; \vt{0}, \tilde{C}_{Z}), \label{ptilde_con_N}
\end{align}
where $H_{i\mid z_i} = C_{i,[i\mid z_i]}C_{[i\mid z_i]}^{-1}$ and $R_{i\mid z_i} = C_{i} - C_{i,[i\mid z_i]}C_{[i\mid z_i]}^{-1}C_{[i\mid z_i],i}$ with $C_{i,[i\mid z_i]}$ being the base covariance matrix between locations mapped to $a_i$ and to $[a_i\mid z_i]$. Recall that each set $\mc{S}_i$ is of size $k_i$, so that $C_{i,[i\mid z_i]}$ has dimension $k_i \times \sum_{j:a_j\in[a_i\mid z_i]} k_j$.
The notation $C_i$ is a shorthand for $C_{i,i}$ for any $i\in\{1,\dots,M\}$. If $[a_i\mid z_i]=\emptyset$, $H_{i\mid z_i}\vt{w}_{[i\mid z_i]}=\vt{0}\in\R^{k_i}$ and $R_{i\mid z_i}=C_{i}$. Thus, equation \eqref{ptilde_con_N} shows that within-node correlations for reference locations are affected by directional dependence with neighboring nodes and become equal to the corresponding base correlations only when the node has no parents. In vector form, $\tilde{p}(\vt{w}_{\mc{S}} \mid Z)$ is multivariate Gaussian with a covariance matrix $\tilde{C}_{Z}$. Since we limit few parents from inferred directions, the precision matrix $\tilde{C}^{-1}_{Z}$ is sparse. More detailed properties and sparsity of $\tilde{C}^{-1}_{Z}$ are discussed in Supplementary Material S2.3. 

Similarly, for non-reference locations, equation \eqref{ptildeU_cond} becomes
\begin{align}
    \tilde{p}(\vt{w}_{\mc{U}}\mid \vt{w}_{\mc{S}},Z) = \prod_{i=1}^M\prod_{\vt{u}\in\mc{U}_i} N(w(\vt{u});H_{\vt{u}\mid z_i}\vt{w}_{[\vt{u}\mid z_i]},R_{\vt{u}\mid z_i}) = N(\vt{w}_{\mc{U}}; H_{\mc{U}\mid Z}\vt{w}_{\mc{S}}, R_{\mc{U}\mid Z}), \label{ptildeU_N}
\end{align}
with $H_{\vt{u}\mid z_i} = C_{\vt{u},[\vt{u}\mid z_i]}C_{[\vt{u}\mid z_i]}^{-1}$ and  $R_{\vt{u}\mid z_i} = C_{\vt{u}} - C_{\vt{u},[\vt{u}\mid z_i]}C_{[\vt{u}\mid z_i]}^{-1}C_{[\vt{u}\mid z_i],\vt{u}}$. In vector form, the conditional density of $\vt{w}_{\mc{U}}$ given $\{\vt{w}_{\mc{S}}, Z\}$ is multivariate Gaussian with mean $H_{\mc{U}\mid Z}\vt{w}_{\mc{S}}$ and covariance matrix $R_{\mc{U}\mid Z}$. Matrix $H_{\mc{U}\mid Z}$ is of size $|\mc{U}|\times k$ with $|\mc{U}|$ being the number of locations in $\mc{U}$. Block-diagonal matrix $R_{\mc{U}\mid Z}$ has the $i$th block matrix that is again diagonal with elements $R_{\vt{u}\mid z_i}$ for all $\vt{u}\in\mc{U}_i$ for $i=1,\dots,M$. If $[b_i\mid z_i]=a_i\cup[a_i\mid z_i]$ and $[a_i\mid z_i]=\{a_j, a_l\}$, $[\vt{u}\mid z_i]$ becomes $\{a_i, a_j, a_l\}$ for any $\vt{u}\in \mc{U}_i$, yielding a $1\times (k_i+k_j+k_l)$ matrix $C_{\vt{u},[\vt{u}\mid z_i]}$ whose elements are $\vt{C}(\vt{u},\vt{s} \mid \bm{\theta})$ for all $\vt{s}\in S_i\cup S_j\cup S_l$. Also, the $i$th block-row of $H_{\mc{U}\mid Z}$ corresponding to locations in $\mc{U}_i$ will have non-zero blocks in columns corresponding to $a_i$, $a_j$, and $a_l$.

We obtain a Gaussian BAG (G-BAG) such that $w(\cdot) \sim \mbox{G-BAG}(0,\tilde{\vt{C}}(\bm{\theta}))$ whose finite dimensional densities are specified by plugging equations \eqref{ptilde_con_N} -- \eqref{ptildeU_N} in equation \eqref{ptildeL}. The covariance function $\tilde{\vt{C}}$ for G-BAG is 
\begin{align}
    \tilde{\vt{C}}(\bm{l}_1,\bm{l}_2\mid \bm{\theta}) = \sum_{Z}\tilde{\vt{C}}(\bm{l}_1,\bm{l}_2\mid Z,\bm{\theta})q(Z) \label{marginal_Ctilde}
\end{align}
for any two locations $\bm{l}_1,\bm{l}_2\in\mc{D}$. For any given $Z$, $\tilde{\vt{C}}(\bm{l}_1,\bm{l}_2\mid Z,\bm{\theta})$ is induced as a covariance function of a GP. Since equations \eqref{ptilde_con_N} and \eqref{ptildeU_N} are all Gaussian, the density $\tilde{p}(\vt{w}_{\mc{L}}\mid  Z)$ is also Gaussian for any finite subset $\mc{L} \subset \mc{D}$, leading to a GP with the covariance function $\tilde{\vt{C}}(\bm{l}_1,\bm{l}_2\mid Z,\bm{\theta})$ in the following form:
\begin{align}
    &\tilde{\vt{C}}(\bm{l}_1,\bm{l}_2 \mid Z,\bm{\theta}) \nonumber \\
    &=\left\{\begin{array}{ll}
         \tilde{C}_{\vt{s}_i,\vt{s}_j},& \text{if }\bm{l}_1=\vt{s}_i\in\mc{S} \text{ and } \bm{l}_2=\vt{s}_j\in\mc{S} \\
         H_{\bm{l}_1 \mid z_i}\tilde{C}_{[\bm{l}_1\mid z_i],s_j}, &\text{if } \bm{l}_1\in \mc{U}_i \text{ and } \bm{l}_2=\vt{s}_j\in\mc{S} \\
         \bm{1}(\bm{l}_1=\bm{l}_2)R_{\bm{l}_1\mid z_i} + H_{\bm{l}_1\mid z_i}\tilde{C}_{[\bm{l}_1\mid z_i],[\bm{l}_2\mid z_j]}H_{\bm{l}_2\mid z_j}^T, &\text{if } \bm{l}_1\in \mc{U}_i \text{ and } \bm{l}_2 \in\mc{U}_j
    \end{array}\right. 
    \label{Ctilde}
\end{align}
where $\tilde{C}_{P,Q}$ is a submatrix of $\tilde{C}_{Z}$ corresponding to locations in sets $P$ and $Q$, and $\bm{1}(\cdot)$ is the indicator function. As $\tilde{\vt{C}}(\cdot,\cdot\mid  Z,\bm{\theta})$ is nonstationary, the marginal covariance function $\tilde{\vt{C}}(\cdot,\cdot\mid \bm{\theta})$ is as well. Derivation and properties of equation \eqref{Ctilde} are summarized in Supplementary Material S2.4. 

G-BAGs can use any covariance function as a base covariance $\vt{C}$; G-BAGs capture directional dependence through mixtures of DAGs, improving flexibility over a base covariance that need not incorporate directional dependence. Since more complex covariance functions have more parameters, complicating estimation and interpretation, we are encouraged to choose simpler covariance functions as a base. As a flexible default, we focus on a nonseparable stationary spatiotemporal covariance in \cite{gneiting_nonseparable_2002}. For a space-time lag $(\bm{h},u)\in \R^{d+1}$, the covariance function is
\begin{align}
    \vt{C}(\bm{h},u\mid {\bm{\theta}}) = \frac{\sigma^2}{(a|u|+1)}\exp\left\{-\frac{c\|\bm{h}\|}{(a|u|+1)^{\kappa/2}}\right\} \label{Gneitingcov}
\end{align}
with $|\cdot|,\|\cdot\|$ denoting one-, two-dimensional Euclidean distance, and $\bm{\theta}=(a,c,\kappa,\sigma^2)$ where $a>0$ and $c>0$ are temporal and spatial decays, respectively, $\kappa \in [0,1]$ is a space-time interaction parameter, and $\sigma^2$ is the variance. Equation \eqref{Gneitingcov} reduces to a separable function when $\kappa=0$. This function has been criticized as inappropriate for modeling covariance of dynamic processes such as winds \citep{stein_space-time_2005} due to its full symmetry defined as $\vt{C}(\vt{h}, u) = \vt{C}(-\vt{h}, u) = \vt{C}(\vt{h}, -u) = \vt{C}(-\vt{h}, -u)$. However, our model can induce asymmetric (or directional, interchangeably) nonstationarity in the resulting covariance even with a fully symmetric stationary base covariance.

We now consider a simplified setting to elucidate the asymmetric features of $\tilde{\vt{C}}(\cdot,\cdot\mid \bm{\theta})$. 

\begin{proposition}
Consider three reference locations $\vt{s}_1$, $\vt{s}_2$, $\vt{s}_3$, each of which forms a partition block. $\vt{s}_1$ is equidistant from $\vt{s}_2$ and $\vt{s}_3$; $\|\bm{h}_2\| = \|\bm{h}_3\|$ and $|u_2| = |u_3|$ where $(\bm{h}_2, u_2)$ is a space-time lag between $\vt{s}_1$ and $\vt{s}_2$, and $(\bm{h}_3, u_3)$ between $\vt{s}_1$ and $\vt{s}_3$. Suppose only two $Z$'s have positive probabilities, $q(Z_1) = \pi$ and $q(Z_2) = 1-\pi$, where $Z_1$ yields a DAG with one directed edge from $\vt{s}_1$ to $\vt{s}_2$, and $Z_2$ with one directed edge from $\vt{s}_1$ to $\vt{s}_3$. Assume an isotropic base covariance function. Then $q(Z_1) > q(Z_2)$ implies $\tilde{\vt{C}}(\vt{s}_1, \vt{s}_2\mid \bm{\theta}) > \tilde{\vt{C}}(\vt{s}_1, \vt{s}_3\mid \bm{\theta})$ with $\tilde{\vt{C}}(\cdot, \cdot \mid \bm{\theta})$ in equation \eqref{marginal_Ctilde}.
\label{prop:nonstationary}
\end{proposition}

Proposition \ref{prop:nonstationary} shows $\tilde{\vt{C}}(\cdot,\cdot\mid \bm{\theta})$ assigns higher covariance to pairs of locations that align with more likely directions. Figure S3 in Supplementary Material S2.5 shows directional behaviors of $\tilde{\vt{C}}(\cdot, \cdot\mid \bm{\theta})$ by comparing contour lines of $\tilde{\vt{C}}(\vt{s}_1, \vt{s}_2\mid \bm{\theta})$ and $\tilde{\vt{C}}(\vt{s}_1, \vt{s}_3\mid \bm{\theta})$ for different lags and values of $\pi$. As expected, the Euclidean distance $(\|\bm{h}_3\|, |u_3|)$ should be shorter than $(\|\bm{h}_2\|, |u_2|)$ to attain the same level of correlation, and this gap between the distances enlarges as $\pi$ increases. This directional nonstationarity is also observed in other settings. Figure S4 in Supplementary Material S2.5 presents heat maps of $\tilde{\vt{C}}(\cdot, \cdot\mid \bm{\theta})$ over multiple partitioning schemes in $\mc{D}=[0,1]^3$ and shows that the induced nonstationary covariance becomes directional with time components; the temporal dimension allows to identify the direction of dependence on each axis. Supplementary Material S2.5 provides more detailed setups and discussions regarding Proposition \ref{prop:nonstationary} and Figures S3 -- S4. 

\begin{figure}
\centering
\includegraphics[width=\textwidth]{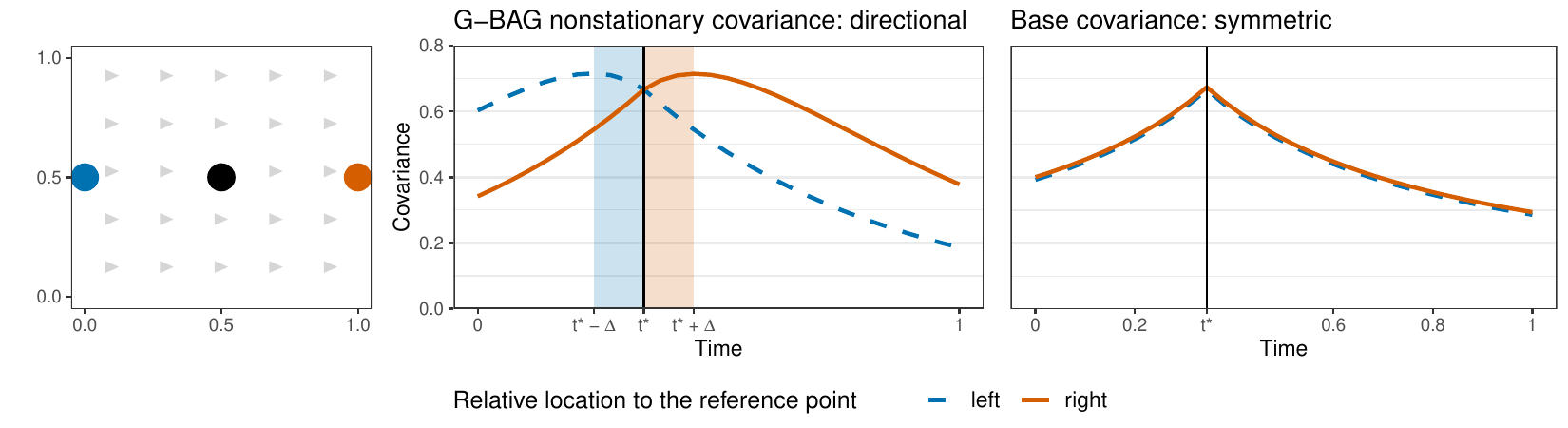}
\caption{Covariances of a reference point at (0.5, 0.5, $t^*$) to (i) locations on its left and to (ii) locations on its right at varying times. A DAG with arrows from west is assumed. The black vertical line indicates the time point $t^*$ of the reference point. The covariance from the reference point to the locations on the left or on the right is represented in a blue dashed line or a vermilion solid line, respectively.}
\label{sim:cov2}
\end{figure}

We further examine the directionality in G-BAGs' nonstationary covariance functions and show that covariances flow along directed edges. We use the base covariance function in equation \eqref{Gneitingcov} with $a = 2$, $c = 0.8$, $\kappa = 0$, and $\sigma^2 = 1$. With $\mc{S}$ on a $3 \times 3 \times 30$ grid in $\mc{D}=[0,1]^3$, we let each grid point be a partition block and assume directed edges from west over the partition. As drawn with gray arrows on the left plot of Figure \ref{sim:cov2}, we can imagine steady winds from west over time. Then we compute covariances of a location in black at (0.5, 0.5, $t^*$) to (i) locations on its left and to (ii) those on its right at different time points. As expected, the right plot of Figure \ref{sim:cov2} shows that stationarity produces the same covariance values for both groups of locations, peaked at time $t^*$ and decreasing as time lag increases. In contrast, the G-BAG induced directional covariance function gives separate curves for the two groups. In the middle plot of Figure \ref{sim:cov2}, we observe that the black point at time $t^*$ has the maximum covariance with locations on its right at time $t^*+\Delta$, while it achieves the maximum covariance with locations on its left in the past ($t^*-\Delta$). Moreover, at each time lag in the future, locations on the right have uniformly higher covariance values with the black point than locations on the left. The results are intuitive as winds are coming from left to right. The locations on the left affect the black point first, and in turn, the black point affects the locations on the right. Hence, covariance persists longer in the future in directions towards which winds blow. Similar asymmetric behavior of G-BAG nonstationary covariance is also found with a different choice of base covariance functions and spatiotemporal parents for each partition block, as illustrated in Supplementary Material S2.5. Additionally, we show in Supplementary Material S2.5 that the G-BAG nonstationary covariance is flexible enough to mimic asymmetric spatiotemporal behavior specified by other existing methods such as the so-called Lagrangian reference frame \citep{salvana_lagrangian_2021}.

\subsection{Bayesian Hierarchical Model with G-BAGs} 

Consider a general regression model
\begin{align}
y(\vt{t}) = x(\vt{t})^T\bm{\beta} + w(\vt{t}) + \epsilon(\vt{t}), \label{reg}
\end{align}
where $y(\vt{t})$ is a response, $x(\vt{t})$ is a vector of point-referenced predictors, $w(\vt{t})$ is a latent spatiotemporal process over domain $\mc{D}$, and $\epsilon(\vt{t}) \sim N(0, \tau^2)$ is a measurement error. Over sampling points $\mc{T}=\{\vt{t}_1,\dots,\vt{t}_n\}$, equation \eqref{reg} is expressed in vector form as $\vt{y}=X\bm{\beta}+\vt{w}+\bm{\epsilon}$ with $\vt{y} = (y(\vt{t}_1),\dots,y(\vt{t}_n))^T$, a $n\times p$ matrix $X$ having $x(\vt{t}_i)^T$ as its $i$th row, $\vt{w} = (w(\vt{t}_1),\dots,w(\vt{t}_n))^T$ and $\bm{\epsilon}=(\epsilon(\vt{t}_1),\dots,\epsilon(\vt{t}_n))^T \sim N(\vt{0},\tau^2I_n)$. We assume $w(\cdot)\sim \mbox{G-BAG}(0,\tilde{\vt{C}}(\bm{\theta}))$ a priori as in Section \ref{sec:GBAG}. We complete a Bayesian specification with priors for all unknowns. The joint posterior distribution for $\left\{ \vt{w}_{\mc{S}}, \vt{w}_{\mc{U}}, \{ z_i \}_{i=1}^M, \bm{\beta}, \tau^2, \bm{\theta}\right\}$ is proportional to: $N(\vt{y};X\bm{\beta}+\vt{w}, \tau^2I_n)\times N(\vt{w}_{\mc{S}};\vt{0}, \tilde{C}_{Z}) \times N(\vt{w}_{\mc{U}};H_{\mc{U}\mid Z}\vt{w}_{\mc{S}}, R_{\mc{U}\mid Z}) \times \prod_{i=1}^M \mbox{Cat}(z_i;\bm{\pi}_{i}) \times N(\bm{\beta};\bm{\mu}_{\beta},V_{\beta}) \times \mbox{IG}(\tau^2;a_{\tau}, b_{\tau}) \times p_{\theta}$, with $\mc{U} = {\mc{T}}\setminus \mc{S}$, $\mbox{Cat}$ and $\mbox{IG}$ denoting Categorical and inverse Gamma distribution, respectively, and $p_{\theta}$ denoting a prior for base covariance parameters. We leave the choice of $p_{\theta}$ general to allow user-specified base covariance $\vt{C}$. The vector $\bm{\pi}_i = (\pi_{i1},\dots,\pi_{iK})^T\in [0,1]^K$ stores the prior probability for each possible value of $z_i$ and $\sum_{h=1}^K \pi_{ih} = 1$. As a uniform default, we set $\pi_{ih}=1/K$. A MCMC sampler is described in Supplementary Material S2.6. Supplementary Material S2.7 proves the sampler has computational complexity of order $n$ at each iteration. In addition, posterior consistency of mixture weights in G-BAGs as well as in general BAGs is discussed in Supplementary Material S2.8. 

\section{Applications on Simulated Data}
\label{sec:simu}

\subsection{Simulation Setup}

We conduct simulation studies to evaluate performance of G-BAGs in prediction and learning DAGs. Our spatiotemporal domain is $\mc{D}=[0,1]^3$ with locations $\vt{t}=(t_1,t_2,t_3)\in\mc{D}$. We consider situations in which our G-BAG model is correctly specified or misspecified. For G-BAG models, we use the base covariance function in equation \eqref{Gneitingcov}.

To evaluate performance gains by inferring a DAG, we compare with a fixed-DAG model, cubic-meshed GPs \citep{peruzzi_highly_2022}, on the same domain partition. A cubic-meshed GP fixes a DAG with three edges: one from the left, one from below, and the other from the past, over the domain tessellation; it is implemented in \texttt{R} package \texttt{meshed}. We also consider two models based on stochastic partial differential equations (SPDEs) \citep{lindgren_explicit_2011}. The first, which we label SPDE-stationary, uses a stationary spatiotemporal covariance function. The second, which we label SPDE-nonstationary, uses a nonstationary covariance function with a spatially varying range such that $\rho(\vt{t}) \propto \exp\{\psi_1 + \psi_2t_1 + \psi_3t_2 + \psi_4(t_1-1)(t_2-1)\}$. Both SPDE models are implemented using the \texttt{R-INLA} package whose basic interface only supports separable spatiotemporal models and thus assume an autoregressive model of order 1 for temporal correlations. From simulated data, we build a training set by subsetting 80\% of the data locations and use the remaining 20\% as a test set for out-of-sample predictions. Both G-BAG and the fixed-DAG model use the training set as a reference set. See Supplementary Material S3 for general considerations to design a reference set.

\subsection{Fitted G-BAG is Correctly Specified}
\label{sec:sim1}

We let $y(\vt{t}) = x(\vt{t})\beta+w(\vt{t})+\epsilon(\vt{t})$ with $x(\vt{t}) \sim N(0,0.1^2)$ and $\epsilon(\vt{t}) \sim N(0,\tau^2)$, on a 40$\times$40$\times$8 regular grid in $\mc{D}$ for 12,800 spatiotemporal locations. Parameters $\beta$ and $\tau^2$ are set at 2 and 0.01, respectively. We partition each of the spatial axes into 6 irregular intervals and the temporal axis into 8 regular intervals, resulting in $M=6 \times 6 \times 8 = 288$ partition blocks. Wind effects that vary in space and time are simulated with directions from west, northwest, north, and northeast as depicted in Figure S8a in Supplementary Material S3.1. We fit a G-BAG model using the same domain partition and the same bag of directions as the data generating model. Twenty five synthetic data sets were generated with $\bm{\theta}_1 = (5, 0.5, 0.9, 2)$ and another 25 data sets with $\bm{\theta}_2 = (10, 0.1, 0.2, 2)$ indicating $a$, $c$, $\kappa$, and $\sigma^2$ in order. G-BAG and fixed-DAG results are based on 1,000 posterior samples out of 17,000 MCMC draws by discarding the first 10,000 samples as burn-in and saving every 7th sample in the subsequent samples. Empirical investigation of posterior draws suggests convergence and adequate mixing. 

\begin{table}[htbp]
\centering
    \begin{tabular}{ccccc}
    \hline 
         &  \multirow{2}{*}{G-BAG} & \multirow{2}{*}{Fixed DAG} & SPDE- & SPDE-  \\ [-3pt]
         &   &  & stationary & nonstationary  \\ \hline
   $\beta=2$ & \textbf{2.004} (0.018) & 2.013 (0.036) & 2.005 (0.056) & 2.005 (0.055) \\
   $\tau^2=0.01$ &\textbf{0.010} ($<$0.001) & 0.016 (0.002) & 0.179 (0.019) & 0.178 (0.019) \\
   $\sigma^2 =2$ & \textbf{1.874} (0.268) & 1.783 (0.705) & -- & -- \\
   $a=5$ & \textbf{6.163} (0.345) & 7.785 (0.345) & -- & -- \\
   $c=0.5$ & \textbf{0.540} (0.078) & 3.263 (0.559) & -- & -- \\
   $\kappa=0.9$ & 0.570 (0.094) & \textbf{0.947} (0.092) & -- & -- \\ 
   RMSPE & \textbf{0.191} (0.003) & 0.406 (0.027) & 0.481 (0.029) & 0.481 (0.029) \\
   MAPE  & \textbf{0.151} (0.002) & 0.274 (0.015) & 0.323 (0.017) & 0.323 (0.017) \\
   95\% CI coverage & \textbf{0.949} (0.005) & 0.945 (0.009) & 0.925 (0.008) & 0.925 (0.009) \\ 
   95\% CI width & \textbf{0.745} (0.007) & 1.724 (0.081) & 1.801 (0.093) & 1.797 (0.093) \\ \hline
   \end{tabular}
   \caption{Simulation results when G-BAG is the true data generating model with parameters $\bm{\theta}_1$. Mean and standard error in parenthesis are calculated over 25 synthetic data sets.}
   \label{tab:sim1_theta1}
\end{table}

The inferred DAG by posterior probabilities from the fitted G-BAG model accurately recovers the true DAG as illustrated in Figure S8 in Supplementary Material S3.1. Boundary partition blocks whose true DAG induces no parent often miss the true arrows but properly with high uncertainty. Prediction and estimation performance of different models are summarized in Table \ref{tab:sim1_theta1} for $\bm{\theta}_1$ and Table S2 for $\bm{\theta}_2$ in Supplementary Material S3.1. The root mean square prediction error (RMSPE), mean absolute prediction error (MAPE), empirical coverage probability and width of 95\% posterior predictive credible intervals (CIs) all indicate large predictive gains of the G-BAG model in this scenario. The nonstationary SPDE barely improves on the stationary SPDE in terms of prediction. These results suggest that when directional associations are present, (a) the nonstationarity specified via a location-specific spatial range does not suffice, and (b) it is recommended to use G-BAGs that explicitly model directional dependence structures. 

The improved performance of G-BAG is also confirmed in parameter estimation. G-BAG accurately estimates the parameters using weakly informative priors. Although the estimate of $\kappa$ seems far from the true value, the true $\kappa$ is included in 95\% CIs by G-BAG for 96\% of the replicates. In contrast, the fixed-DAG and the SPDE models considerably overestimate $\tau^2$. As a result, none of the 95\% CIs of $\tau^2$ by the fixed-DAG and the SPDE models include the true value over the replicates. Furthermore, in the fixed-DAG model, the estimate of $\sigma^2$ across 25 data sets is subject to large uncertainty, and the empirical coverage probability of 95\% CIs of $\sigma^2$ is as low as 0.440. Discussions about Table \ref{tab:sim1_theta1} similarly apply to Table S2. 

We examine robustness of G-BAGs to parent assignment schemes. Our default G-BAG assumes that each partition block $\mc{S}_i$ at time $t$ selects a spatial parent by direction and chooses the partition block covering the same region as $\mc{S}_i$ at $t-1$ as a temporal parent. We fit another G-BAG model, G-BAG*, with another reasonable parent assignment scheme where each partition block at $t$ selects a spatial parent at $t$ by direction and a temporal parent at $t-1$ from the same direction, while all the other modeling assumptions remain the same. Table S2 shows that G-BAG* is as accurate as the default G-BAG in terms of parameter estimation and predictions, even when the data are generated under the default parent assignment scheme.

\subsection{Fitted G-BAG is Misspecified}

We generate data on a regular lattice of size $193\times 193\times 59$ according to $y(\vt{t}) = w(\vt{t})+\epsilon(\vt{t})$ with $\epsilon(\vt{t})\sim N(0,\tau^2)$ and $\tau^2=0.1$. We create $w(\vt{t})$ with directed edges from north fixed over $1 \times 193 \times 59$ partition blocks to mimic a surface under constant wind directions. The base covariance function used in data generation is $\vt{C}(\bm{h},u \mid \bm{\theta}, \nu)$ $=$ $\sigma^2/\{2^{\nu-1}\Gamma(\nu)(a|u|+1)\}\left\{-c\|\bm{h}\|/(a|u|+1)^{\kappa/2}\right\}^{\nu}K_{\nu}\left(-c\|\bm{h}\|/(a|u|+1)^{\kappa/2}\right)$ where $\Gamma(\cdot)$ is the gamma function, $\nu$ is the smoothness parameter of space, and  , as introduced in \cite{gneiting_nonseparable_2002}. Fixing $\nu=1.5$, we generate 25 synthetic data sets with $\bm{\theta}_3=(5, 20, 1, 150)$ and another 25 with $\bm{\theta}_4=(10, 20, 1, 150)$ indicating $a$, $c$, $\kappa$, and $\sigma^2$ in order. Figure S10 in Supplementary Material S3.2 shows the latter mimics faster wind speed than the former from north to south due to the higher temporal decay.

Since the generated data have almost 2.2 million locations, for the purpose of replication, we take a subset of data of size $n$ = 18,750 by retaining $25\times 25 \times 30$ grid points. We fit a misspecified G-BAG model with respect to the domain partition, the bag, and the base covariance function. We implemented the G-BAG model on an axis-parallel partition of size $M=2 \times 6 \times 30=360$. Based on visual inspection of the data, we placed $K=4$ directions from northwest, north, northeast, and east in the bag. The base covariance function in equation \eqref{Gneitingcov} was utilized. For G-BAG and fixed-DAG models, we drew 7,000 MCMC samples, of which 5,000 samples were discarded as burn-ins, and every second sample was saved in the remaining 2,000 draws. Convergence and mixing were adequate.

G-BAG enables inferences on directions along which correlations move across the domain, which are found robust to different parent assignment schemes. Over 85\% of the partition blocks assign the highest posterior probability to the true direction north out of four directions in the bag under both the default G-BAG and G-BAG* with directional temporal parents. The average posterior probability of the north arrow in such blocks is as high as 0.7 under G-BAG and 0.87 under G-BAG*. In terms of prediction errors, all models show similar performance. MAPEs are 0.358, 0.368, 0.368, and 0.376, and RMSPEs are 0.450, 0.462, 0.462, and 0.475 on average by fixed-DAG, G-BAG, G-BAG*, and SPDE-nonstationary models, respectively. Despite misspecifications, we have well-calibrated posteriors under G-BAG models as predictive CIs have a good empirical coverage probability. G-BAG* and the fixed-DAG model have respective empirical coverage probabilities of 0.950 and 0.964 on average for 95\% posterior predictive CIs. The average empirical coverage probability of SPDE models is 0.895. Slightly better predictive performance of the fixed-DAG model than the G-BAG models can be attributable to (a) the data generating DAG being fixed and (b) three parents in the fixed-DAG versus two parents in G-BAGs. It is encouraging that G-BAGs have comparable performance to the fixed-DAG model using less information from parent nodes, by choosing the most appropriate parents based on inferred directions. Prior choices, convergence diagnostics, detailed prediction results and visualization of inferred directions in partition blocks are available in Supplementary Material S3.2. An additional misspecified simulation case in which data are generated from other existing methods of constructing a directional spatiotemporal covariance, not relying on DAGs, can also be found in Supplementary Material S3.2.

\section{Air Quality Data Analysis in California, the United States} \label{sec:anal-CA}

Among many ambient pollutants, fine particulate matter that is 2.5 microns or less in diameter (PM2.5) is the primary concern due to its abundance and association to adverse health effects including heart and lung diseases and associated premature deaths \citep{kloog_long-_2013, gutierrez-avila_cardiovascular_2018}. However, PM2.5 monitoring sites are sparsely located. Each of the 165 monitoring stations of the United States Environmental Protection Agency (EPA) in California as of 2020 cover a large average area of 2,570 $km^2$ and thus cannot be used for accurate estimation of local levels of PM2.5. Such local effects are of great interest during seasonal wildfires.

The spatiotemporal spread of PM2.5 is heavily affected by winds. A large body of literature has included wind-relevant variables in an attempt to predict PM2.5 \citep{wu_exposure_2006, calder_dynamic_2008, wang_effects_2015, preisler_statistical_2015, kleine_deters_modeling_2017, aguilera_santa_2020}. However, there is a fundamental issue in summarizing wind direction at discrete times. The lack of representative values for wind direction leads different data sources to provide different summaries; EPA provides average wind direction, while the National Oceanic and Atmospheric Administration provides direction of the fastest wind. Moreover, due to the volatile nature of winds, naive averages of such wind directions over a given time period may rarely be meaningful. Therefore, we assume a G-BAG prior on the latent process of log transformed PM2.5 to overcome these difficulties and implicitly incorporate wind effects. Since G-BAGs enable different prevailing wind directions to be selected for different regions of a domain, we can flexibly mimic the potential volatility of wind directions which may explain associated local covariance in PM2.5.  

The year 2020 was the largest wildfire season in California's modern history. During fire events, dramatically poor air quality is witnessed due to wildfire emissions in which PM2.5 is the primary pollutant \citep{liu_airborne_2017}. This environmental risk has made California outstanding in its widespread use of low-cost sensors such as PurpleAir; more than half of the PurpleAir sensors in the United States are concentrated in California as of February, 2020 \citep{desouza_distribution_2021}. Therefore, we analyze daily PM2.5 levels in California during fire seasons (August 1 to October 22, 2020) using EPA and PurpleAir measurements. For improved comparability to regulatory monitors, PurpleAir measurements are calibrated using recommended practice by EPA \citep{barkjohn_development_2021}. Details are found in Supplementary Material S4. 

The model is $y(\vt{t}) = x(\vt{t})\beta + w(\vt{t}) + \epsilon(\vt{t})$ where $x(\vt{t})$ is the Euclidean distance to the nearest fire on each day at a spatiotemporal location $\vt{t}$, and $y(\vt{t})$ is the mean centered log(PM2.5). We expect $x(\vt{t})$ to capture elevated PM2.5 level due to proximity to wildfires. The covariate $x(\vt{t})$ is standardized to have mean 0 and standard deviation 1. A total of 110,473 irregular spatiotemporal locations are observed, of which 20\% are held out for validation and the rest serves as reference locations. We additionally predict at 274,564 new locations on a fine grid. The directed edges from west, northwest, north, and northeast are chosen in a bag because California lies within the area of prevailing westerlies. California over 83 days is partitioned by $M=16 \times 20 \times 83$ = 26,560 rectangular cuboids, and each covers approximately $55^2km^2$ a day corresponding to 0.5$^{\circ}$ spatial resolution. We analyze 1,000 posterior samples after 15,000 burn-in and 15 thinning. Based on convergence diagnostics, we confirm adequate convergence and mixing of G-BAG. We compare to fixed-DAG and SPDE-nonstationary models. Justification of using Euclidean distance, detailed prior specifications, and convergence diagnostics are given in Supplementary Material S4.

\begin{figure}[t]
\centering
\includegraphics[height=0.47\textheight]{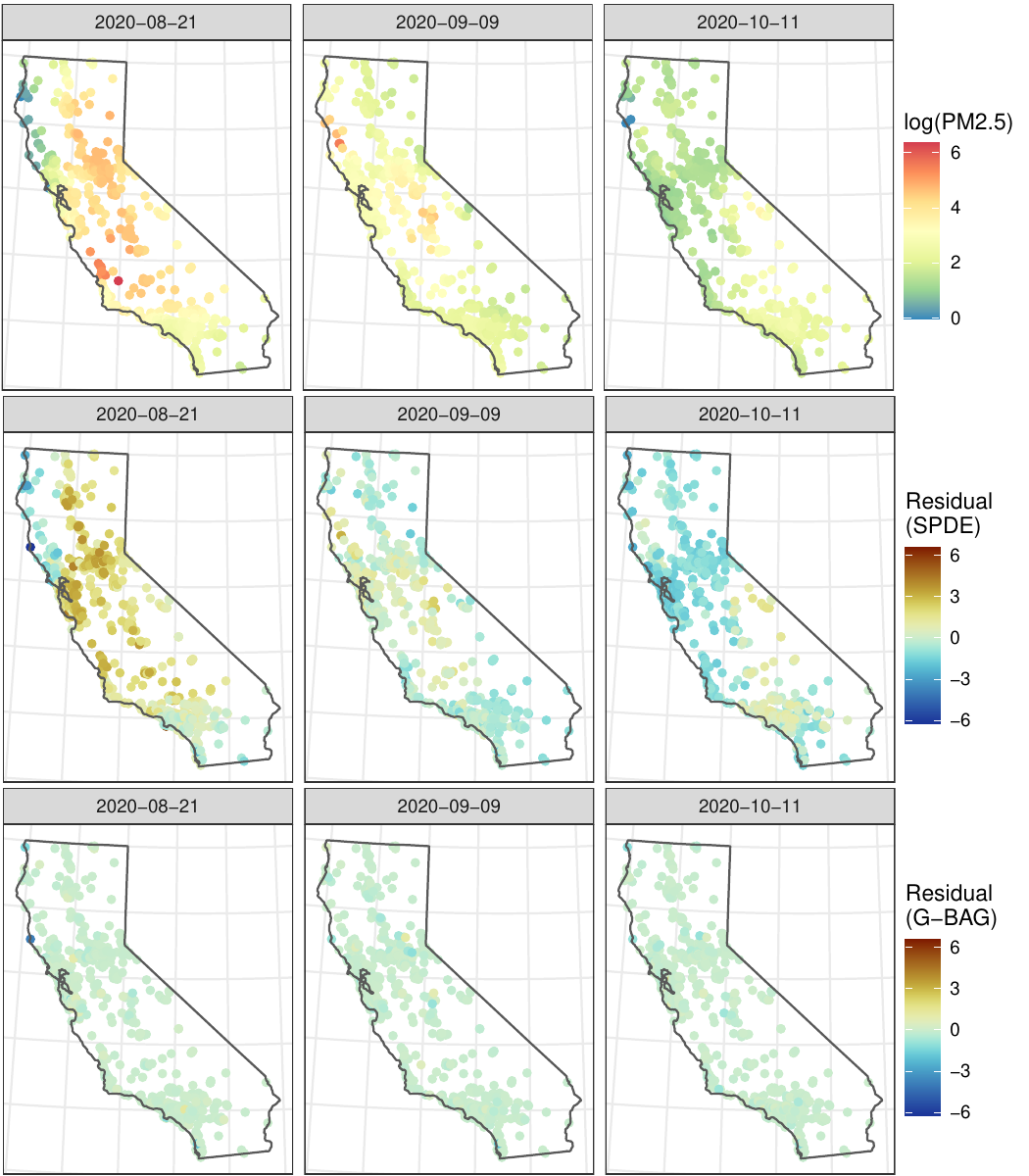}
\caption{Comparison of residuals by SPDE-nonstationary (middle row) and by G-BAG (bottom row) on randomly chosen dates. Residuals are computed for both training and test data as $\vt{y}$-$\hat{\vt{w}}$ where $\hat{\vt{w}}$ indicates fitted latent spatiotemporal random effects by two models. The observed log(PM2.5) in California (top row) is given for reference.}
\label{anal:CA_res}
\end{figure}

Table \ref{tab:anal_CA} summarizes model fitting results. G-BAG and the fixed-DAG model produce similar parameter estimates, and the effect of the distance to the nearest fire appears insignificant in that the 95\% CIs of $\beta$ from both models contain zero. Although the fitted $\beta$ is significantly negative from SPDE-nonstationary, it is likely misled due to unexplained space-time variations after fitting spatiotemporal random effects. Figure \ref{anal:CA_res} illustrates clear spatiotemporal patterns in residuals of the SPDE model. SPDE-nonstationary underestimates PM2.5 when the true PM2.5 is high and overestimates when the true PM2.5 is low. Much larger estimate for $\tau^2$ by the SPDE model than by G-BAG also corroborates remaining variability. These suggest that the nonstationarity via location-specific spatial range is insufficient to fully characterize the cause and removal of air pollution affected by wind directions in California over this time period.

Table \ref{tab:anal_CA} shows that G-BAG yields smaller errors than the other two models and better accuracy of uncertainty quantification in prediction than the fixed-DAG model. A detailed comparison is illustrated in Figure S16 in Supplementary Material S4 by which we confirm that the G-BAG model improves prediction by a large margin at each time point especially over the fixed-DAG model. We achieved up to 50\% reduction in prediction errors. These results signify that the unknown DAG approach and construction of nonstationarity via inferred directions help enhance prediction performance in this application.

\begin{table}[htbp]
    \centering
    \begin{tabular}{cccc}
    \hline
    &  \multirow{2}{*}{G-BAG} & \multirow{2}{*}{Fixed DAG} & SPDE-  \\ [-3pt]
    &   &  & nonstationary  \\ \hline             
   \multirow{2}{*}{$\beta$}  & 0.003 & 0.008 & $-$0.038 \\ [-3pt]
   & ($-$0.011, 0.016) & ($-$0.003, 0.021) & ($-$0.054, $-$0.021) \\ 
   \multirow{2}{*}{$\tau^2$} & 0.011 & 0.011 & 0.079 \\ [-3pt]
   & (0.011, 0.011) & (0.011, 0.011) & (0.076, 0.081) \\
   $\sigma^2$ & 3.781 (3.600, 3.990) & 4.410 (4.410, 4.410) & -- \\
   $a$ & 3.099 (2.963, 3.241) & 1.262 (1.262, 1.262) & -- \\
   $c$ & 0.009 (0.008, 0.009) & 0.010 (0.010, 0.010) & -- \\
   $\kappa$ & 0.011 (0.000, 0.041) & 0.152 (0.152, 0.152) & -- \\ 
   RMSPE & \textbf{0.296} & 0.343 & 0.343\\
   MAPE  & \textbf{0.154} & 0.213 & 0.174\\
   95\% CI coverage & 0.963 & 0.969 & \textbf{0.961}\\ 
   95\% CI width & 1.504 & 1.794 & \textbf{1.216} \\ \hline
   \end{tabular}
   \caption{Posterior summaries and prediction performance measures of G-BAG, fixed-DAG, and SPDE-nonstationary models on California PM2.5 data. Posterior mean and 95\% CI in parenthesis are provided for each parameter.}
   \label{tab:anal_CA}
\end{table}

The predicted surfaces of log(PM2.5) and observations are presented in Figure S17 in Supplementary Material S4. We find that west and north are the dominant prevailing wind directions causing associations in PM2.5 in California from August to October, 2020. The directions from west and north are selected in more than 70\% of the partitioned regions. The bottom row of Figure S17 shows that fluctuations in areas with elevated PM2.5 levels ($>35 \mu g/m^3$) progress from west and north directions, as accurately captured by the posterior mode. Between latitude 36$^{\circ}$N and 40$^{\circ}$N and longitude 121$^{\circ}$W and 124$^{\circ}$W, cleaner air started to flow in from the west on September 16 and expanded through the north--northwest within in a few days. Figure S17 also shows that G-BAG can produce plausible predicted values even in regions lacking data with appropriately increased prediction uncertainty. 

\begin{figure}[htbp]
\centering
\includegraphics[width=\textwidth]{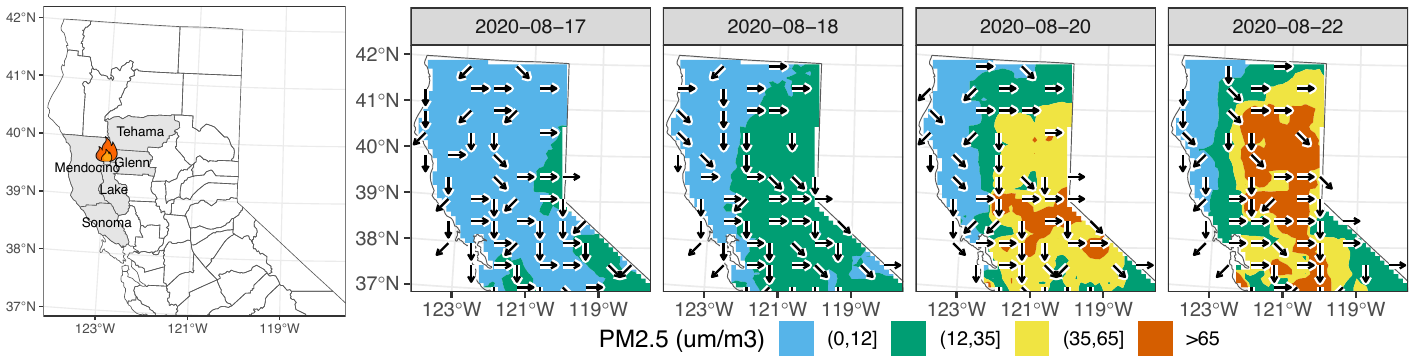}
\caption{Left: Northern California map with counties Mendocino, Lake, Glenn, Tehama, and Sonoma. The fire symbol marks where the August Complex started. Right: Predicted surfaces of discretized PM2.5 in northern California by G-BAG after the start of the August Complex. The posterior mode of wind directions is overlaid.}
\label{anal:CA_pred_AugustComplex}
\end{figure}

We examine mid August of 2020 with a particular interest in the August Complex, the largest wildfire in California's history. After ignition on August 16 around shared boundaries of Mendocino, Lake, Glenn, and Tehama counties, effects of the wildfire on PM2.5 started to appear on August 18. See predicted PM2.5 surfaces by G-BAG in northern California in the right panel of Figure \ref{anal:CA_pred_AugustComplex}, compared to the county map of the same region in the left panel. Due to westerlies, the left side of Mendocino enjoyed better air quality than the right side where PM2.5 from fires accumulated and exceeded 12 $\mu g/m^3$, the annual standard of EPA. Further elevated PM2.5 over the 24-hour EPA standard (35 $\mu g/m^3$) stayed on the same side until August 20 mainly due to directions from northwest and north. On August 22, however, the west Mendocino also had to face poorer ambient conditions because directions from northeast arose around [38$^\circ$N, 40$^\circ$N] at 123$^\circ$W. We also observe the air quality in Sonoma was damaged on August 22 due to constant winds from the north. 

We investigate sensitivity of G-BAGs to the domain partitioning and bag of directions. The first alternative G-BAG (G-BAG1) coarsens the partitioning, keeping every other setting the same; California over 83 days is now partitioned into $9\times 11 \times 83$ = 8,217 rectangular cuboids, each of which covers $\sim 100^2km^2$ per day. The second alternative G-BAG (G-BAG2) chooses directed edges from west, southwest, south, and southeast. 

Table S5 in Supplementary Material S4 demonstrates robustness of G-BAGs to the domain partitioning and bag of directions with respect to parameter estimation and prediction accuracy. All G-BAG models conclude that the coefficient $\beta$ of the distance to the nearest fire is not significantly different from zero and estimate $\tau^2$ to be 0.011 with tight CIs, yielding negligible differences in MAPEs. We also confirmed in Figure S14 in Supplementary Material S4 that G-BAG1 and G-BAG2 show no spatiotemporal patterns in their residual spaces of $\vt{y}$-$\hat{\vt{w}}$ which are almost identical to the residuals of G-BAG in Figure \ref{anal:CA_res}. Although G-BAG1 increases RMSPE by 0.01 compared to the original G-BAG, the RMSPE is still lower than RMSPEs from fixed DAG or SPDE-nonstationary. 

Figure S20 in Supplementary Material S4 presents predicted surfaces of log(PM2.5) by three G-BAG models over the same period as Figure S17. Although G-BAG1 with coarse partitioned regions shows slightly more blocky predictions than the original G-BAG or G-BAG2, the predicted surfaces from all G-BAG models look alike, demonstrating the dominant movement of PM2.5 from west to east. Moreover, despite different partitions used in G-BAG and G-BAG1, the percentages of partitioned regions that chose each direction as the most likely direction are highly similar; (38\%, 14\%, 35\%, 13\%) for the directions from west, northwest, north, and northeast in G-BAG and (41\%, 12\%, 34\%, 13\%) in G-BAG1. Overall, this sensitivity analysis suggests that G-BAGs are highly robust to bag of directions and moderately robust to the domain partitioning. 

\section{Discussion} \label{sec:dis}

We introduced a class of nonstationary processes based on allowing unknown directed edges over a domain partition. We leave great freedom and creativity of designing partition blocks for future users, which can be of varying sizes/shapes and be adaptive to concentration of observations over a domain. That being said, careful consideration of a computational budget and a desirable resolution of directional inferences is required for a reasonable and practical domain partition. For instance, having too many observations in each block, which results in crude partitioning overall, is against our aim to infer directional dependence over the domain and puts high burden on computation (see Supplementary Material S2.7). Conversely, for a fine resolution of directional inference, one may choose to place many partition blocks and even let each partition block contain only one location. However, the computational cost will increase as the number of partition blocks increases. Even with unlimited computational budget, pursuing an excessively high resolution of directional inference is not recommended as such fine partitioning will entail great uncertainty in inferred directions due to little information available from each neighboring block. Therefore, balancing between a desired resolution of directional inference and the number of locations in each partition block is important under a limited computational budget. In some circumstances, placing more small partition blocks in regions with dense data and fewer large partition blocks in regions with sparse data would be beneficial, allowing more detailed inferences on directions in the densely observed regions and less uncertain inferences on directions in the sparsely observed regions.

Our methods have broad applicability. For example, BAGs can simulate realistic air pollution processes under varying wind dynamics, while providing mechanistic insights on how air quality behaves with abnormal events such as fires. This contributes to a body of literature assessing the impacts of the spread of smoke plumes on air quality \citep{wu_exposure_2006, preisler_statistical_2015, aguilera_santa_2020}. More realistic dynamic exposure models can in turn provide improved estimates of individual-level exposures, of critical use in assessing health effects of pollutants based on spatially misaligned data.

In future research, it is of substantial interest to build on our initial BAGs framework in several directions in order to relax modeling assumptions. One way is to reduce sensitivity to the domain partitioning by including several different possibilities within a single ensemble analysis. In addition, we can relax distributional assumption. For example, the GP base process can be replaced with a Student-t process to produce a t-BAGs framework that is heavy-tailed and more robust to outliers. Spatial dependence in directional arrows for neighboring spatial regions can be incorporated, and extensions to multivariate cases can be considered. With additional data such as wind speed, we can also model wind intensity and let covariance be a function of direction and intensity. Another interesting path for future research is to allow arbitrary directions by angles, potentially extending the finite mixture to an infinite mixture. The resulting process will still be valid with the angles constrained on a subset of degrees such as $(0^{\circ}, 180^{\circ}]$.

\section*{Acknowledgement} 
We are grateful for the financial support from the National Institute of Environmental Health Sciences through grants R01ES027498, R01ES028804, and R01ES035625 and from the European Research Council under the European Union’s Horizon 2020 research and innovation programme (grant agreement No 856506). We would like to thank David Buch for helpful discussions.  

\newpage
\begin{center}
{\spacingset{1} \LARGE\bf Supporting Materials for\\``\articletitle''}
\end{center}
\setcounter{section}{0}
\setcounter{figure}{0}
\setcounter{table}{0}
\setcounter{proposition}{0}
\setcounter{theorem}{0}
\setcounter{remark}{0}
\renewcommand{\thesection}{S\arabic{section}}
\renewcommand{\thefigure}{S\arabic{figure}}
\renewcommand{\thetable}{S\arabic{table}}
\renewcommand{\theproposition}{S\arabic{proposition}}
\renewcommand{\thetheorem}{S\arabic{theorem}}
\renewcommand{\theremark}{S\arabic{remark}}

\section{Useful Links} \label{s:links}

An \texttt{R} package \texttt{bags} is available at \url{https://github.com/jinbora0720/bags}. The code to reproduce all analyses in this paper is provided at \url{https://github.com/jinbora0720/GBAGs}. 

\section{Spatiotemporal Process Modeling Using BAGs}

\subsection{Potential Pitfalls of Fixed-DAG Models}
\label{s:pitfall}


\begin{figure}[hbpt]
\centering
\includegraphics[width=0.85\textwidth]{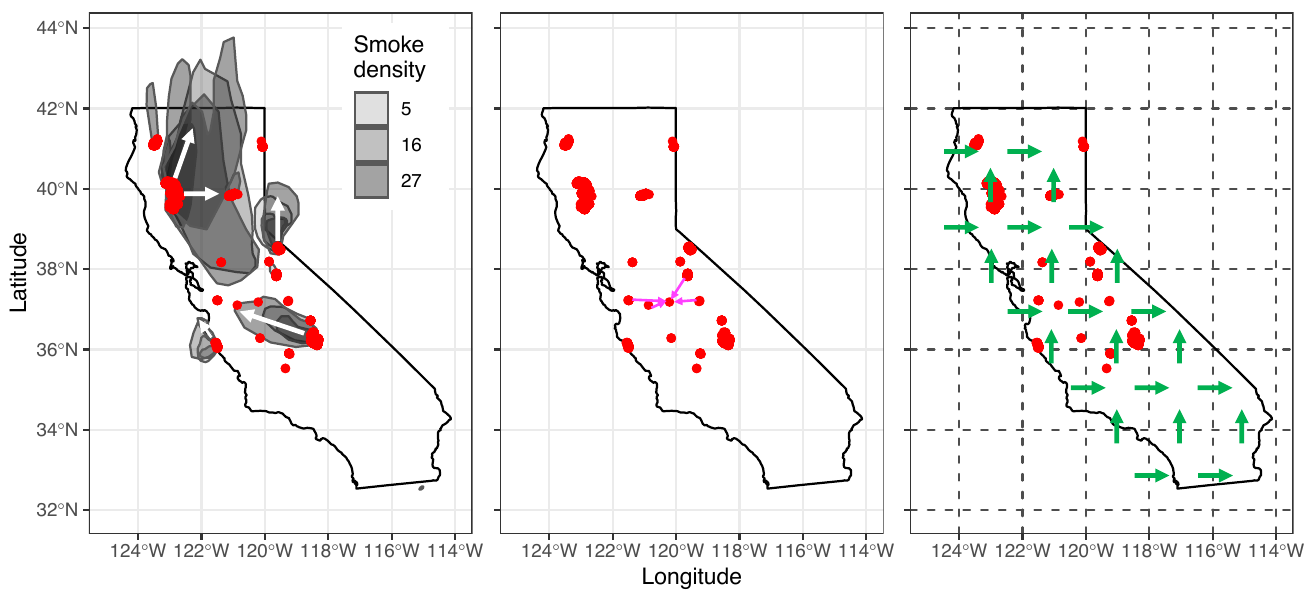}
\caption{Left: forest fire centers as red dots and heavy smoke areas in California, September 4, 2020. White arrows of arbitrary lengths visually assist to indicate directions of smoke spread. Center: four nearest neighbors from NNGP connected to a location at around (37$^{\circ}$N, 120$^{\circ}$W) by pink arrows. Right: a cubic mesh from a meshed GP as green arrows with dashed regions by a rectangular partition overlaid.}
\label{mot:CAfire2}
\end{figure}

\begin{proof}[Proof of Proposition~\ref{prop:KL}]
Consider a random vector $\vt{w}$ with a base density $p$, separated into $M$ disjoint sets of locations. Then $p^*_0(\vt{w}) = \prod_{i=1}^M p(\vt{w}_i\mid \vt{w}_{[i]_0})$ where $\vt{w}_{[i]_0}$ is a subset of $\vt{w}$ corresponding to $[v_i]_0$. 

We now determine which density between $p^*_1$ and $p^*_2$ approximates $p^*_0$ more accurately in terms of the Kullback-Leibler divergence. The Kullback-Leibler divergence from $p^*_1$ to $p^*_0$ is defined as $\mbox{KL}(p^*_0 \| p^*_1) = \int \log\{p^*_0(\vt{w})/p^*_1(\vt{w})\}p^*_0(\vt{w})d\vt{w}$. Since $p^*_1$ and $p^*_2$ differ only for the $j$th node, 
\begin{align}
    \frac{p^*_1(\vt{w})}{p^*_2(\vt{w})} = \frac{p(\vt{w}_j\mid \vt{w}_{[j]_1})}{p(\vt{w}_j\mid \vt{w}_{[j]_2})} = \frac{p(\vt{w}_j\mid \vt{w}_{[j]_{1\setminus 2}}, \vt{w}_{[j]_{1\cap 2}})}{p(\vt{w}_j\mid \vt{w}_{[j]_{2\setminus 1}}, \vt{w}_{[j]_{1\cap 2}})}, \label{KL_p1:p2}
\end{align}
letting $[v_j]_{1 \cap 2} = [v_j]_{1} \cap [v_j]_{2}$, $[v_j]_{1\setminus 2} = [v_j]_{1} \setminus [v_j]_{2}$, and $[v_j]_{2\setminus 1} = [v_j]_{2} \setminus [v_j]_{1}$. 

With a subset of $\vt{w}$ corresponding to $V_r = V\setminus \{v_j \cup [v_j]_{1} \cup [v_j]_{2}\}$ denoted by $\vt{w}_{r}$, 
\begin{align}
    &\mbox{KL}(p^*_0 \| p^*_2) - \mbox{KL}(p^*_0 \| p^*_1) \nonumber \\ 
    &= \int \log \left\{\frac{p^*_1(\vt{w})}{p^*_2(\vt{w})}\right\} p^*_0(\vt{w})d\vt{w} \nonumber \\
    &= \int \log \left\{\frac{p(\vt{w}_j\mid \vt{w}_{[j]_{1\setminus 2}}, \vt{w}_{[j]_{1\cap 2}})}{p(\vt{w}_j\mid \vt{w}_{[j]_{2\setminus 1}}, \vt{w}_{[j]_{1\cap 2}})}\right\} p^*_0(\vt{w}_j, \vt{w}_{[j]_{1\setminus 2}}, \vt{w}_{[j]_{2\setminus 1}}, \vt{w}_{[j]_{1\cap 2}}, \vt{w}_{r})d\vt{w} \nonumber \\ 
    &= \int \log \left\{\frac{p(\vt{w}_j\mid \vt{w}_{[j]_{1\setminus 2}}, \vt{w}_{[j]_{1\cap 2}})}{p(\vt{w}_j\mid \vt{w}_{[j]_{2\setminus 1}}, \vt{w}_{[j]_{1\cap 2}})}\right\} p^*_0(\vt{w}_j, \vt{w}_{[j]_{1\setminus 2}}, \vt{w}_{[j]_{2\setminus 1}}, \vt{w}_{[j]_{1\cap 2}})d\vt{w}_j d\vt{w}_{[j]_{1\setminus 2}} d\vt{w}_{[j]_{2\setminus 1}} d\vt{w}_{[j]_{1\cap 2}} \nonumber \\
    &=\int -\log \left\{\frac{p(\vt{w}_j\mid \vt{w}_{[j]_{2\setminus 1}}, \vt{w}_{[j]_{1\cap 2}})}{p(\vt{w}_j\mid \vt{w}_{[j]_{1\setminus 2}}, \vt{w}_{[j]_{1\cap 2}})}\right\} p^*_0(\vt{w}_j\mid \vt{w}_{[j]_{1\setminus 2}}, \vt{w}_{[j]_{2\setminus 1}}, \vt{w}_{[j]_{1\cap 2}}) \times \nonumber \\
    &~~~~~~~~~~~p^*_0(\vt{w}_{[j]_{1\setminus 2}}, \vt{w}_{[j]_{2\setminus 1}}, \vt{w}_{[j]_{1\cap 2}})d\vt{w}_j d\vt{w}_{[j]_{1\setminus 2}} d\vt{w}_{[j]_{2\setminus 1}} d\vt{w}_{[j]_{1\cap 2}} \nonumber \\
    &> -\log \left[\int \left\{\frac{p(\vt{w}_j\mid \vt{w}_{[j]_{2\setminus 1}}, \vt{w}_{[j]_{1\cap 2}})}{p(\vt{w}_j\mid \vt{w}_{[j]_{1\setminus 2}}, \vt{w}_{[j]_{1\cap 2}})}\right\} p(\vt{w}_j\mid \vt{w}_{[j]_{1\setminus 2}}, \vt{w}_{[j]_{1\cap 2}}) p^*_0(\vt{w}_{[j]_{2\setminus 1}}, \vt{w}_{[j]_{1\cap 2}})d\vt{w}_j d\vt{w}_{[j]_{2\setminus 1}} d\vt{w}_{[j]_{1\cap 2}} \right] \nonumber \\
    &= -\log \left\{\int p(\vt{w}_j\mid \vt{w}_{[j]_{2\setminus 1}}, \vt{w}_{[j]_{1\cap 2}}) p(\vt{w}_{[j]_{2\setminus 1}}, \vt{w}_{[j]_{1\cap 2}})d\vt{w}_j  d\vt{w}_{[j]_{2\setminus 1}} d\vt{w}_{[j]_{1\cap 2}}\right\} = 0\nonumber
\end{align}
by equation \eqref{KL_p1:p2} and Jensen's inequality with a strictly convex function $-\log(\cdot).$ Therefore, building conditioning sets based on the wrong directions worsens the accuracy in approximating the target density $p^*_0$. 
\end{proof}

Figure \ref{mot:CAfire2} demonstrates how fixed DAGs will poorly explain the local dynamics in the spread of wildfire smoke and air pollutants. The two given examples of fixed DAGs include parents from wrong directions. In the center panel, three out of four NNGP neighbors are outside the smoke above the location at (37$^{\circ}$N, 120$^{\circ}$W) and may be undesirably non-informative despite proximity. Similarly, in the right panel, a cubic-meshed GP model fails to capture locally varying associations due to the fixed and repeated pattern in its DAG. This mismatch between edges in fixed DAGs and true directions in data will be inevitable and critical for inference if directional dependence in the data is important. 

\subsection{Coherent Processes With Unknown DAGs}
\label{s:proof}

\begin{proof}[Proof of Proposition~\ref{prop:jointdensity}]
The joint density in equation \eqref{ptilde_cond} is proper because there exists an associated DAG $\tilde{\mc{G}}_{\mc{S}}(Z)$ whose nodes are $A$ and edges are identified by $Z$. Then
\begin{align*}
    \int \tilde{p}(\vt{w}_{\mc{S}}) d\vt{w}_{\mc{S}} 
    &= \int \left\{\sum_{Z}\tilde{p}(\vt{w}_{\mc{S}}\mid Z)q(Z)\right\} d\vt{w}_{\mc{S}} \\
    &= \sum_{Z} \left\{\int\tilde{p}(\vt{w}_{\mc{S}}\mid Z)d\vt{w}_{\mc{S}}\right\}q(Z) \\
    &=\sum_{Z}q(Z)\\&=\sum_{z_1=1}^K\cdots\sum_{z_M=1}^K\{q_1(z_1)\times\cdots \times q_M(z_M)\}\\
    &=\sum_{z_1=1}^K q_1(z_1)\times\cdots\times\sum_{z_M=1}^K q_M(z_M) =1
\end{align*}
because $\sum_{z_i=1}^K q_i(z_i)=1$ for all $i=1,\dots,M$. Hence, $\tilde{p}(\vt{w}_{\mc{S}})$ is a proper joint density. 
\end{proof}

\begin{proof}[Proof of Proposition~\ref{prop:uniqueDAG}]
Suppose there exists a DAG $\mc{G}'$ such that $\mc{G}' \neq \mc{G}$ and $\mc{G}'$ is Markov equivalent to the estimated DAG $\mc{G}$. By definition of Markov equivalence class, the skeleton of $\mc{G}'$ is equal to the skeleton of $\mc{G}$. The skeleton of an arbitrary graph is obtained with the same vertices and edges, by replacing any directed edges with undirected ones. Since only one direction is available in the bag on each axis to ensure acyclicity, $\mc{G}'$ must be $\mc{G}$. Therefore, the Markov equivalence class of $\mc{G}$ is of size 1. 
\end{proof}

\begin{proof}[Proof of Proposition~\ref{prop:Kolmogorov}]
Kolmogorov consistency conditions indicate consistency with (a) permutation of indices and (b) marginalization. First, we show that $\tilde{p}(\vt{w}_{\mc{L}}) = \tilde{p}(\vt{w}_{\mc{L}_{\pi}})$ for any permutation $\mc{L}_{\pi}$ of $\mc{L}$. Let $\mc{U}_{\mc{L}_{\pi}} = \mc{L}_{\pi}\setminus\mc{S}$ with the fixed reference set $\mc{S}$. Given a fixed partition,
\begin{align*}
    \tilde{p}(\vt{w}_{\mc{L}}) &= \int \left\{ \sum_{Z}\tilde{p}(\vt{w}_{\mc{U}_{\mc{L}}}\mid \vt{w}_{\mc{S}},Z)\tilde{p}(\vt{w}_{\mc{S}}\mid Z)q(Z)\right\}\prod_{\vt{s}_i\in\mc{S}\setminus\mc{L}}d\{w(\vt{s}_i)\} \\
    &= \int \sum_{Z}\left( \prod_{i=1}^M \left[ \prod_{\vt{u}\in\mc{U}_{\mc{L}_i}}p\{\vt{w}(\vt{u})\mid \vt{w}_{i},\vt{w}_{[i\mid z_i]}\}p(\vt{w}_i\mid \vt{w}_{[i\mid z_i]})q_i(z_i)\right]\right)\prod_{\vt{s}_i\in\mc{S}\setminus\mc{L}}d\{w(\vt{s}_i)\} \\
    &= \int \sum_{Z}\left( \prod_{i=1}^M \left[ \prod_{\vt{u}\in\mc{U}_{\mc{L}_{\pi_i}}}p\{\vt{w}(\vt{u})\mid \vt{w}_{i},\vt{w}_{[i\mid z_i]}\}p(\vt{w}_i\mid \vt{w}_{[i\mid z_i]})q_i(z_i)\right]\right)\prod_{\vt{s}_i\in\mc{S}\setminus\mc{L}_{\pi}}d\{w(\vt{s}_i)\} \\
    &=\int \left\{ \sum_{Z}\tilde{p}(\vt{w}_{\mc{U}_{\mc{L}_{\pi}}}\mid \vt{w}_{\mc{S}},Z)\tilde{p}(\vt{w}_{\mc{S}}\mid Z)q(Z)\right\}\prod_{\vt{s}_i\in\mc{S}\setminus\mc{L}_{\pi}}d\{w(\vt{s}_i)\} =\tilde{p}(\vt{w}_{\mc{L}_{\pi}})
\end{align*}
because $\mc{U}_{\mc{L}}$ and $\mc{U}_{\mc{L}_{\pi}}$, $\mc{S}\setminus\mc{L}$ and $\mc{S}\setminus\mc{L}_{\pi}$ share the same collection of elements, respectively. Second, we show that $\tilde{p}(\vt{w}_{\mc{L}}) = \int \tilde{p}(\vt{w}_{\mc{L}\cup\{\bm{l}_0\}})dw(\bm{l}_0)$ for a new location $\bm{l}_0 \in \mc{D}\setminus \mc{L}$. Take $\mc{L}_1 = \mc{L}\cup\{\bm{l}_0\}$. We consider two cases: when $\bm{l}_0\in\mc{S}$ and when $\bm{l}_0\notin\mc{S}$. If $\bm{l}_0\in\mc{S}$, then $\mc{U}_{\mc{L}}=\mc{L}\setminus\mc{S} = \mc{L}_1\setminus\mc{S}= \mc{U}_{\mc{L}_1}$ and $\mc{S}\setminus\mc{L} = (\mc{S}\setminus\mc{L}_1)\cup\{\bm{l}_0\}$, resulting in
\begin{align*}
    \int \tilde{p}(\vt{w}_{\mc{L}\cup\{\bm{l}_0\}})dw(\bm{l}_0) &= \int \left[\int \left\{ \sum_{Z}\tilde{p}(\vt{w}_{\mc{U}_{\mc{L}_1}}\mid \vt{w}_{\mc{S}},Z)\tilde{p}(\vt{w}_{\mc{S}}\mid Z)q(Z)\right\}\prod_{\vt{s}_i\in\mc{S}\setminus\mc{L}_1}d\{w(\vt{s}_i)\} \right]dw(\bm{l}_0) \\
    &=\int \left\{ \sum_{Z}\tilde{p}(\vt{w}_{\mc{U}_{\mc{L}}}\mid \vt{w}_{\mc{S}},Z)\tilde{p}(\vt{w}_{\mc{S}}\mid Z)q(Z)\right\}\prod_{\vt{s}_i\in\mc{S}\setminus\mc{L}}d\{w(\vt{s}_i)\} =\tilde{p}(\vt{w}_{\mc{L}}).
\end{align*}
If $\bm{l}_0\notin\mc{S}$, then $\mc{U}_{\mc{L}}\cup\{\bm{l}_0\}=\mc{U}_{\mc{L}_1}$ and $\mc{S}\setminus\mc{L} = \mc{S}\setminus\mc{L}_1$, yielding 
\begin{align*}
    &\int \tilde{p}(\vt{w}_{\mc{L}\cup\{\bm{l}_0\}})dw(\bm{l}_0) \\
    = & \int \left[\int \left\{ \sum_{Z}\tilde{p}(\vt{w}_{\mc{U}_{\mc{L}_1}}\mid \vt{w}_{\mc{S}},Z)\tilde{p}(\vt{w}_{\mc{S}}\mid Z)q(Z)\right\}\prod_{\vt{s}_i\in\mc{S}\setminus\mc{L}_1}d\{w(\vt{s}_i)\} \right]dw(\bm{l}_0)\\
    = & \int \left(\int \left[ \sum_{Z}\tilde{p}\{w(\bm{l}_0)\mid \vt{w}_{\mc{S}},Z\}\tilde{p}(\vt{w}_{\mc{U}_{\mc{L}}}\mid \vt{w}_{\mc{S}},Z)\tilde{p}(\vt{w}_{\mc{S}}\mid Z)q(Z)\right]\prod_{\vt{s}_i\in\mc{S}\setminus\mc{L}}d\{w(\vt{s}_i)\} \right)dw(\bm{l}_0) \\
    = & \int \sum_{Z} \left[\int  \tilde{p}\{w(\bm{l}_0)\mid \vt{w}_{\mc{S}},Z\}dw(\bm{l}_0)\right] \tilde{p}(\vt{w}_{\mc{U}_{\mc{L}}}\mid \vt{w}_{\mc{S}},Z)\tilde{p}(\vt{w}_{\mc{S}}\mid Z)q(Z)\prod_{\vt{s}_i\in\mc{S}\setminus\mc{L}}d\{w(\vt{s}_i)\} \\
    = & \int \left\{\sum_{Z} 1\times \tilde{p}(\vt{w}_{\mc{U}_{\mc{L}}}\mid \vt{w}_{\mc{S}},Z)\tilde{p}(\vt{w}_{\mc{S}}\mid Z)q(Z)\right\}\prod_{\vt{s}_i\in\mc{S}\setminus\mc{L}}d\{w(\vt{s}_i)\} =\tilde{p}(\vt{w}_{\mc{L}})
\end{align*}
where $\tilde{p}(\vt{w}_{\mc{U}_{\mc{L}_1}}\mid \vt{w}_{\mc{S}},Z) =\tilde{p}\{w(\bm{l}_0)\mid \vt{w}_{\mc{U}_{\mc{L}}},\vt{w}_{\mc{S}},Z\}\tilde{p}(\vt{w}_{\mc{U}_{\mc{L}}}\mid \vt{w}_{\mc{S}},Z)= \tilde{p}\{w(\bm{l}_0)\mid \vt{w}_{\mc{S}},Z\}\tilde{p}(\vt{w}_{\mc{U}_{\mc{L}}}\mid \vt{w}_{\mc{S}},Z)$ due to the conditional independence assumption at non-reference locations given the reference set. A finite sum is interchangeable with an integral.
\end{proof}

\subsection{Conditional Precision Matrix of Reference Locations in G-BAGs}
\label{s:Ctildez}

We discuss properties of $\tilde{C}^{-1}_{Z}$, the precision matrix of $\vt{w}_{\mc{S}}$ conditional on $Z$. Simple linear algebra implies that $\tilde{C}^{-1}_{Z}$ should be $(I_k-H_{Z})^TR_{Z}^{-1}(I_k-H_{Z})$ with the identity matrix $I_k$ of size $k$ and a block-diagonal matrix $R_{Z} = \mbox{diag}(R_{1\mid z_1},\cdots,R_{M\mid z_M})$. $H_{Z}$ is a $k\times k$ sparse block matrix whose $i$th block-row has $H_{i\mid z_i}$ for locations mapped to $[a_i\mid z_i]$ and zero otherwise. For instance, if $[a_i\mid z_i] = \{a_j, a_l\}$ for $j<l$ then $H_{i\mid z_i}=[H_{i,j}; H_{i,l}]$. Given a bag of directions and a partition, it is always possible to find an ordering of $\{1,\dots,M\}$ such that $[a_i\mid z_i]\subset\{a_1,\dots,a_{i-1}\}$ for $i=1,\dots,M$, resulting in a lower triangular matrix $H_{Z}$ with zero diagonals. This renders $|I_k-H_{Z}| = 1$, and thus $|\tilde{C}_{Z}| = \prod_{i=1}^M|R_{i\mid z_i}|.$ 

Furthermore, the $(i,j)$th block of $\tilde{C}^{-1}_{Z}$ for $i\geq j$ is 
\begin{align}
    \tilde{C}^{-1}_{Z}(i,j) &= \left\{
    \begin{array}{ll}
     R_{i|z_i}^{-1}+\sum_{l:[a_l|z_l]\ni a_i}H_{l,i}^TR_{l|z_l}^{-1}H_{l,i}, &\text{if } i=j\\
     -R_{i|z_i}^{-1}H_{i,j}, &\text{if } a_j\in[a_i|z_i] \\
     H_{l,i}^TR_{l|z_l}^{-1}H_{l,j}, &\text{if } \{a_i,a_j\}\subset [a_l|z_l] \\
     \vt{0}, &\text{otherwise}
    \end{array} \right. 
    \label{s:Ctildeinv}
\end{align} 
with $\tilde{C}^{-1}_{Z}(j,i) = (\tilde{C}^{-1}_{Z}(i,j))^T$ by symmetry. The off-diagonal sparsity structure of $\tilde{C}^{-1}_{Z}$ is governed by the undirected moral graph of $\tilde{\mc{G}}_{\mc{S}}(Z)$, inducing a non-zero block at $(i,j)$ (a) if $a_i$ is a parent of $a_j$ or vice versa, or (b) if $a_i$ and $a_j$ have a common child $a_l$, i.e., $\{a_i,a_j\} \subset [a_l \mid z_l]$.


\subsection{Conditional Nonstationary Covariance Function of G-BAGs}
\label{s:condCtilde}

The conditional nonstationary covariance function in equation \eqref{Ctilde} is computed as follows. Functions cov$_{\tilde{p}}$ and $E_{\tilde{p}}$ implicitly depend on $\bm{\theta}$. When $\bm{l}_1=\vt{s}_i\in\mc{S}$ and $\bm{l}_2=\vt{s}_j\in\mc{S}$, the result is trivial. When $\bm{l}_1$ is not in $\mc{S}$ but $\bm{l}_2$ is, so that $\bm{l}_1\in\mc{U}_i$ and $\bm{l}_2=\vt{s}_j\in\mc{S}$, 
\begin{align*}
    \mbox{cov}_{\tilde{p}}&\{w(\bm{l}_1), w(\vt{s}_j)\mid Z\} \\ &=  E_{\tilde{p}}[\mbox{cov}_{\tilde{p}}\{w(\bm{l}_1), w(\vt{s}_j)\mid \vt{w}_{\mc{S}}, Z\}\mid Z] +  \mbox{cov}_{\tilde{p}}[E_{\tilde{p}}\{w(\bm{l}_1)\mid \vt{w}_{\mc{S}}, Z\}, E_{\tilde{p}}\{w(\vt{s}_j)\mid \vt{w}_{\mc{S}}, Z\}\mid Z] \\
    &= E_{\tilde{p}}(0\mid Z) +  \mbox{cov}_{\tilde{p}}\{H_{\bm{l}_1\mid z_i}\vt{w}_{[\bm{l}_1\mid z_i]}, w(\vt{s}_j)\mid Z\}  =H_{\bm{l}_1\mid z_i}\tilde{C}_{[\bm{l}_1\mid z_i],s_j}.
\end{align*}
When neither $\bm{l}_1$ nor $\bm{l}_2$ is in $\mc{S}$, which is $\bm{l}_1\in\mc{U}_i$ and $\bm{l}_2\in\mc{U}_j$ for some $i,j\in \{1,\dots,M\}$,
\begin{align*}
    \mbox{cov}_{\tilde{p}}&\{w(\bm{l}_1), w(\bm{l}_2)\mid Z\} \\ &=  E_{\tilde{p}}[\mbox{cov}_{\tilde{p}}\{w(\bm{l}_1), w(\bm{l}_2)\mid \vt{w}_{\mc{S}}, Z\}\mid Z] +  \mbox{cov}_{\tilde{p}}[E_{\tilde{p}}\{w(\bm{l}_1)\mid \vt{w}_{\mc{S}}, Z\}, E_{\tilde{p}}\{w(\bm{l}_2)\mid \vt{w}_{\mc{S}}, Z\}\mid Z] \\
    &= \bm{1}(\bm{l}_1=\bm{l}_2)R_{\bm{l}_1\mid z_i} +  \mbox{cov}_{\tilde{p}}(H_{\bm{l}_1\mid z_i}\vt{w}_{[\bm{l}_1\mid z_i]}, H_{\bm{l}_2\mid z_j}\vt{w}_{[\bm{l}_2\mid z_j]}\mid Z) \\
    &=\bm{1}(\bm{l}_1=\bm{l}_2)R_{\bm{l}_1\mid z_i} + H_{\bm{l}_1\mid z_i}\tilde{C}_{[\bm{l}_1\mid z_i],[\bm{l}_2\mid z_j]}H_{\bm{l}_2\mid z_j}^T.
\end{align*}

Assume that the base covariance function $\vt{C}(\bm{\theta})$ is continuous. Due to domain partitioning in BAGs, $\tilde{\vt{C}}(\bm{l}_1,\bm{l}_2\mid Z, \bm{\theta})$ is continuous if $\bm{l}_1$ and $\bm{l}_2$ are in the same partitioned region and discontinuous if the locations span different regions. See Supplementary Material Section B of \cite{peruzzi_highly_2022} for a proof. 

\subsection{Directional nonstationarity of G-BAGs}
\label{s:nonstationarity}

\begin{figure}[htbp]
\centering
\includegraphics[width=0.25\textwidth]{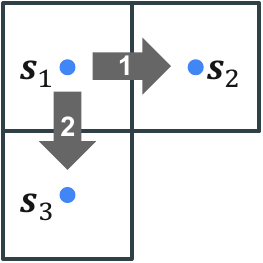}
\caption{Example of the setup in Proposition \ref{prop:nonstationary}.}
\label{fig:prop4}
\end{figure}

\begin{proof}[Proof of Proposition~\ref{prop:nonstationary}]
Let reference locations $\vt{s}_1$, $\vt{s}_2$, and $\vt{s}_3$ form partition blocks 1, 2, and 3, respectively. Define direction 1 to assign a directed edge from $\vt{s}_1$ to $\vt{s}_2$ and direction 2 to assign a directed edge from $\vt{s}_1$ to $\vt{s}_3$. Then $Z_1 = (z_1=1,z_2=1,z_3=1)$ and $Z_2 = (z_1=2,z_2=2,z_3=2)$. Figure \ref{fig:prop4} illustrates example setups of reference locations, partitioning, and directed edges relevant to this proposition. We drop $\bm{\theta}$ for brevity in this proof. By equation \eqref{marginal_Ctilde}, $\tilde{\vt{C}}(\vt{s}_1, \vt{s}_i) = \pi\tilde{\vt{C}}(\vt{s}_1, \vt{s}_i \mid Z_1) + (1-\pi)\tilde{\vt{C}}(\vt{s}_1, \vt{s}_i \mid Z_2)$ for $i=2,3$. Using equations \eqref{Ctilde}, \eqref{s:Ctildeinv}, and simple algebra for $3\times 3$ matrices, 
\begin{align}
    \tilde{\vt{C}}(\vt{s}_1, \vt{s}_2 \mid Z_1) =-\frac{d}{ab-d^2}, \quad \tilde{\vt{C}}(\vt{s}_1, \vt{s}_3 \mid Z_2) =-\frac{h}{eg-h^2} \label{proofs:condcov}
\end{align} where \[\tilde{C}^{-1}_{Z_1} = \begin{bmatrix} a & d & 0 \\ d & b & 0 \\ 0 & 0 & c\end{bmatrix}, \quad \tilde{C}^{-1}_{Z_2} = \begin{bmatrix} e & 0 & h \\ 0 & f & 0 \\ h & 0 & g\end{bmatrix}\] whose elements are identified as $a = R_{1|z_1=1}^{-1}+H^2_{2|z_2=1}R^{-1}_{2|z_2=1}$, $b = R^{-1}_{2|z_2=1}$, $d = -R^{-1}_{2|z_2=1}H_{2|z_2=1}$, $e = R_{1|z_1=2}^{-1}+H^2_{3|z_3=2}R^{-1}_{3|z_3=2}$, $g = R^{-1}_{3|z_3=2}$, and $h = -R^{-1}_{3|z_3=2}H_{3|z_3=2}$. Simplifying the right hand side of two equations in \eqref{proofs:condcov}, we observe $\tilde{\vt{C}}(\vt{s}_1, \vt{s}_2 \mid Z_1)=\vt{C}(\vt{s}_1, \vt{s}_2)$ and $\tilde{\vt{C}}(\vt{s}_1, \vt{s}_3 \mid Z_2)=\vt{C}(\vt{s}_1, \vt{s}_3)$ for the base covariance function $\vt{C}$. In addition, $\tilde{\vt{C}}(\vt{s}_1,\vt{s}_2 \mid Z_2)$ is zero because $\vt{s}_1$ and $\vt{s}_2$ are not a parent of the other or have a common child under $Z_2$. Similarly, $\tilde{\vt{C}}(\vt{s}_1,\vt{s}_3 \mid Z_1)$  is also zero. Putting all together, $\tilde{\vt{C}}(\vt{s}_1, \vt{s}_2) = \pi\vt{C}(\vt{s}_1, \vt{s}_2)$ and $\tilde{\vt{C}}(\vt{s}_1, \vt{s}_3) = (1-\pi)\vt{C}(\vt{s}_1, \vt{s}_3)$. Hence, $\tilde{\vt{C}}(\vt{s}_1, \vt{s}_2) - \tilde{\vt{C}}(\vt{s}_1, \vt{s}_3)$ is a function of $\pi$ and the base covariance function. If $\vt{C}(\cdot, \cdot)$ is isotropic and a lag between $\vt{s}_1$ and $\vt{s}_2$ is identical to that between $\vt{s}_1$ and $\vt{s}_3$, then $\tilde{\vt{C}}(\vt{s}_1, \vt{s}_2) - \tilde{\vt{C}}(\vt{s}_1, \vt{s}_3) = (2\pi-1)c^*$ with $c^* = \vt{C}(\vt{s}_1, \vt{s}_2) = \vt{C}(\vt{s}_1, \vt{s}_3) > 0$. Therefore, $\tilde{\vt{C}}(\vt{s}_1, \vt{s}_2) > \tilde{\vt{C}}(\vt{s}_1, \vt{s}_3)$ if $\pi$ is larger than 0.5.
\end{proof}

\begin{figure}
\centering
\includegraphics[width=0.95\textwidth]{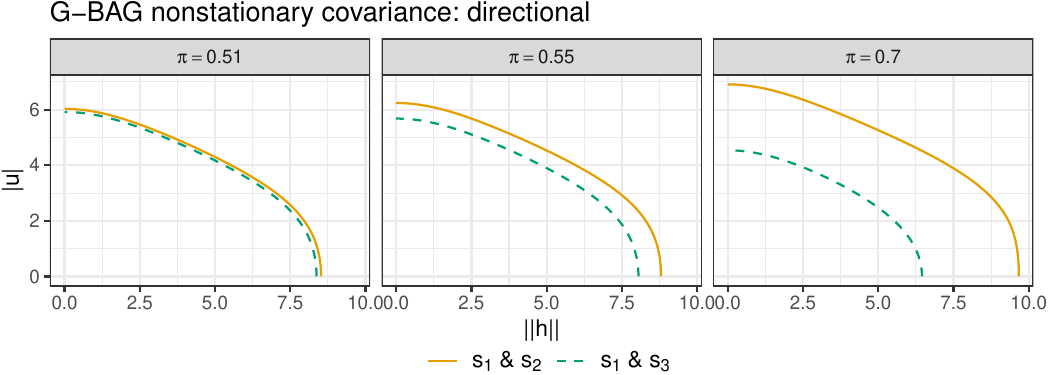}
\caption{Comparison of G-BAG induced covariance for two pairs of locations, $\vt{s}_1 \& \vt{s}_2$ and $\vt{s}_1 \& \vt{s}_3$, as specified in Proposition \ref{prop:nonstationary}. Contour lines are drawn at covariance of 0.1.}
\label{sim:cov0}
\end{figure}

Figure \ref{sim:cov0} compares contour lines of $\tilde{\vt{C}}(\vt{s}_1, \vt{s}_2\mid \bm{\theta})$ and $\tilde{\vt{C}}(\vt{s}_1, \vt{s}_3\mid \bm{\theta})$ for different lags and values of $\pi$. The following fully symmetric covariance function in \cite{apanasovich_cross-covariance_2010} is chosen as a base covariance for illustration: $\vt{C}(\bm{h},u \mid \bm{\theta})$ $=$ $\sigma^2/\{(a|u|^2+1)^{\kappa}(a\|\bm{h}\|^2+1)^{\kappa/2}\}\exp\left\{-c\|\bm{h}\|^2/(a|u|^2+1)^{\kappa}-c|u|^2\right\}$ with $\sigma^2 = 1$, $a = 0.07$, $c = 0.01$, and $\kappa = 1$ to show resulting asymmetry is uniquely due to BAGs. Figure \ref{sim:cov0} indeed shows directional behaviors in the induced covariance: the Euclidean distance of spatiotemporal lags should be smaller in the less likely direction (dashed lines in Figure \ref{sim:cov0}) than in the more likely direction (solid lines) to attain the same level of correlation, and the gap between the distances increases as $\pi$ increases. 

\begin{figure}
\centering
\includegraphics[width=\textwidth]{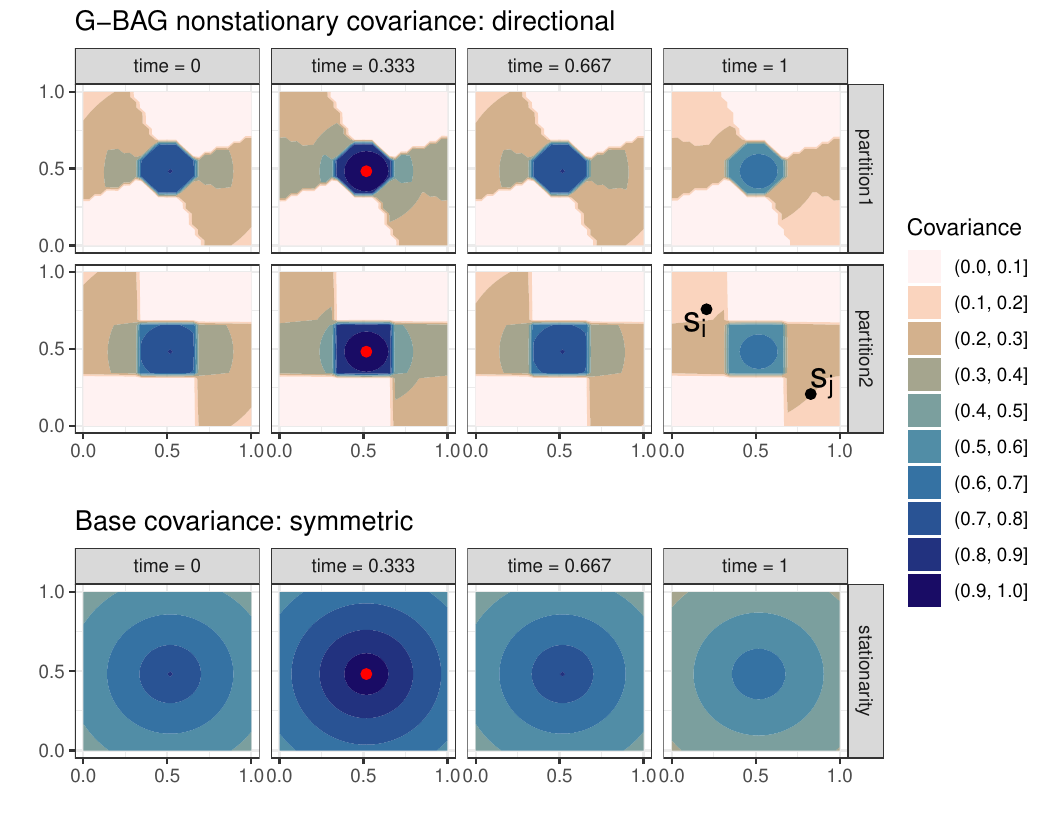}
\caption{Covariance heat maps between a reference point which is the red dot around (0.5, 0.5, 0.333) and the others. G-BAG nonstationary covariances are depicted with two partitioning schemes (first two rows), using the base covariance on the bottom row.}
\label{sim:cov1}
\end{figure}

We visualize asymmetric behaviors of $\tilde{\vt{C}}(\cdot,\cdot\mid \bm{\theta})$ in other settings as well. The reference set $\mc{S}$ is a $30\times 30\times 4$ grid on $\mc{D}=[0,1]^3$ divided into $M=3 \times 3 \times 4 = 36$ partition blocks. With a bag of three arrows coming from west, northwest, and north, we assume that only three DAGs have a positive probability. Each DAG consists of only one of the three directed edges across the whole domain, with the true probabilities $\mbox{pr}(z_1 = 1, \dots, z_{36} = 1) = 0.5$, $\mbox{pr}(z_1 = 2, \dots, z_{36} = 2) = 0.4$, and $\mbox{pr}(z_1 = 3, \dots, z_{36} = 3) = 0.1$ where west, northwest, and north are enumerated as directions 1, 2, and 3, respectively. We use the base covariance function in equation \eqref{Gneitingcov} and fix $a = 0.7$, $c = 0.8$, $\kappa = 0$, and $\sigma^2 = 1$. The resulting G-BAG nonstationary covariances from two partitioning schemes are depicted in Figure \ref{sim:cov1}. Identically at each time, the first partitioning on the top row has an octagon in the middle with fan-shaped arms, and the second partitioning on the bottom row consists of axis-parallel rectangles. Each row of Figure \ref{sim:cov1} is a covariance heat map between the reference point (red point in the middle at time $= 0.333$) and other locations. 

The induced covariance is orientational. The reference point has higher covariance on west -- east and northwest -- southeast axes than other axes, while the base covariance produces the same value at the same space-time lag regardless of axes. With time components, however, the induced nonstationary covariance becomes directional. The time coordinate allows identification of the direction from west over east and northwest over southeast on the corresponding axes, as implied in the second row of Figure \ref{sim:cov1}. In the second row, covariance from the reference point to $\vt{s}_j$ is higher than that to $\vt{s}_i$ despite the same space-time lag. This is because the path from the reference point to $\vt{s}_j$ aligns with the DAG with northwest arrows, whereas $\vt{s}_i$ requires southeast arrows not specified in any true DAGs. 

\begin{figure}
\centering
\includegraphics[width=\textwidth]{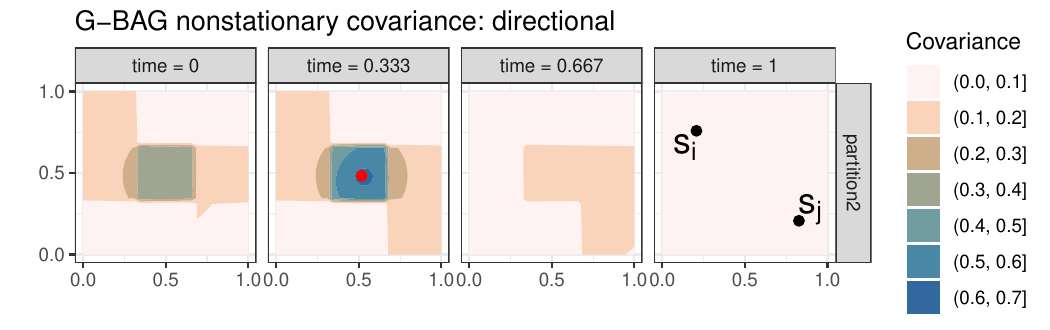}
\caption{Reproduction of G-BAG covariance under partition2 in Figure \ref{sim:cov1} with an exponential base covariance function and a new parent assignment scheme.}
\label{sim:cov1_newparent}
\end{figure}

\begin{figure}
\centering
\includegraphics[width=0.5\textwidth]{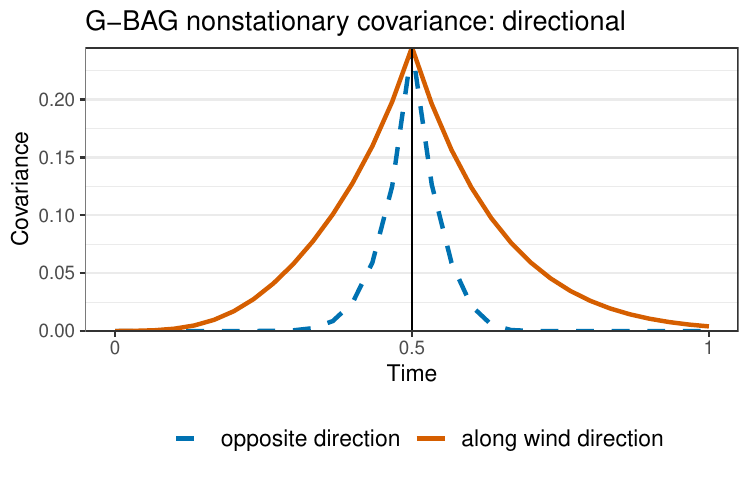}
\caption{Covariances of the point at (0.5, 0.5, 0.5) to (i) locations along the wind direction and to (ii) locations to the opposite direction, represented in a vermilion solid line and a blue dashed line, respectively.}
\label{sim:cov2_newparent}
\end{figure}

We now show that such spatiotemporal asymmetry in G-BAG covariance can be induced even from a purely spatial and symmetric base covariance function with an appropriate parent assignment scheme. We redraw a G-BAG covariance in Figure \ref{sim:cov1} with changes in base covariance function and parent assignment scheme: in Figure \ref{sim:cov1_newparent}, we choose an exponential base covariance with variance equal to 1 and decay equal to 0.8, and a directional temporal parent is employed, which is a previous partition block of a spatial parent. The default temporal parent is a previous partition block of a given block (refer to ``the simplifying assumption'' in Section \ref{sec:unknownDAG}) and thus non-directional. Figure \ref{sim:cov1_newparent} presents that the resulting covariance is higher along the path of prevailing arrows, namely west and northwest. We create another figure similar to Figure \ref{sim:cov2}. With $\mc{S}$ on a $31 \times 3 \times 31$ grid in $\mc{D}=[0,1]^3$, we let each grid point be a partition block and assume constant westerly wind over the spatiotemporal domain. We compute covariance values of the location at (0.5, 0.5, 0.5) to (i) locations along the wind direction and to (ii) those towards the opposite direction. For comparison, at each time, we choose one location from group (i) and another from group (ii) that are equally spatially apart from (0.5, 0.5, 0.5). Identically to Figure \ref{sim:cov1_newparent}, we assume the exponential base covariance function and the new parent assignment scheme. Figure \ref{sim:cov2_newparent} illustrates that locations along the wind direction have larger G-BAG covariance to the reference point (0.5, 0.5, 0.5) than those towards the opposite direction despite the same spatiotemporal lag. In conclusion, both Figures \ref{sim:cov1_newparent} and \ref{sim:cov2_newparent} suggest that the directional nonstationarity in G-BAGs with stronger covariance along the path of DAGs is not specific to a particular parent assignment scheme, and that G-BAGs can transform a purely spatial base covariance to spatiotemporal covariance with directional temporal parents.

\begin{figure}[htbp]
    \centering
    \includegraphics[width = \textwidth]{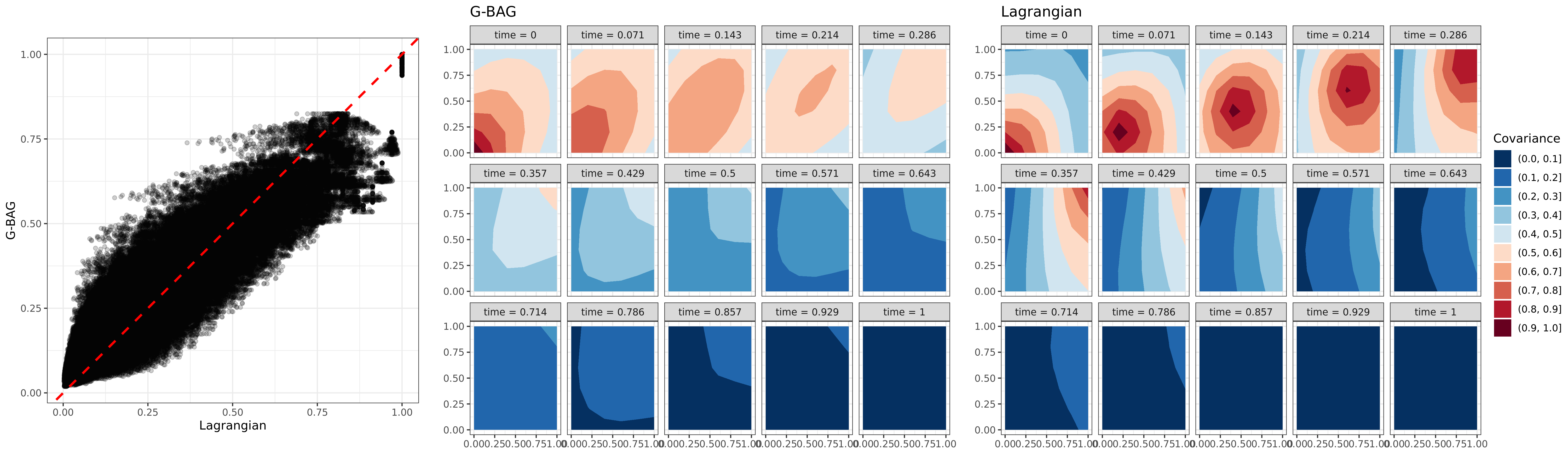}
    \caption{Comparison of Lagrangian and G-BAG covariances. Left: scatter plot of Lagrangian vs. G-BAG covariance values. The red dashed line is a 45-degree line for reference. Center and right: Contour plots of G-BAG and Lagrangian covariance at (0,0,0) to the other locations.}
    \label{sim:cov3}
\end{figure}

Lastly, we show that G-BAG is able to capture directional dependence specified by other existing methods such as Lagrangian spatiotemporal nonstationary covariances \citep{salvana_lagrangian_2021}. Here we mimic a Lagrangian covariance function based on a spatially varying parameters model defined as 
\begin{align*}
    C((\bm{l}_s, l_t), (\bm{l}'_s, l'_t)) = &\sigma(\bm{l}_s-\vt{v}l_t, \bm{l}'_s-\vt{v}l'_t)\mc{M}_{\nu}[\{\bm{l}_s-\bm{l}'_s-\vt{v}(l_t-l'_t)\}^T\\ &\times D(\bm{l}_s-\vt{v}l_t, \bm{l}'_s-\vt{v}l'_t)^{-1}\{\bm{l}_s-\bm{l}'_s-\vt{v}(l_t-l'_t)\}]^{1/2}
\end{align*}
where $\bm{l}_s = (l_{s,x}, l_{s,y})^T \in \R^2$ is a spatial coordinate, $l_t \in \R$ is a temporal coordinate, $\vt{v} = (v_x, v_y)^T \in \R^2$ is a constant advection velocity, the matrix $D(\bm{l}_s-\vt{v}l_t, \bm{l}'_s-\vt{v}l'_t)$ is the spatially varying scale parameter with $D(\bm{l}_s, \bm{l}'_s) = 0.5\{D(\bm{l}_s) + D(\bm{l}'_s)\}$, $\sigma(\bm{l}_s-\vt{v}l_t, \bm{l}'_s-\vt{v}l'_t)$ the spatially varying variance defined by $\sigma(\bm{l}_s,\bm{l}'_s) = |D(\bm{l}_s)|^{1/4}|D(\bm{l}'_s)|^{1/4}|D(\bm{l}_s,\bm{l}'_s)|^{-1/2}$, and $\mc{M}_{\nu}$ is the \matern covariance with variance $\sigma_L^2$, decay $\phi_L$, and smoothness $\nu_L$. The matrix $D(\bm{l}_s)$ is parameterized through the following spectral decomposition
\begin{align*}
    D(\bm{l}_s) = 
    \begin{bmatrix}
        \cos\{\phi(\bm{l}_s)\} & -\sin\{\phi(\bm{l}_s)\} \\
        \sin\{\phi(\bm{l}_s)\} & \cos\{\phi(\bm{l}_s)\}
    \end{bmatrix}
    \begin{bmatrix}
        \lambda_1(\bm{l}_s) & 0 \\
        0 & \lambda_2(\bm{l}_s)
    \end{bmatrix}
    \begin{bmatrix}
        \cos\{\phi(\bm{l}_s)\} & \sin\{\phi(\bm{l}_s)\} \\
        -\sin\{\phi(\bm{l}_s)\} & \cos\{\phi(\bm{l}_s)\}
    \end{bmatrix}.
\end{align*}
We let $\phi(\bm{l}_s-\vt{v}l_t) = (l_{s,x}-v_xl_t-0.5) + (l_{s,y}-v_yl_t-0.5) + 3(l_{s,x}-v_xl_t-0.5)^2 - (l_{s,y}-v_yl_t-0.5)^2$, $\lambda_1(\bm{l}_s-\vt{v}l_t) = \exp[\exp\{-(l_{s,x}-v_xl_t)^2\}- \exp\{-(l_{s,y}-v_yl_t)^2\}]$, and $\lambda_2(\bm{l}_s-\vt{v}l_t) = \exp[\sin(l_{s,x}-v_xl_t)\exp\{-(l_{s,y}-v_yl_t)^2\}-\sin(l_{s,y}-v_yl_t)\exp\{-(l_{s,x}-v_xl_t)^2\}]$, following the code (\url{https://github.com/marysalvana/Lagrangian/tree/main}) provided by the authors. We generated the Lagrangian covariance with $\sigma_L^2=1$, $\phi_L=1$, $\nu_L=0.5$, and $\vt{v} = (3.1, 3.1)^T$ over a $6\times 6\times 15$ grid on $\mc{D}=[0,1]^3$. We managed to mimic this Lagrangian covariance by a G-BAG covariance $\tilde{\vt{C}}$ from the base covariance in equation \eqref{Gneitingcov} with $\sigma^2 = 1$, $a=3.5$, $c=0.96$, $\kappa=0.8$, and constant directed edges from southwest over $M=11\times 1 \times 15$ partition blocks slicing the domain from northwest to southeast. Figure \ref{sim:cov3} illustrates that G-BAG covariance closely mimics the asymmetric Lagrangian spatiotemporal covariance. The Pearson correlation coefficient of the Lagrangian and the G-BAG covariances over all locations is about 0.92, which is also supported by the scatter plot on the left panel of Figure \ref{sim:cov3} where covariance values are well concentrated around the 45-degree line. In addition, the contour plot of G-BAG covariance at (0,0,0) to the other locations on the middle panel resembles the Lagrangian counterpart on the right, characterizing wind movement from southwest to northeast.
    
\subsection{Posterior Sampling for Regression Models with G-BAGs}
\label{s:sampler}

A straightforward Markov chain Monte Carlo (MCMC) sampler to obtain posterior samples with a general $\vt{C}$ is provided below:
\begin{itemize}
\item Update $\bm{\beta}$ from 
    \[( \bm{\beta} \mid \textrm{---}) \sim N\left((V_{\beta}^{-1}+X^TX/\tau^2)^{-1}(V_{\beta}^{-1}\bm{\mu}_{\beta}+X^T(\vt{y}-\vt{w})/\tau^2), (V_{\beta}^{-1}+X^TX/\tau^2)^{-1}\right).\]
\end{itemize}

\begin{itemize}
\item Update $\tau^2$ from 
    \begin{align*}
        (\tau^2\mid \textrm{---}) \sim
        \mbox{IG}\left(a_{\tau}+\frac{n}{2}, b_{\tau} + \frac{1}{2}(\vt{y}-X\bm{\beta}-\vt{w})^T(\vt{y}-X\bm{\beta}-\vt{w})\right).
    \end{align*}
\end{itemize}

\begin{itemize}
\item Update $z_i$ from 
    \[\mbox{pr}(z_i=h\mid \vt{w}_{\mc{S}}, \vt{w}_{\mc{U}}) = \frac{\pi_{ih}N(\vt{w}_i;H_{i\mid h}\vt{w}_{[i\mid h]},R_{i\mid h})\prod_{\vt{u}\in\mc{U}_i}N(w(\vt{u});H_{\vt{u}\mid h}\vt{w}_{[\vt{u}\mid h]},R_{\vt{u}\mid h})}{\sum_{l=1}^K\pi_{il}N(\vt{w}_i;H_{i\mid l}\vt{w}_{[i\mid l]},R_{i\mid l})\prod_{\vt{u}\in\mc{U}_i}N(w(\vt{u});H_{\vt{u}\mid l}\vt{w}_{[\vt{u}\mid l]},R_{\vt{u}\mid l})}\]
    for $h=1,\dots,K$ and $i=1,\dots,M$.
\end{itemize}

\begin{itemize}
\item Update $\vt{w}_i$ for $i=1,\dots,M$ from $\{\vt{w}_i\mid  \textrm{---}\} \sim N(\Sigma_{i}\bm{\eta}_{i}, \Sigma_{i})$ where
\begin{align*}
  \Sigma_{i}^{-1} = &\sum_{\vt{u}:[\vt{u}]\ni a_i}H^T_{\vt{u}, i}R_{\vt{u}}^{-1}H_{\vt{u},i} + \sum_{j:[a_j\mid z_j]\ni a_i} H_{j,i}^TR_{j\mid z_j}^{-1}H_{j,i}+R_{i\mid z_i}^{-1} + I^*_{k_i}/\tau^2,\\
  \bm{\eta}_i= &\sum_{\vt{u}:[\vt{u}]\ni a_i}H^T_{\vt{u}, i}R_{\vt{u}}^{-1}(w(\vt{u})-H_{\vt{u}, -i}\vt{w}_{[\vt{u}, -i]}) + \sum_{j:[a_j\mid z_j] \ni a_i}H_{j,i}^T R_{j\mid z_j}^{-1}(\vt{w}_j-H_{j, -i}\vt{w}_{[j, -i]})+\\
  &R_{i\mid z_i}^{-1}H_{i\mid z_i}\vt{w}_{[i\mid z_i]} + (\vt{y}^*_i-X^*_i\bm{\beta})/\tau^2.
\end{align*}
Let $\mc{S}^*_i = \mc{T}\cap\mc{S}_i$ and  $|\mc{S}^*_i| = n_i\leq k_i=|\mc{S}_i|$. The response $\vt{y}^*_i$ is a $k_i\times 1$ vector whose $j$th element is $y(\vt{s}_{i_j})$ if $\vt{s}_{i_j}\in \mc{S}^*_i$ or 0 otherwise, leaving $n_i$ non-zero elements. Similarly, $X_{i}$ is a $k_i\times p$ matrix with zeros at rows corresponding to locations outside $\mc{S}_i^*$ and $I^*_{k_i} = \mbox{diag}\{\bm{1}(\vt{s}_{i_1}\in \mc{S}^*_i), \dots, \bm{1}(\vt{s}_{i_{k_i}}\in \mc{S}^*_i)\}$. Provided that $a_i$ is a parent of $a_j$, $H_{j,i}$ and $H_{j,-i}$ are a submatrix of $H_{j\mid z_j}$ corresponding to locations in $a_i$ and in $[a_j\mid z_j]\setminus a_i$, respectively. The vector $\vt{w}_{[j,-i]}$ is obtained by removing $\vt{w}_i$ from $\vt{w}_{[j\mid z_j]}$. Lastly, if $\vt{u}\in\mc{U}_j$, $R_{\vt{u}}$ becomes $R_{\vt{u}\mid z_j}$, and $H_{\vt{u},i}$ is a submatrix of $H_{\vt{u}\mid z_j}$ by choosing columns that correspond to $\vt{w}_i$ given that $\vt{w}_i\subseteq\vt{w}_{[\vt{u}\mid z_j]}$, and $H_{\vt{u},-i}$ a submatrix of $H_{\vt{u}\mid z_j}$ by choosing columns that correspond to $\vt{w}_{[u,-i]}=\vt{w}_{[\vt{u}\mid z_j]\setminus i}$.
\end{itemize}

\begin{itemize}
\item Update $w(\vt{u})$ for $\vt{u}\in \mc{U}_i$ for $i=1,\dots,M$ from 
    \begin{align*}
        \{w(\vt{u}) \mid x(\vt{u}), y(\vt{u}), \tau^2, \vt{w}_{\mc{S}}, Z\} &\sim N(\sigma_{\vt{u}}^2\mu_{\vt{u}},\sigma_{\vt{u}}^2) \text{ where }\\
        \sigma_{\vt{u}}^2 = (1/R_{\vt{u}\mid z_i}+1/\tau^2)^{-1}, &\quad  \mu_{\vt{u}} = H_{\vt{u}\mid z_i}\vt{w}_{[\vt{u}\mid z_i]}/R_{\vt{u}\mid z_i}+(y(\vt{u})-x(\vt{u})^T\bm{\beta})/\tau^2.
    \end{align*}
\end{itemize}

\begin{itemize}
\item Finally, $\bm{\theta}$ is updated from a Metropolis-Hastings step with target density \[p_{\theta}\times N(\vt{w}_{\mc{S}};\vt{0}, \tilde{C}_{Z})N(\vt{w}_{\mc{U}};H_{\mc{U}\mid Z}\vt{w}_{\mc{S}}, H_{\mc{U}\mid Z}).\] 
For instance, in the covariance function in equation \eqref{Gneitingcov}, $\bm{\theta}$ is $(a, c, \kappa, \sigma^2)$. We use the robust adaptive Metropolis algorithm proposed by \citet{vihola_robust_2012} for $g(a, c, \kappa)$ with target acceptance rate of 0.234. A link function $g$ maps $(a,c,\kappa)$ to $(-\infty, \infty)$ range. Uniform priors are used. A Gibbs step to update $\sigma^2$ can be easily derived with a prior $\sigma^2 \sim \mbox{IG}(a_{\sigma}, b_{\sigma})$.
\end{itemize}

Posterior predictive samples at an arbitrary location $\bm{l}\in\mc{D}$ are obtained as follows: at each iteration, sample $y(\bm{l})\mid \vt{y} \sim N(x(\bm{l})^T\bm{\beta} + w(\bm{l}),\tau^2)$ if $\bm{l}\in\mc{S}\cup\mc{T}$. If $\bm{l}\notin\mc{S}\cup\mc{T}$, then first sample $w(\bm{l})$ from $N(H_{\bm{l}}\vt{w}_{[\bm{l}]},R_{\bm{l}})$ conditioned on $\vt{w}_{\mc{S}}$ and $Z$ if desired, and then sample $y(\bm{l})\mid \vt{y} \sim N(x(\bm{l})^T\bm{\beta} + w(\bm{l}),\tau^2)$.

\subsection{Computational Cost of G-BAGs} 
\label{s:cost}

In this section, we show that G-BAGs have computational complexity of order $n$ at each iteration of the sampler in Section S2.6. For explanatory purposes, we assume $|\mc{S}_i| = |\mc{U}_i|=m$ for $i = 1,\dots,M$. The number of locations in $\mc{S}\cup\mc{U}$ are at maximum $n+k$, in which case $m=(n+k)/(2M)$. Let us assume each node has $J$ parent nodes or less, and all locations in $\mc{U}_i$ share the same parent set. The number of directions available in a bag is $K$. 

Updates of $Z$ involve (a) $N(\vt{w}_i;H_{i\mid h}\vt{w}_{[i\mid h]},R_{i\mid h})$ and (b) $\prod_{\vt{u}\in\mc{U}_i}N\{w(\vt{u});H_{\vt{u}\mid h}\vt{w}_{[\vt{u}\mid h]},R_{\vt{u}\mid h}\}$ for $h=1,\dots,K$. At a fixed $h$, (a) requires $C_{[i\mid h]}^{-1}$ for $H_{i\mid h}$ and $R_{i\mid h}$, and $R_{i\mid h}^{-1}$ for density evaluation. Since $C_{[i\mid h]}$ is of size $Jm \times Jm$ and $R_{i\mid h}$ is of size $m\times m$, inversion of these matrices leads to complexity $\mc{O}((J^3+1)m^3)$. Repeating for each $i$ and $h$, the overall complexity for (a) becomes $\mc{O}(MK(J^3+1)m^3)$. Similarly, (b) requires $C_{[\vt{u}\mid h]}^{-1}$ for $H_{\vt{u}\mid h}$ and $R_{\vt{u}\mid h}$, and $R_{\vt{u}\mid h}^{-1}$ for density evaluation. The $Jm \times Jm$ matrix $C_{[\vt{u}\mid h]}$ is common for all $\vt{u}\in\mc{U}_i$, and $R_{\vt{u}\mid h}$ is a $1 \times 1$ matrix. Thus, the total floating point operations (flops) pertaining to (b) is $MK\{(Jm)^3+m\times 1^3\}$ considering repetitions over $i$ and $h$. Second, with $H$'s and $R^{-1}$'s stored from the first step, posterior updates of $\vt{w}_\mc{S}$ use $Mm^3$ flops in computing $\Sigma_i\in\R^{m\times m}$ for all $i$'s, while updates of $\vt{w}_{\mc{U}}$ use $Mm1^3$ flops in computing $\sigma^2_{\vt{u}}\in\R$ for $Mm$ $\vt{u}$'s. 

Adding all these steps, each iteration has approximate computational complexity of $\mc{O}\{MK(J^3+1)m^3 + MK(J^3m^3+m) + Mm^3+Mm\}$, which is less than or equal to $\mc{O}\{Mm^3(2KJ^3 + K + 2)\}$, as $K$ is typically chosen to be small enough to ensure $(K+1)/m^2 \leq 1$. In an extreme case where $K$ is as large as $n$, the computational complexity becomes $\mc{O}\{M(n/M)^3(2KJ^3+K+2)\}$, which is then approximately $\mc{O}\{n(2KJ^3+K+2)\}$ since it is reasonable to choose the number of partition blocks $M$ proportionally to the sample size $n$. Given that the fixed value of $J$ is relatively small, each iteration has a linear computational complexity in $n$. 

During the Gibbs iterations, $H_{i\mid z_i}$, $R^{-1}_{i\mid z_i}$, $H_{\vt{u}\mid z_i}$, and $R^{-1}_{\vt{u}\mid z_i}$ need to be stored for $i=1,\dots,M$, $z_i=1,\dots,K$, and $m$ $\vt{u}$'s in $\mc{U}_i$. These matrices are of size $m\times Jm$, $m \times m$, $1 \times Jm$, and $1 \times 1$, respectively, causing storage cost $\mc{O}(2Jm^2MK+m^2MK+mMK)$. Given that $K$ and $J$ are relatively small and with $M \propto n$ and $k\approx n$, $\mc{O}[\{2JK+K+K/m\}m^2M] \leq \mc{O}[\{2JK+K+1\}n^2/M] \approx \mc{O}(n)$.

\subsection{Posterior Consistency of Mixture Weights in BAGs}
\label{s:asymptotics}

Consider the data generating model in equation \eqref{reg} with $\bm{\beta} = \vt{0}$: $y(\vt{t}) = w(\vt{t}) + \epsilon(\vt{t})$ with $w(\cdot) \sim \mbox{G-BAG}(0, \tilde{\vt{C}}(\bm{\theta}))$ and $\epsilon(\vt{t}) \sim N(0,\tau^2)$ for $\vt{t} \in \mc{T} = \{\vt{t}_1,\dots,\vt{t}_n\}$. The set of locations $\mc{T}$, a bag of directions of size $K$, and a partition of size $M$ are fixed. We set $\mc{S} = \mc{T}$. For $\vt{y} = (y(\vt{t}_1),\dots,y(\vt{t}_n))^T$, we assume replicates independently and identically distributed according to $\vt{y}_1,\dots,\vt{y}_r \sim \tilde{p}(\vt{y}) = \sum_Z \tilde{p}(\vt{y} \mid Z)q_0(Z)$, where $Z = (z_1,\dots,z_M)^T$ is a vector of directions, $q_0(Z)$ is the unknown true probability of directions $Z$, and $\tilde{p}(\vt{y}\mid Z) = N(\vt{y}; \vt{0}, \tilde{C}_{Z} + \tau^2 I_n)$ with $\tilde{C}_{Z}$ as in equation \eqref{ptilde_con_N}. In this subsection, we use the notation $\tilde{C}_{Z}(\bm{\theta})$ to make the dependence of $\tilde{C}_{Z}$ on $\bm{\theta}$ explicit. Define $\Gamma(Z)$ to be a configuration of parents induced by directed edges $Z$ over the partition. Then there are only $L \leq K^M$ unique configurations, enumerated as $\{\Gamma_1,\dots,\Gamma_L\}$. For instance, if $Z$ and $Z'$ are equal but on a boundary partition block $i$ with $z_i = 1$ and $z'_i = 2$ where both directions 1 and 2 induce no parents to the block $i$ then $\Gamma(Z) = \Gamma(Z')$.

The prior probability of directions $Z$ is $q(Z)$. Updating this prior with information in the data $Y_r = \{\vt{y}_1,\ldots,\vt{y}_r\}$ produces the posterior probability $q(Z \mid Y_r)$. We define probabilities $\Phi = \{\bm{\phi} \in (0,1)^L \mid \sum_{l=1}^L\phi_l =1\}$ of the different configurations $\Gamma_1,\ldots,\Gamma_L$ as deterministic functions of the probabilities of different directions $Z$. The prior for $\bm{\phi}$ is induced via $\phi_l  = \sum_{Z} \bm{1}\{\Gamma(Z) = \Gamma_l\}q(Z)$, and the posterior via $\phi_l  = \sum_{Z} \bm{1}\{\Gamma(Z) = \Gamma_l\}q(Z \mid Y_r)$ for $l=1,\dots,L.$ Proposition \ref{prop:phi} shows that $\sum_{l=1}^L \phi_l = 1$. 

For $\bm{\phi} \in \Phi$, define a probability measure 
\begin{align}
P_{\phi}(\cdot) = \sum_{l=1}^L \phi_l F_l(\cdot), \label{eq:mixturemeasure}
\end{align}
over $(\R^n,\mathcal{Y})$ with $\mathcal{Y}$ the Borel subsets of $\R^n$, and $F_l$ the probability measure corresponding to $N(\vt{y}; \vt{0}, \tilde{C}_l(\bm{\theta}) + \tau^2I_n)$ in G-BAGs with base covariance parameters $\bm{\theta}$ and nugget $\tau^2$. For any given $\bm{\theta}$, the matrix $\tilde{C}_l(\bm{\theta})$ is $\tilde{C}_{Z}(\bm{\theta})$ for any $Z$ such that $\Gamma(Z) = \Gamma_l$ because $\tilde{C}_{Z}(\bm{\theta})$ is a function of $H_Z$ and $R_Z$ whose elements are determined by the configuration of parents $\Gamma(Z)$. 

Assuming data $y_1,\ldots,y_r$ are generated independently from $P_{\phi_0}$ with some $\bm{\phi}_0\in\Phi$, we show in Theorem \ref{thm:consistency} that the posterior distribution of the probabilities $\bm{\phi}$ concentrates around $\bm{\phi}_0$ as $r$ increases, that is, the posterior is consistent.

\begin{theorem}
There exists $\Phi_0 \subseteq \Phi$ such that $pr(\bm{\phi} \in \Phi_0) = 1$ and for all $\bm{\phi}_{0}\in \Phi_0$, if $\vt{y}_1,\dots,\vt{y}_r \sim P_{\phi_0}$ then for all $\epsilon > 0$
\begin{align*}
    \lim_{r\rightarrow \infty} \mbox{pr}\{\bm{\phi} \in \tilde{B}(\bm{\phi}_0, \epsilon) \mid \vt{y}_1,\dots,\vt{y}_r\} = 1 \quad P_{\phi_0}\text{-almost surely} 
\end{align*}
with $\tilde{B}(\bm{\phi}_0, \epsilon) = \{\bm{\phi} \in \Phi \mid \|\bm{\phi} - \bm{\phi}_0\| < \epsilon\}$ where $\|\cdot\|$ denotes the Euclidean distance.
\label{thm:consistency}
\end{theorem}



Theorem \ref{thm:consistency} is adapted from Theorem 2.4 in \cite{miller_consistency_2022} and is applicable to general BAGs with appropriate $F_l$'s to represent various base processes. 
    
\subsubsection{Proofs}

\begin{proposition}
Let $\Phi = \{\bm{\phi} \in (0,1)^L \mid \sum_{l=1}^L\phi_l = 1 \}$. The prior for $\phi_l$ is induced via $\phi_l=\sum_{Z} \bm{1}\{\Gamma(Z) = \Gamma_l\}q(Z)$, and the posterior via $\phi_l=\sum_{Z} \bm{1}\{\Gamma(Z) = \Gamma_l\}q(Z \mid Y_r)$. Then $\sum_{l=1}^L\phi_l = 1$. \label{prop:phi}
\end{proposition}

\begin{proof}[Proof of Proposition~\ref{prop:phi}]
Notice that each $\Gamma(Z)$ is equal to one and only one $\Gamma_l$. Therefore, the prior for $\phi_l$ satisfies 
\begin{align*}
    \sum_{l=1}^L\phi_l &= \sum_{l=1}^L\sum_{Z} \bm{1}\{\Gamma(Z) = \Gamma_l\}q(Z) = \sum_{Z} q(Z) = 1.
\end{align*}
The last equality is proven as part of the proof of Proposition \ref{prop:jointdensity}.
Similarly, the posterior for $\phi_l$ satisfies
\begin{align*}
    \sum_{l=1}^L\phi_l &= \sum_{l=1}^L\sum_{Z} \bm{1}\{\Gamma(Z) = \Gamma_l\}q(Z \mid Y_r) \\
    &= \sum_{Z} q(Z \mid Y_r) \\
    &= \sum_{Z} \int q(Z \mid W_r, Y_r) \tilde{p}(W_r \mid Y_r)dW_r \\
    &= \int \sum_{Z} q(Z \mid W_r) \tilde{p}(W_r \mid Y_r)dW_r = 1
\end{align*}
with $W_r = \{\vt{w}_1,\dots,\vt{w}_r\}$. Recall the summation $\sum_Z$ is over $K^M$ possible $Z$'s, enumerated as $\{Z_1,\dots,Z_{K^M}\}$. The last equality is then due to 
\begin{align*}
    \sum_{Z} q(Z \mid W_r) &= \sum_{s=1}^{K^M} q(Z_s \mid W_r) \\
    &= \sum_{s=1}^{K^M} \prod_{i=1}^M \mbox{pr}(z_i = z_{s_i} \mid W_r) \\
    &= \prod_{i=1}^M \sum_{z_{s_i}=1}^{K} \mbox{pr}(z_i = z_{s_i} \mid W_r) = 1
\end{align*}
as $\sum_{z_{s_i}=1}^{K} \mbox{pr}(z_i = z_{s_i} \mid W_r) = 1$ for all $i=1,\dots,M$.
\end{proof}

\begin{proof}[Proof of Theorem~\ref{thm:consistency}]
This theorem results from Doob's theorem on posterior consistency \citep{miller_detailed_2018}. Doob's theorem is based on two conditions: measurability and identifiability.

Since the projection $\bm{\phi} \rightarrow \phi_l$ is measurable, and $F_l(A)$ is a probability measure on a measurable set $A$, the function $\bm{\phi} \rightarrow P_{\phi}(A) = \sum_{l=1}^L \phi_l F_l(A)$ is measurable for every measurable $A\subseteq \mc{Y}$. 

In terms of identifiability, $P_{\phi}$ does not have identical mixture components and is not invariant with respect to permutation of component labels. Let us elaborate. We assume $\bm{\theta}$ and $\tau^2$ known. Then every mixture component is fully determined by $\tilde{C}_l(\bm{\theta})$. We want to show if $\tilde{C}_l(\bm{\theta})$'s are distinct for all $l=1,\dots,L$. Let $\Gamma(Z) = \Gamma_l$ and $\Gamma(Z') = \Gamma_{l'}$ for some $Z, Z'$. Suppose $\tilde{C}_{Z}(\bm{\theta}) = \tilde{C}_{Z'}(\bm{\theta})$. Then every $(i,j)$ block of $\tilde{C}^{-1}_{Z}(\bm{\theta})$ should be equal to the $(i,j)$ block of $\tilde{C}^{-1}_{Z'}(\bm{\theta})$ for $1 \leq j < i \leq M$. By equation \eqref{s:Ctildeinv}, the identical $(i,j)$ blocks in $\tilde{C}^{-1}_{Z}(\bm{\theta})$ and $\tilde{C}^{-1}_{Z'}(\bm{\theta})$ imply two scenarios: (1) every pair of nodes $a_i$ and $a_j$ have the same relationship in $Z$ and $Z'$ (``parent-child'': $a_j$ is a parent of $a_i$, ``parents'': $a_i$ and $a_j$ have a common child $a_l$ for some $l\in\{1,\dots,M\}$ such that $j<i<l$, or not connected), or (2) without loss of generality, for some $(i,j)$ pairs, the nodes $a_j$ and $a_i$ have the parent-child relationship in $Z$, while they are parents of $a_l$ in $Z'$, but the numeric values, namely $-R_{i|z_i}^{-1}H_{i,j}$ and $H_{l,i}^TR_{l|z'_l}^{-1}H_{l,j}$, happen to be the same, given that $-R_{i|z_i}^{-1}H_{i,j}$ and $H_{l,i}^TR_{l|z'_l}^{-1}H_{l,j}$ yield non-zero blocks. 

In the first scenario, we show that the parents configurations induced by $Z$ and $Z'$ should be equal to each other, that is, $\Gamma_l = \Gamma_{l'}$. Suppose $\Gamma_l \neq \Gamma_{l'}$. Then there exists $i\in\{1,\dots,M\}$ such that $[a_i \mid z_i] \neq [a_i \mid z'_i]$. Since every $(i,j)$ relationship matches in $Z$ and $Z'$, however, the parent-child relationship for $i$ should also match, implying if $a_j \in [a_i\mid z_i]$ then $a_j \in [a_i\mid z'_i]$ and vice versa. Hence, $[a_i\mid z_i] = [a_i\mid z'_i]$, which proves $\Gamma_l = \Gamma_{l'}$ by contradiction. 

We show that the second scenario never occurs. For any triples $(i,j,l)$ such that $1 \leq j < i < l \leq M$ with two edges among them, potential relationships that satisfy the parent-child relationship for $a_j$ and $a_i$ can only be either (a) $a_j$ is a parent of $a_l$ or (b) $a_i$ is a parent of $a_l$, due to ordering of indices. However, under (a), $\tilde{C}^{-1}_{Z}(l,i)$ is a zero block, while $\tilde{C}^{-1}_{Z'}(l,i)$ returns a non-zero block $-R_{l|i}^{-1}H_{l,i}$. Similarly, under (b), $\tilde{C}^{-1}_{Z}(l,j)$ is a zero block, and $\tilde{C}^{-1}_{Z'}(l,j)$ returns a non-zero block $-R_{l|j}^{-1}H_{l,j}$. These (a) and (b) cases both yield some blocks that cannot match in $\tilde{C}^{-1}_{Z}(\bm{\theta})$ and $\tilde{C}^{-1}_{Z'}(\bm{\theta})$, contradicting the original supposition that $\tilde{C}_{Z}(\bm{\theta}) = \tilde{C}_{Z'}(\bm{\theta})$. The same discussion trivially carries over if $a_l$ is not connected to $a_i$ or $a_j$ in $Z$. 

Therefore, by the first scenario, we now conclude if $\tilde{C}_{Z}(\bm{\theta}) = \tilde{C}_{Z'}(\bm{\theta})$ then $\Gamma_l = \Gamma_{l'}$; each $\Gamma_l$ gives a distinct $\tilde{C}_l(\bm{\theta})$. Since $\tilde{C}_l(\bm{\theta})$'s are fixed as locations are fixed, $F_l$'s are all distinct and fixed. With these $F_l$'s, $P_{\phi} = P_{\phi'}$ implies $\bm{\phi} = \bm{\phi}'$: $P_{\phi} = P_{\phi'}$ $\Leftrightarrow$ $\sum_{l=1}^L\phi_lF_l = \sum_{l=1}^L\phi'_lF_l$ $\Rightarrow$ $\phi_l = \phi'_l$ for all $l = 1,\dots,L$. 

Since the two conditions are satisfied, we can apply Doob's theorem which yields $\mbox{pr}(\bm{\phi} \in \tilde{B}(\bm{\phi}_0, \epsilon) \mid Y_r) \rightarrow 1$ almost surely as $r \rightarrow \infty$.
\end{proof}




\section{Applications on Simulated Data}
\label{s:simu}

BAGs infer directions over $\mc{S}$ by construction. If a certain region which is sparsely covered by observed locations is of considerable interest in terms of directional inferences, then we suggest placing reference locations in that region. The inferred directions in such a region will naturally have higher uncertainty than if data were denser. Computational cost of G-BAGs increases as the size of $\mc{S}$ increases, which discourages enlarging $k$ much beyond $n$. Therefore, we choose to use the data locations as reference locations throughout simulations and real applications, following standard practice for reference set choice in the literature  \citep{finley_efficient_2019, peruzzi_highly_2022, dey_graphical_2022}.

\subsection{Fitted G-BAG is Correctly Specified}
\label{s:sim1}

\begin{figure}
\centering
\begin{subfigure}[htbp]{\textwidth}
\centering
\includegraphics[width=0.95\textwidth]{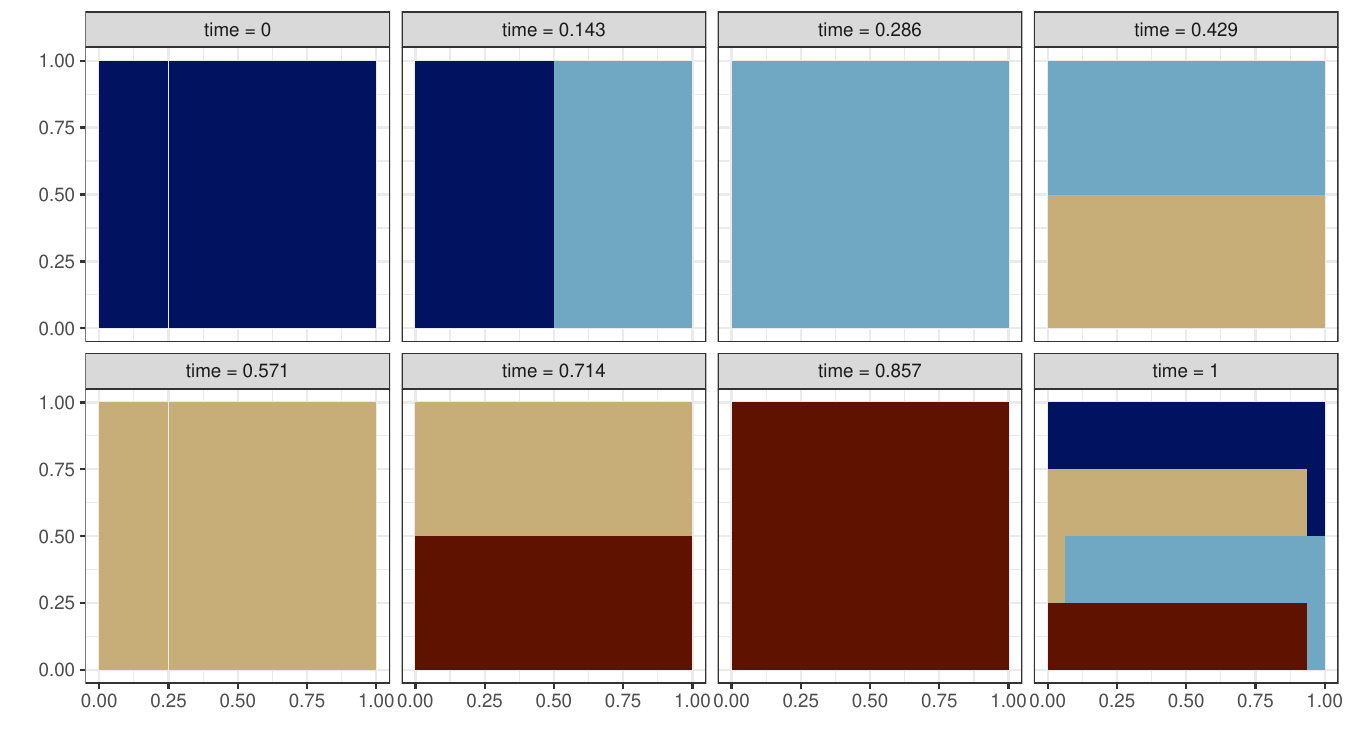}
\caption{True directions}
\label{sim1:truewind}
\end{subfigure} \\~\\~\\
\begin{subfigure}[htbp]{\textwidth}
\centering
\includegraphics[width=0.95\textwidth]{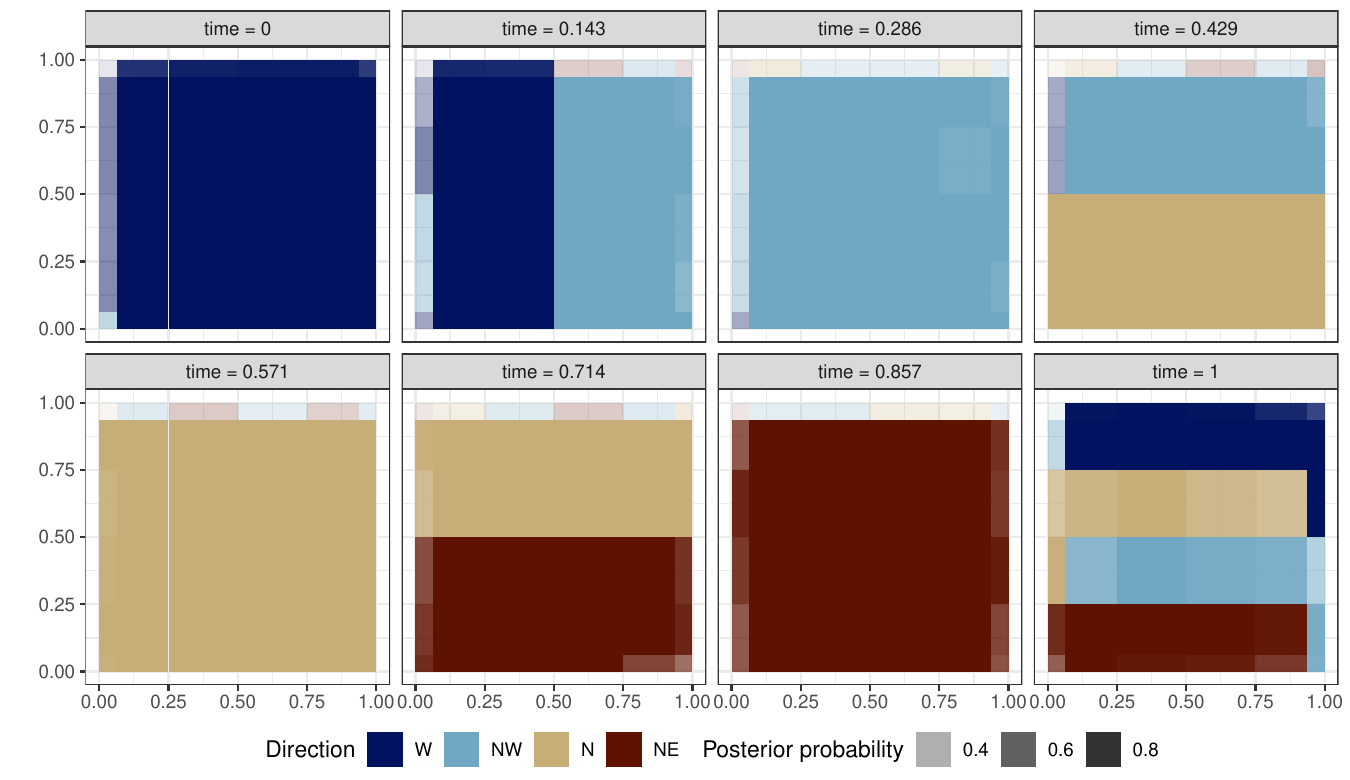}
\caption{Inferred directions}
\label{sim1:estwind}
\end{subfigure}
\caption{Each time frame depicts the whole spatial domain and each color corresponds to (a) the true directions generating spatiotemporal dependence and (b) directions estimated by the G-BAG model. In (b), the inferred direction in each partition block is the directed edge with the highest posterior probability. Higher probability values correspond to more opaque color shading. Results are based on 25 data sets generated from the G-BAG model with $\bm{\theta}_1$. Direction W, west; NW, northwest; N, north; NE, northeast.}
\label{fig:sim1}
\end{figure}

The true directions shift rapidly across the space-time domain and are depicted in Figure \ref{sim1:truewind}. To fit a G-BAG model, a vague normal prior $N(0,10^2)$ is specified for $\beta$. IG priors are specified for nugget ($\tau^2$) and partial sill ($\sigma^2$) such that $\tau^2 \sim \mbox{IG}(2,0.1)$ and $\sigma^2 \sim \mbox{IG}(2,1)$. The covariance parameters have weakly informative uniform priors: $a\sim \mbox{Unif}(4,8)$ and $c\sim \mbox{Unif}(0.158, 0.789)$ under $\bm{\theta}_1$ and $a\sim \mbox{Unif}(7.330, 14.667)$ and $c\sim \mbox{Unif}(0.075, 0.373)$ under $\bm{\theta}_2$. In both cases, $\kappa \sim \mbox{Unif}(0,1)$. The lower and upper bounds for $a$ and $c$ correspond to various base correlation values ranging from 0.1 to 0.9 at varying temporal and spatial distances, respectively. For instance, $a\sim \mbox{Unif}(7.330, 14.667)$ lets correlation drop to 0.1 roughly between half maximal and maximal temporal distance. In the fixed-DAG model, its default priors $\sigma^2, \tau^2 \sim \mbox{IG}(2,1)$, $\beta \sim N(0, 10^2)$ are used except for the covariance decay parameters which have the common uniform prior whose lower bound is the minimum of lower bounds for $a$ and $c$ in the G-BAG model, and an upper bound is the maximum of upper bounds for $a$ and $c$. SPDE models implemented in the \texttt{R-INLA} package use penalized complexity priors \citep{simpson_penalising_2017}. In an autoregressive model of order 1, the autocorrelation parameter is assumed larger than 0.05 with prior probability 0.99, and the precision of white noise is assumed larger than 1 with prior probability 0.01. For SPDE-stationary models, the spatial standard deviation is larger than 1.5 with prior probability 0.01, and the range parameter is smaller than 1.8 or 9 when the truth is $\bm{\theta}_1$ or $\bm{\theta}_2$, respectively. In SPDE-nonstationary models, $\psi$'s are modeled independently and identically with $N(0,0.3^{-1})$ prior.

\begin{figure}[htbp]
\centering
\includegraphics[width=0.8\textwidth]{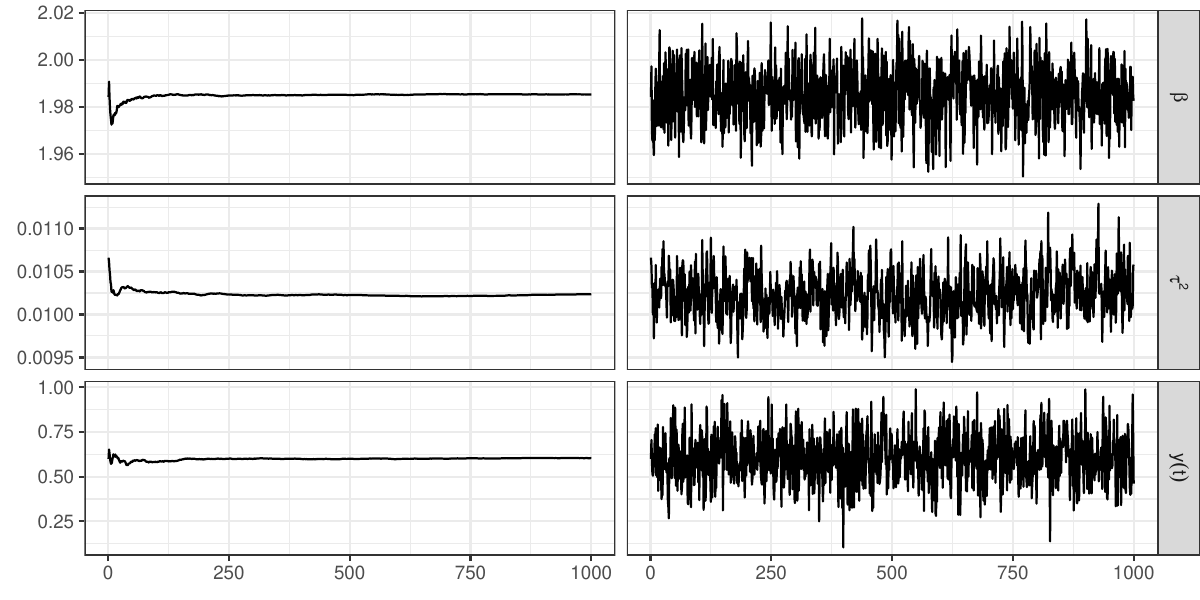}
\caption{Running mean plots on the left column and trace plots on the right column for $\beta$, $\tau^2$, and $y(\vt{t})$ at an arbitrary location. Each column shows the results from a randomly selected data set out of 25 replicates using $\bm{\theta}_2$.}
\label{sim1:converg}
\end{figure}

MCMC convergence and mixing were investigated for model parameters, predicted $y(\vt{t})$, and estimated $Z$ using trace plots, running mean plots, effective sample size, and hypothesis tests. Trace plots in Figure \ref{sim1:converg} show convergence to stationarity and adequate mixing for $\beta$, $\tau^2$, and predicted $y(\vt{t})$ under G-BAG. Running mean plots show that we stopped properly after the running means had stabilized. The effective sample size was computed using the \texttt{R} package \texttt{coda} \citep{plummer_coda_2006} and reported in Table \ref{tab:sim1_ess}. We confirm that estimates of effective sample size for model parameters and predicted $y(\vt{t})$ are adequately high. A chi-squared diagnosis test for convergence of a categorical variable was conducted as suggested in \cite{deonovic_convergence_2017} independently for each partitioned region to assess convergence of $z_i$. The test determines whether there is a significant discrepancy in the frequency distribution between two portions of posterior draws. We compared the first 35\% and the last 35\% of a chain. The average rejection rate of the null hypothesis that indicates no discrepancy between the beginning and the end of the chain across 25 replicates in both cases of $\bm{\theta}_1$ and $\bm{\theta}_2$ was 0.037 and 0.032, respectively, at a significance level of 0.05. Convergence diagnostics altogether suggest appropriate convergence of the parameters and variables of interest. 

\begin{table}[htbp]
    \centering
    \begin{tabular}{ccccccc}
    \hline
    & $\beta$ & $\tau^2$ & $\sigma^2 \times c$ & $a$ & $\kappa$ & $y(\vt{t})$\\ \hline
   $\bm{\theta}_1$ & 935 & 196 & 329 & 137 & 139 & 1,011 \\
   $\bm{\theta}_2$ & 1,000 & 429 & 357 & 205 & 224 & 1,012 \\ \hline
   \end{tabular}
   \caption{The average of effective sample size estimates for 1,000 posterior draws across 25 replicates with $\bm{\theta}_1$ and $\bm{\theta}_2$. The effective sample size estimates are aggregated for predicted $y(\vt{t})$ over 2,560 locations.}
   \label{tab:sim1_ess}
\end{table}

\begin{table}[htbp]
    \centering
    \begin{tabular}{cccccc}
    \hline
         &  \multirow{2}{*}{G-BAG} & \multirow{2}{*}{G-BAG*} & \multirow{2}{*}{Fixed-DAG} & SPDE- & SPDE-  \\ [-3pt]
     &   &  &  & stationary & nonstationary  \\ \hline 
   \multirow{2}{*}{$\beta=2$} & \textbf{2.003} & 2.003 & 2.014 & 2.019 & 2.019 \\ [-3pt]
   & (0.010) & (0.011) & (0.026) & (0.027) & (0.027) \\
   \multirow{2}{*}{$\tau^2=0.01$} & \textbf{0.010} & 0.010 & 0.015 & 0.151 & 0.150 \\ [-3pt]
   & ($<$0.001) & ($<$0.001) & (0.002) & (0.026) & (0.026) \\
   \multirow{2}{*}{$\sigma^2 =2$} & 1.635 & 1.649 & \textbf{1.891} & \multirow{2}{*}{--} & \multirow{2}{*}{--}\\ [-3pt]
   & (0.315) & (0.296) & (0.618) &  &  \\
   \multirow{2}{*}{$a=10$} & \textbf{10.773} & 10.994 & 14.138 & \multirow{2}{*}{--} & \multirow{2}{*}{--} \\ [-3pt]
   & (0.756) & (0.132) & (0.880) &  &  \\
   \multirow{2}{*}{$c=0.1$} & \textbf{0.124} & 0.125 & 3.107 & \multirow{2}{*}{--} & \multirow{2}{*}{--} \\ [-3pt]
   & (0.023) & (0.025) & (0.505) & & \\
   \multirow{2}{*}{$\kappa=0.2$} & \textbf{0.340} & 0.496 & 0.930 & \multirow{2}{*}{--} & \multirow{2}{*}{--} \\ [-3pt]
   & (0.107) & (0.023) & (0.118) & &  \\ 
   \multirow{2}{*}{RMSPE} & \textbf{0.129} & 0.129 & 0.357 & 0.429 & 0.428 \\ [-3pt]
   & (0.002) & (0.002) & (0.035) & (0.040) & (0.041) \\
   \multirow{2}{*}{MAPE}  & \textbf{0.103} & 0.103 & 0.229 & 0.274 & 0.274 \\ [-3pt]
   & (0.002) & (0.002) & (0.017) & (0.022) & (0.022) \\
   \multirow{2}{*}{95\% CI coverage} & \textbf{0.950} & 0.949 & 0.945 & 0.926 & 0.927 \\ [-3pt]
   & (0.006) & (0.006) & (0.008) & (0.009) & (0.009) \\ 
   \multirow{2}{*}{95\% CI width} & \textbf{0.504} & 0.505 & 1.570 & 1.651 & 1.649 \\ [-3pt]
   & (0.005) & (0.005) & (0.145) & (0.140) & (0.148) \\ \hline
   \end{tabular}
   \caption{Simulation results when G-BAG is the true data generating model with $\bm{\theta}_2$. Mean and standard error in parenthesis are calculated over 25 synthetic data sets.}
   \label{tab:sim1_theta2}
\end{table}

\subsection{Fitted G-BAG is Misspecified}
\label{s:sim2}

\begin{figure}[htbp]
\centering
\includegraphics[width=\textwidth]{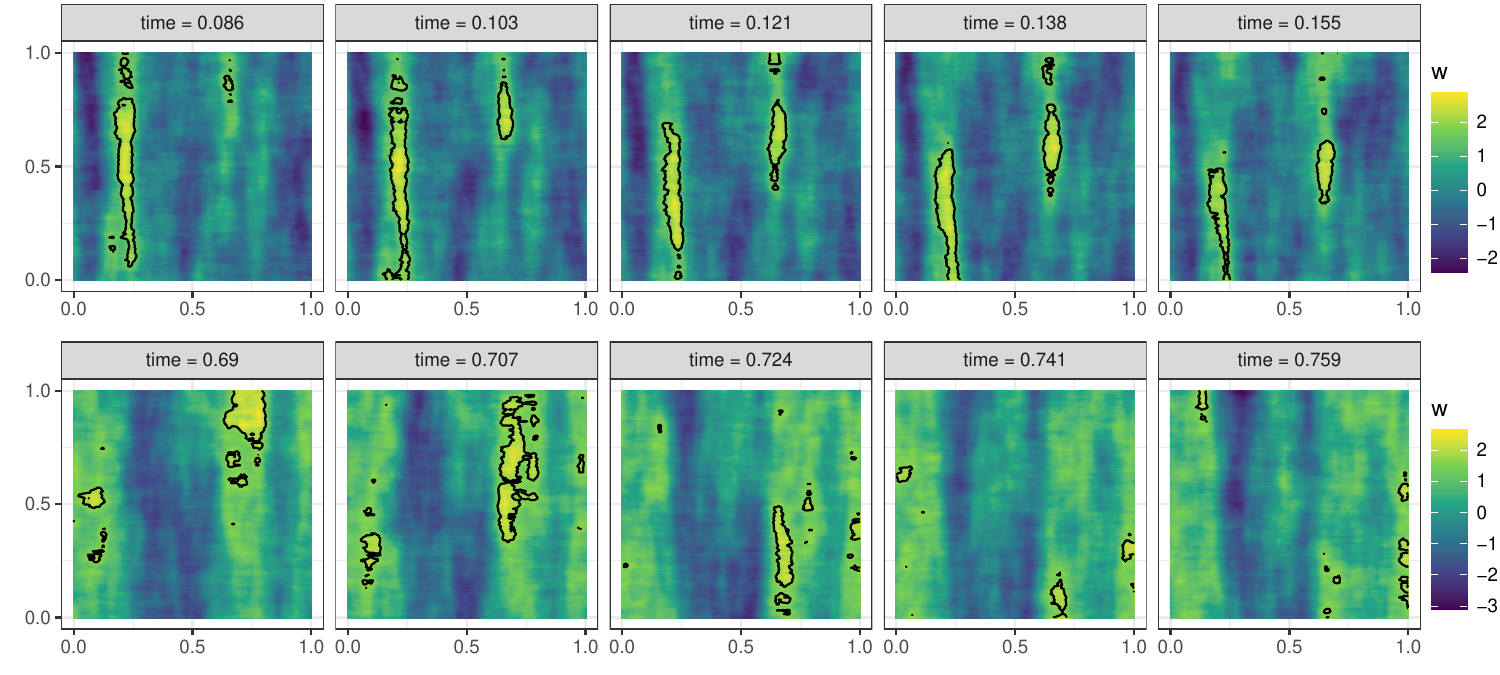}
\caption{Simulated spatiotemporal field $\vt{w}$ with $\bm{\theta}_3$ (top row) and $\bm{\theta}_4$ (bottom row) at different five consecutive time points. Contour lines are drawn at $w(\vt{t})=1.7$.}
\label{sim2:truew}
\end{figure}

We generate 25 synthetic data sets with $\bm{\theta}_3 = (5,20,1,150)$ and another 25 with $\bm{\theta}_4 = (10,20,1,150)$. Figure \ref{sim2:truew} illustrates simulated examples of the true $\vt{w}$ with $\bm{\theta}_3$ and $\bm{\theta}_4$. Directional dependence from north to south and different speed by $\bm{\theta}$'s are evident across the spatiotemporal domain. 

In the G-BAG model, prior distributions for other parameters are identical as before except $a\sim \mbox{Unif}(0.1,9)$ or $a\sim \mbox{Unif}(0.1,15)$ under $\bm{\theta}_3$ or $\bm{\theta}_4$, respectively, while $\kappa \sim \mbox{Unif}(0,1)$, $c\sim \mbox{Unif}(0.363, 21.183)$ in both cases. The prior for decay parameters in the fixed-DAG model changes accordingly. The penalized complexity priors are modified for SPDE-stationary so that the range parameter is assumed to be smaller than 0.1 with prior probability 0.01, while the spatial standard deviation is higher than 1 with prior probability 0.01. 

\begin{figure}[htbp]
\centering
\includegraphics[width=\textwidth]{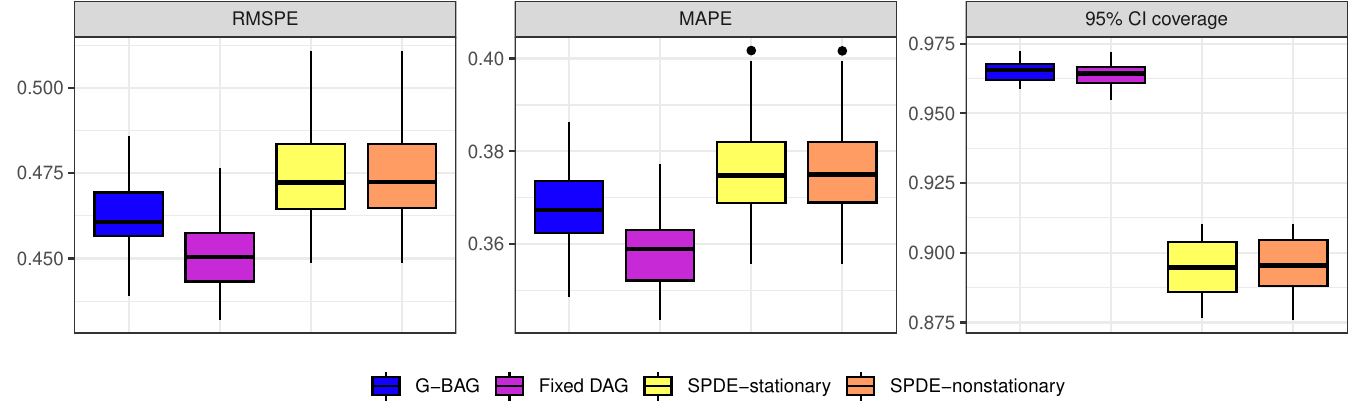}
\caption{Simulation results when the fitted G-BAG is misspecified, and the true covariance parameters are $\bm{\theta}_3$. Box plots over 25 synthetic data sets are presented.}
\label{sim2:pred}
\end{figure}

\begin{figure}[htbp]
\centering
\includegraphics[width=\textwidth]{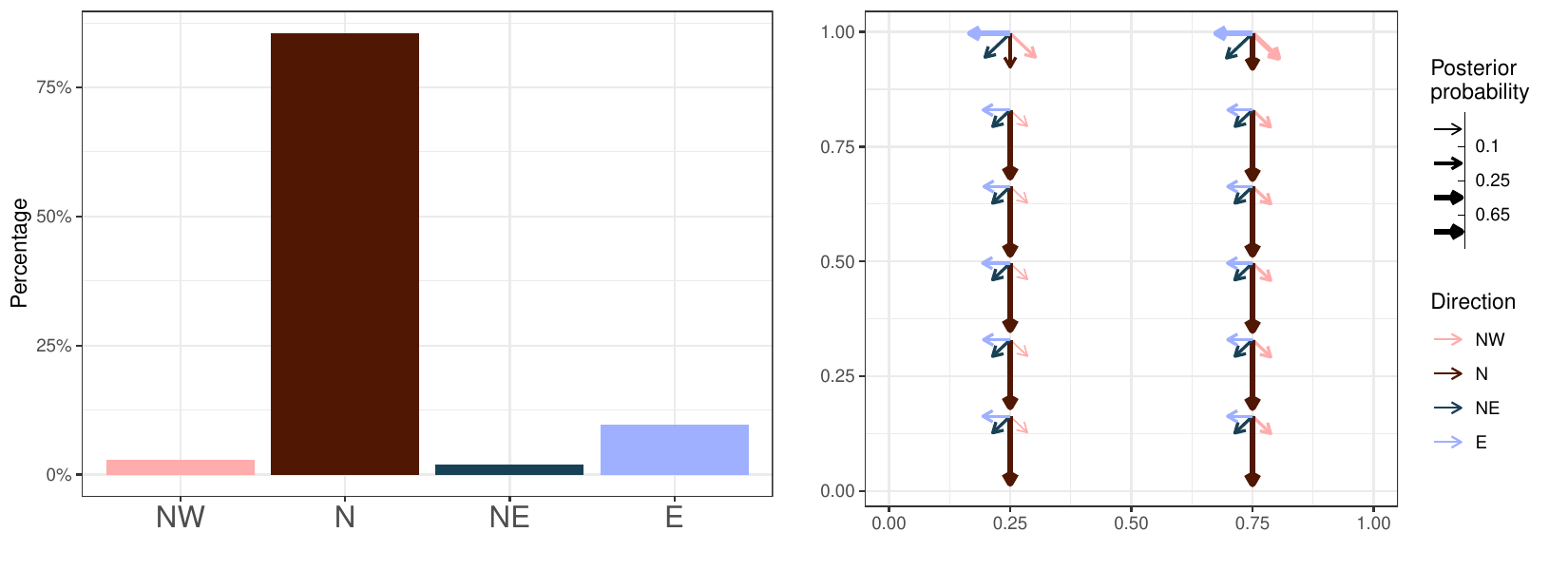}
\caption{Left: Histogram of wind directions chosen by the highest posterior probability in each partition block. Right: Average posterior probabilities of wind directions in the bag across 30 time points. The right plot illustrates the spatial domain $[0,1]^2$ with 12 partition blocks, within each of which arrows have different length and thickness proportionally to their posterior probabilities. This result is averaged over 25 synthetic data with $\bm{\theta}_3$.}
\label{sim2:wind}
\end{figure}

In this simulation study, we focus on inferences of predicted $y(\vt{t})$ and estimated $Z$. Hence, we briefly comment on convergence of MCMC for these variables in the G-BAG model. The average estimate of effective sample size of predicted $y(\vt{t})$ at 3,750 locations from 1,000 posterior draws is 1,009 and 1,011 in simulations with $\bm{\theta}_3$ and $\bm{\theta}_4$, respectively. For $Z$, we do not reject the null hypothesis at 0.05 significance level in about 95\% of the partition blocks on average across 25 replicates in both cases of $\bm{\theta}_3$ and $\bm{\theta}_4$. Based on these diagnostics, we conclude proper convergence of the two variables of interest. We present summaries of predictive performance and inferred directions across 25 synthetic data sets with $\bm{\theta}_3$ in Figures \ref{sim2:pred} -- \ref{sim2:wind}. Comparison of numerical prediction summaries between G-BAG and G-BAG* is provided in Table \ref{tab:sim2_theta3}. Results from simulations with $\bm{\theta}_4$ are highly similar and thus omitted. 

\begin{table}[htbp]
    \centering
    \begin{tabular}{ccc}
    \hline
         &  G-BAG & G-BAG* \\ \hline 
   RMSPE & 0.462 (0.010) & 0.462 (0.010) \\
   MAPE  & 0.368 (0.008) & 0.368 (0.008) \\
   95\% CI coverage & 0.965 (0.004) & 0.950 (0.004)\\ 
   95\% CI width & 1.960 (0.018) & 1.810 (0.024)\\ \hline
   \end{tabular}
   \caption{Prediction summaries when the fitted G-BAGs are misspecified. The true covariance parameters $\bm{\theta}_3$. Mean and standard error in parenthesis are calculated over 25 synthetic data sets.}
   \label{tab:sim2_theta3}
\end{table}

We run another simulation study in which data are generated from a full GP with an asymmetric spatiotemporal covariance function. We employ equation (12) of \cite{stein_space-time_2005}, which is 
\begin{align*}
    C((\bm{l}_s, l_t), (\bm{l}'_s, l'_t)) = \frac{\pi\alpha^2}{2^{\nu+|l_t-l'_t|-1}\Gamma(\nu+|l_t-l'_t|+1)}(\alpha r)^{\nu+|l_t-l'_t|}K_{\nu+|l_t-l'_t|}(\alpha r)
\end{align*}
with $r = \|\bm{l}_s-\bm{l}'_s-\epsilon (l_t-l'_t)\vt{v}\|$, where $\bm{l}_s \in \R^2$ is a spatial coordinate, $l_t \in \R$ is a temporal coordinate, $\vt{v} \in \R^2$ is a unit vector, $K_{\nu}$ is the modified Bessel function of the second kind of order $\nu$, and $\alpha \geq 0$, $\nu > 0$, $\epsilon >0$. The vector $\vt{v}$ controls wind directions, and $\epsilon$ controls wind speed. Under this full GP model with the Stein's covariance, which we call GP-Stein, we can only consider small data. Hence, we make $25 \times 25 \times 10$ grid locations in $\mc{D}=[0,1]^3$ and create their covariance matrix $\Sigma$ with $\alpha = 0.5$, $\nu = 0.5$, $\epsilon = 0.2$, and $\vt{v} = (0.447, 0.894)^T$. The selected unit vector $\vt{v}$ lets wind blow from south-southwest. We then generate 25 data sets from a multivariate Gaussian distribution with mean zero and covariance $\Sigma + \tau^2I$ with $\tau^2 = 0.01$. 

We fit G-BAG, fixed-DAG, and SPDE-nonstationary models. For G-BAG, we put west, southwest, and south directions in a bag and assign the following priors for covariance parameters: $\tau^2 \sim \mbox{IG}(2, 0.1)$, $\sigma^2 \sim \mbox{IG}(2, 1)$, $a \sim \mbox{Unif}(0.053, 19)$, $c \sim \mbox{Unif}(0.036, 2.118)$, and $\kappa \sim \mbox{Unif}(0, 1)$. The uniform distributions for $a$ and $c$ are selected so that the base correlation is as low as 0.05 and as high as 0.95 at the maximum temporal and spatial distance, respectively. The fixed-DAG model shares the same prior distributions except for $a, c \sim \mbox{Unif}(0.036, 19)$. For G-BAG and fixed-DAG models, we partition the domain into $3 \times 3 \times 10$ axis-parallel blocks. In the SPDE-nonstationary model, $\psi$'s are assigned $N(0,0.3^{-1})$ priors, the autocorrelation parameter is assumed larger than 0.05 with prior probability 0.99, and the precision of white noise is assumed larger than 1 with prior probability 0.01. G-BAG and fixed-DAG results are based on 1,000 posterior samples out of 10,000 MCMC draws by discarding the first 2,000 samples as burn-in and saving every 8th sample in the subsequent samples.

We evaluate predictive performance on randomly selected 20\% of data locations. Since no code is available to estimate unknown parameters in the Stein's covariance, we use kriging values with the true covariance parameters for prediction and consider them only as an (unobtainable in practice) gold standard. Nevertheless, Table \ref{tab:sim3} shows that prediction errors (RMSPE and MAPE) by G-BAG and fixed-DAG are within one standard error of average prediction errors by GP-Stein. Only SPDE-nonstationary exhibits elevated prediction errors and slight under-coverage of 95\% CIs. 
    
\begin{figure}[htbp]
    \centering
    \includegraphics[width = \textwidth]{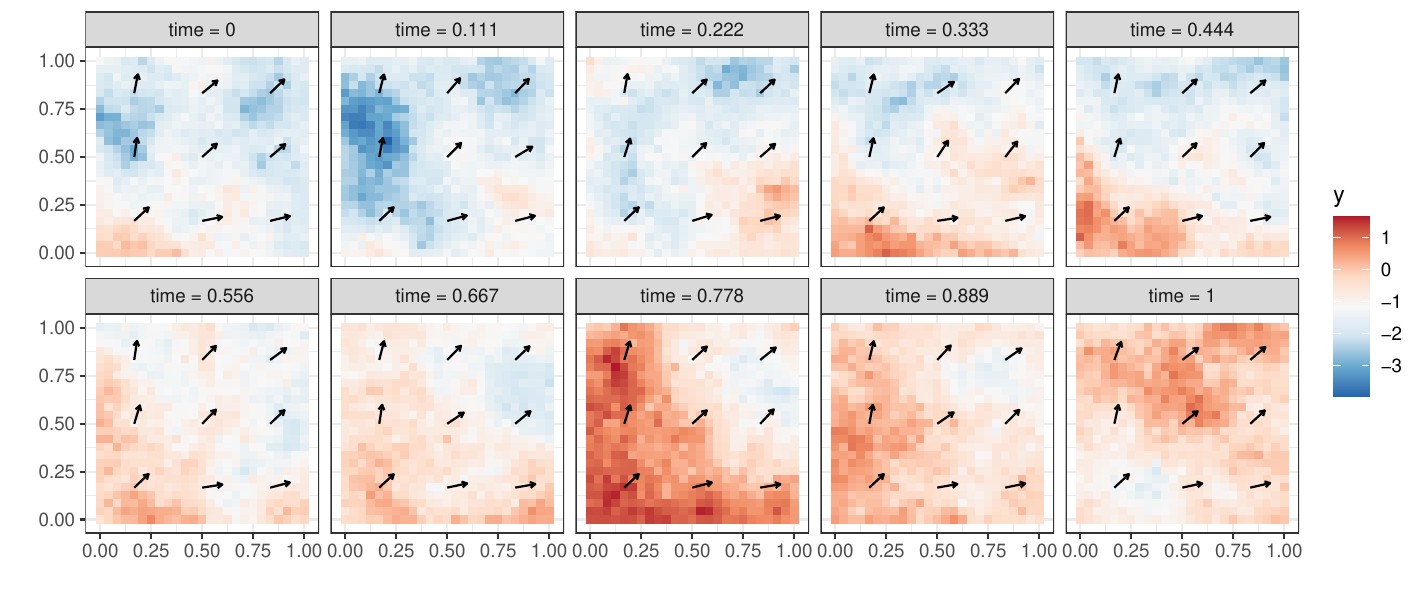}
    \caption{Simulated and G-BAG predicted $\vt{y}$ from a randomly selected replicate. Inferred wind directions by G-BAG averaged over all replicates are overlaid.}
    \label{sim3:predy}
\end{figure}

An additional advantage of G-BAG is to infer direction dependence of data. We present inferred wind directions averaged over all replicates in Figure \ref{sim3:predy}. The inferred directions, particularly in partition blocks on the upper two rows, are between south and southwest, matching the true direction (south-southwest) as specified by $\vt{v}$. Hence, we conclude that G-BAG's prediction is comparable to the true model when data are generated from a general asymmetric spatiotemporal covariance and can help comprehend the asymmetric covariance whose parameters may not always be straightforward to interpret.

\begin{table}[t!]
        \centering
        \begin{tabular}{ccccc}
        \hline
             &  GP-Stein & G-BAG & Fixed-DAG & SPDE-nonstationary \\ \hline 
       RMSPE & 0.200 (0.005) & 0.205 (0.005) & 0.202 (0.005) & 0.212 (0.005)\\
       MAPE  & 0.159 (0.004) & 0.163 (0.005) & 0.161 (0.005) & 0.169 (0.005)\\
       95\% CI coverage & -- & 0.951 (0.005) & 0.957 (0.005) & 0.930 (0.007)\\ 
       95\% CI width & -- & 0.808 (0.010) & 0.823 (0.009) & 0.770 (0.009) \\ \hline
       \end{tabular}
       \caption{Prediction summaries when data are generated from GP-Stein. Mean and standard error in parenthesis are calculated over 25 synthetic data sets.}
       \label{tab:sim3}
\end{table}

\section{Air Quality Analysis in California, the United States}
\label{s:anal-CA}

PurpleAir PM2.5 sensors require corrections for comparability with regulatory monitors. We thus calibrate our PM2.5 data following the United States EPA's practice \citep{barkjohn_development_2021}: $PM_{2.5} = 5.75 + 0.52 PA_{CF1} - 0.09RH$ if $PA_{CF1} \leq 343 \mu g/m^3$; $PM_{2.5} = 2.97 + 0.46 PA_{CF1} + 3.93\times 10^{-4} PA^2_{CF1}$ otherwise. $PM_{2.5}$ is the calibrated value used in our analyses, $PA_{CF1}$ is the PurpleAir measurements with a correction factor labeled as $CF=1$, and $RH$ is relative humidity.

\begin{figure}[htbp]
    \centering
    \includegraphics[width = 0.75\textwidth]{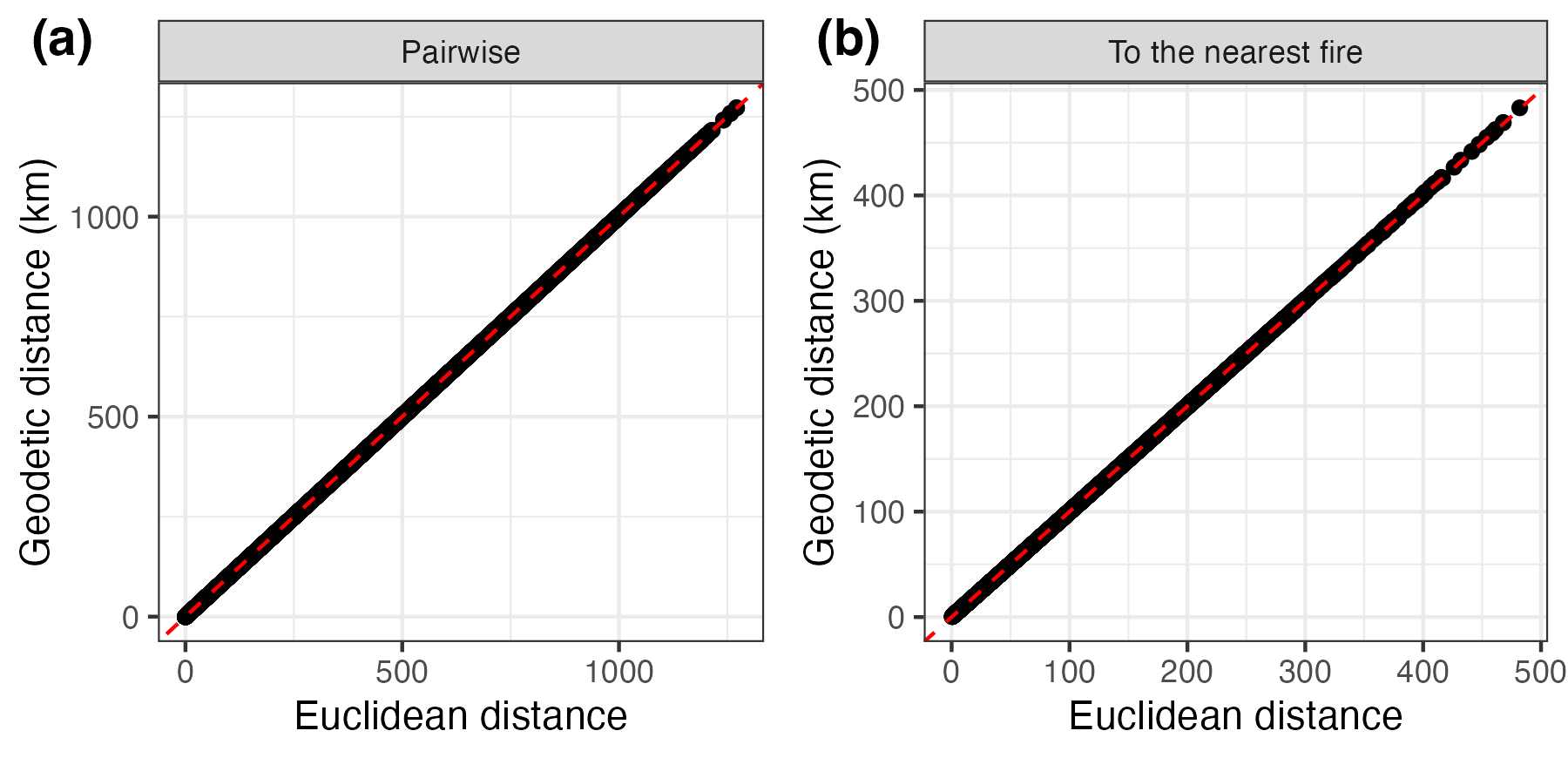}
    \caption{Comparison of Euclidean distance and geodetic distance (a) across all spatially unique pairs of locations and (b) to the nearest fire on a day at each location. The red dashed line is a 45-degree line.}
    \label{anal:CA_compare_dist}
\end{figure}

We also convert longitude and latitude to the Universal Transverse Mercator (UTM) coordinates. These coordinates are measured east and north in kilometers. Regardless of projections, distortions in distances are inevitable, while Euclidean distance on the projected coordinates can serve as an accurate approximation to geodetic distance for relatively small spatial domains such as California \citep{banerjee_hierarchical_2014}. We compare geodetic distance with longitude and latitude and Euclidean distance with easting and northing in Figure \ref{anal:CA_compare_dist}. Geodetic and Euclidean distances for all location pairs in panel (a) and to the nearest fire at each location in panel (b) are almost identical with average absolute differences of 0.337 and 0.129, respectively. We used \texttt{rdist.earth} to compute geodetic distance and \texttt{rdist} functions for Euclidean distance in \texttt{fields} package in \texttt{R}.

We assume $\tau^2 \sim \mbox{IG}(2,0.1)$, $\sigma^2 \sim \mbox{IG}(2,1)$, $a,c \sim \mbox{Unif}(0,1000)$, $\kappa \sim \mbox{Unif}(0,1)$, and $\beta \sim N(0,10^2)$. The G-BAG and fixed-DAG models have the same priors for all parameters. The location-specific spatial range is assumed in the SPDE model as $\rho(\vt{t}) \propto \exp(\psi_1 + \psi_2t'_1 + \psi_3t'_2)$ where $t'_1$ and $t'_2$ are scaled eastings and northings in $[0,1]$, respectively. Penalized complexity priors are identical as in Section \ref{sec:sim1} except independent $N(0,1^2)$ priors for $\psi_1$, $\psi_2$, and $\psi_3$.

\begin{figure}[htbp]
\centering
\includegraphics[width=0.9\textwidth]{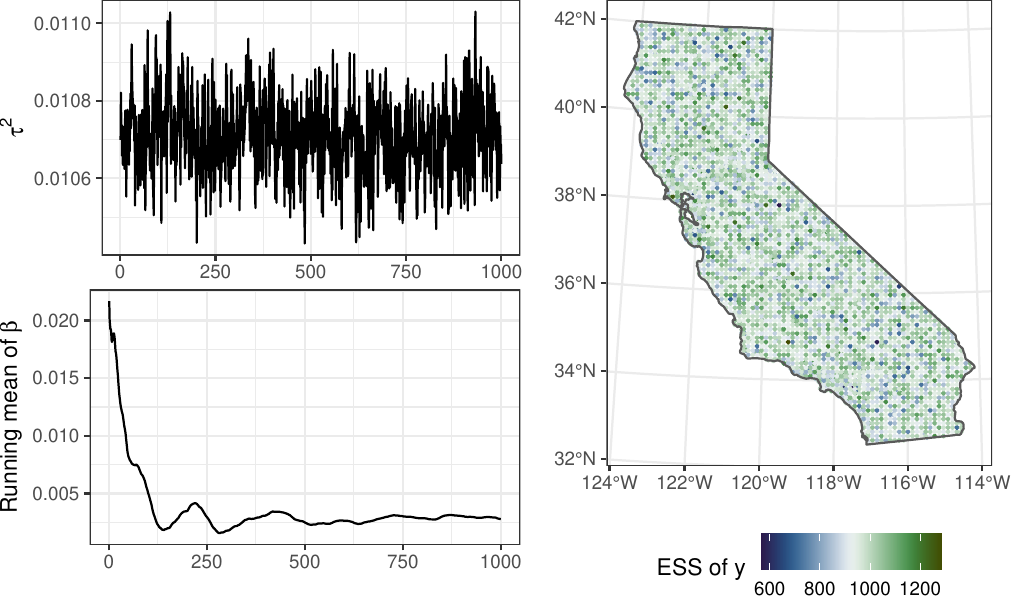}
\caption{Convergence diagnostics of G-BAG for California data. The top left panel shows the trace plot of $\tau^2$, the bottom left panel shows the running mean of $\beta$, and the right panel illustrates a map of ESS of $y(\vt{t})$ for a randomly selected day.}
\label{anal:CA_converge}
\end{figure}

Out of 30,000 iterations, first 15,000 iterations were discarded and every 15th draw was saved for the final analysis. To assess convergence of MCMC, we investigated empirical diagnostic tools such as trace plots, running mean plots, and ESS. Results for the nugget $\tau^2$, regression coefficient $\beta$, and estimated and predicted $y(\vt{t})$ are presented in Figure \ref{anal:CA_converge}. The trace plot in the top left panel provides evidence of convergence to stationarity in $\tau^2$ posterior samples. The running mean against iterations in the bottom left panel shows we stopped appropriately after the stabilization of the running mean of $\beta$. The map of ESS of $y(\vt{t})$ on a random day in the right panel also demonstrates properly high values across California. At most of locations, ESS computed using the \texttt{R} package \texttt{posterior} is above 700 from 1,000 posterior draws. The independent chi-squared tests for $z_i$'s concluded convergence at about 95\% of the partitioned regions by comparing the frequency distribution between the first 35\% and the last 35\% of the chain at a significance level of 0.05. These diagnostics of convergence imply adequate convergence of the parameters and variables we make inferences on. 

\begin{figure}[!ht]
\centering
\includegraphics[height=0.35\textheight]{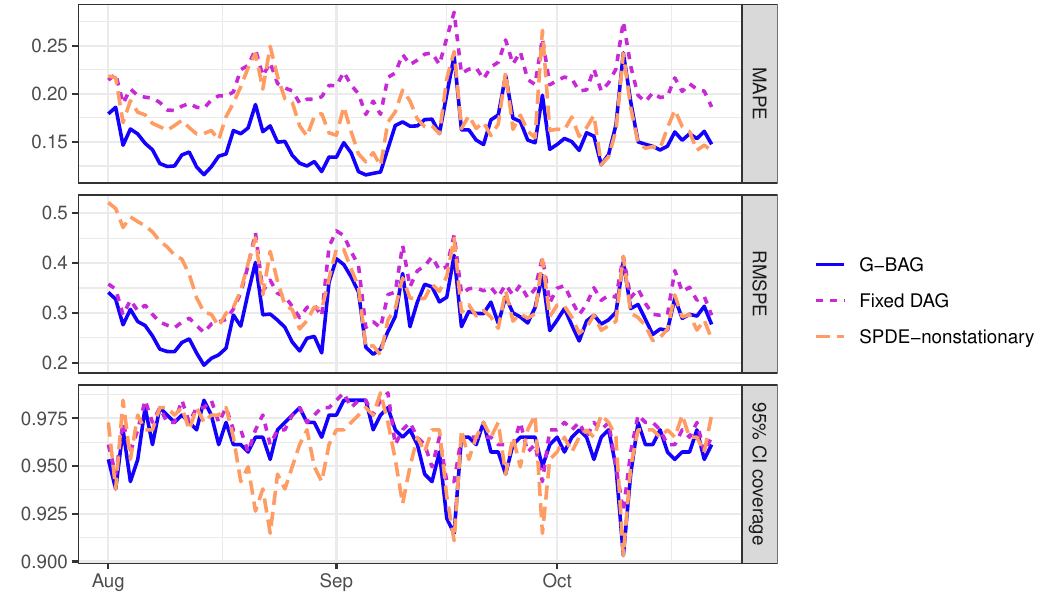}
\caption{Comparison of daily prediction performance measures by G-BAG, fixed-DAG, and SPDE-nonstationary models for California data. From top to bottom, MAPE, mean absolute prediction error; RMSPE, root mean square prediction error; CI, posterior predictive credible intervals.}
\label{anal:CA_pred}
\end{figure}

\begin{figure}[htbp]
\centering
\includegraphics[width=0.56\textwidth]{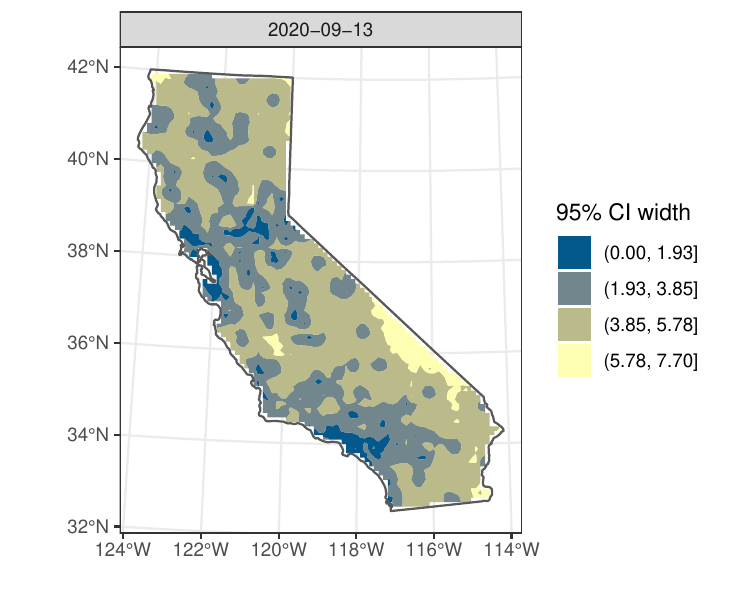}
\caption{The 95\% posterior predictive credible interval widths on November 13, 2020. The surfaces bear much resemblance at any time point.}
\label{anal:CA_pred_ci}
\end{figure}

Figure \ref{anal:CA_pred} presents prediction performance measures under all models at each day and suggests improved prediction of G-BAG compared to alternatives. Prediction uncertainty of G-BAG on a randomly selected day is illustrated in Figure \ref{anal:CA_pred_ci} from which we conclude that G-BAG properly increases uncertainty in regions with lack of observed data.

\begin{figure}[htbp]
\centering
\includegraphics[width=\textwidth]{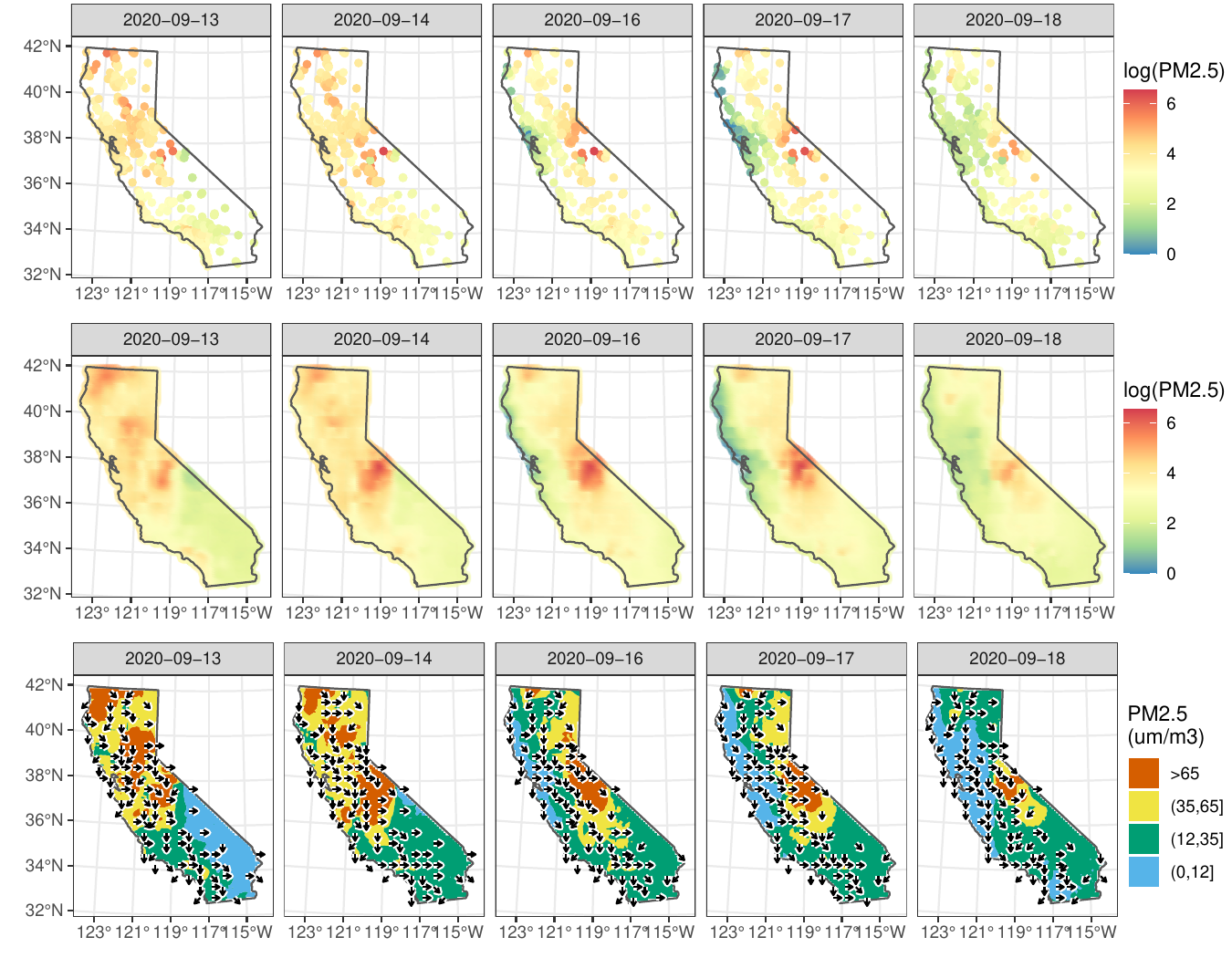}
\caption{Observed (top row) and predicted (middle) log(PM2.5) by G-BAG in California. On the bottom, the posterior mode of wind directions is overlaid on discretized predicted results for PM2.5 based on EPA standards.}
\label{anal:CA_pred_final}
\end{figure}

\begin{table}[htbp]
    \centering
    \begin{tabular}{cccc}
    \hline
    & G-BAG & G-BAG1 & G-BAG2 \\ \hline             
   \multirow{2}{*}{$\beta$}  & 0.003 & 0.007 & 0.003 \\ [-3pt]
   & ($-$0.011, 0.016) & ($-$0.007, 0.021) & ($-$0.011, 0.016) \\ 
   \multirow{2}{*}{$\tau^2$} & 0.011 & 0.011 & 0.011 \\ [-3pt]
   & (0.011, 0.011) & (0.010, 0.011) & (0.011, 0.011) \\
   \multirow{2}{*}{$\sigma^2$} & 3.781 & 4.726 & 3.802 \\ [-3pt]
   & (3.600, 3.990) & (4.429, 4.963) & (3.636, 3.976) \\
   \multirow{2}{*}{$a$} & 3.099 & 2.584 & 3.091 \\ [-3pt]
   & (2.963, 3.241) & (2.470, 2.728) & (2.963, 3.219) \\
   \multirow{2}{*}{$c$} & 0.009 & 0.008 & 0.009 \\ [-3pt]
   & (0.008, 0.009) & (0.008, 0.008) & (0.008, 0.009) \\
   \multirow{2}{*}{$\kappa$} & 0.011 & 0.017 & 0.011 \\ [-3pt]
   & (0.000, 0.041) & (0.000, 0.062) & (0.000, 0.041) \\ 
   RMSPE & 0.296 & 0.306 & 0.295 \\
   MAPE  & 0.154 & 0.155 & 0.153 \\
   95\% CI coverage & 0.963 & 0.966 & 0.964 \\ 
   95\% CI width & 1.504 & 1.570 & 1.502 \\ \hline
   \end{tabular}
   \caption{Posterior summaries and prediction performance measures of different G-BAG models on California PM2.5 data. Posterior mean and 95\% CI in parenthesis are provided for each parameter. G-BAG1 has $9\times 11 \times 83$ partitioning, and G-BAG2 uses west, southwest, south, and southeast directions.}
   \label{tab:sensitivity_anal_CA}
\end{table}

\begin{figure}[htbp]
\centering
\includegraphics[height=0.47\textheight]{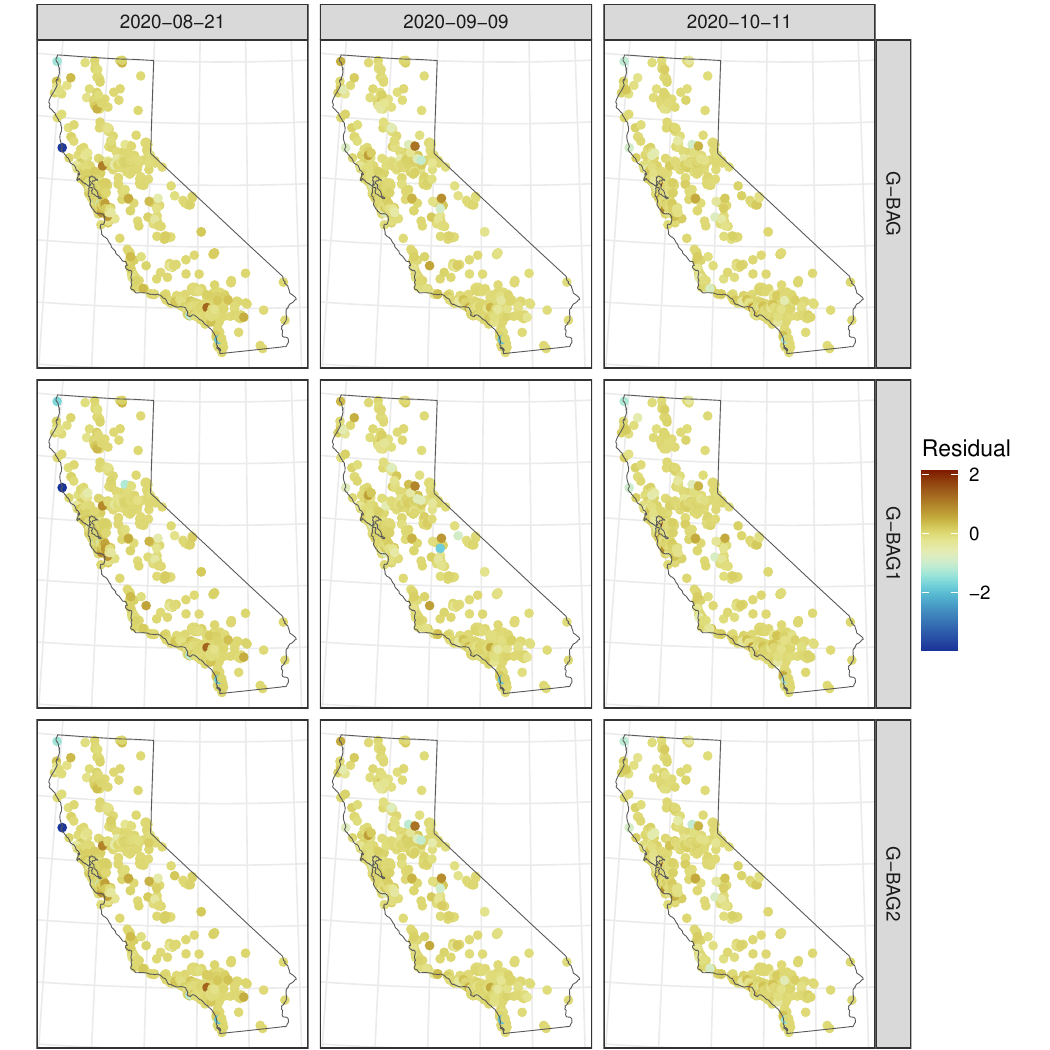}
\caption{Comparison of residuals, $\vt{y}$-$\hat{\vt{w}}$, by G-BAG models on the same dates as Figure \ref{anal:CA_res}.}
\label{anal:CA_sensitivity_res}
\end{figure}

\begin{figure}[htbp]
\centering
\includegraphics[height=0.47\textheight]{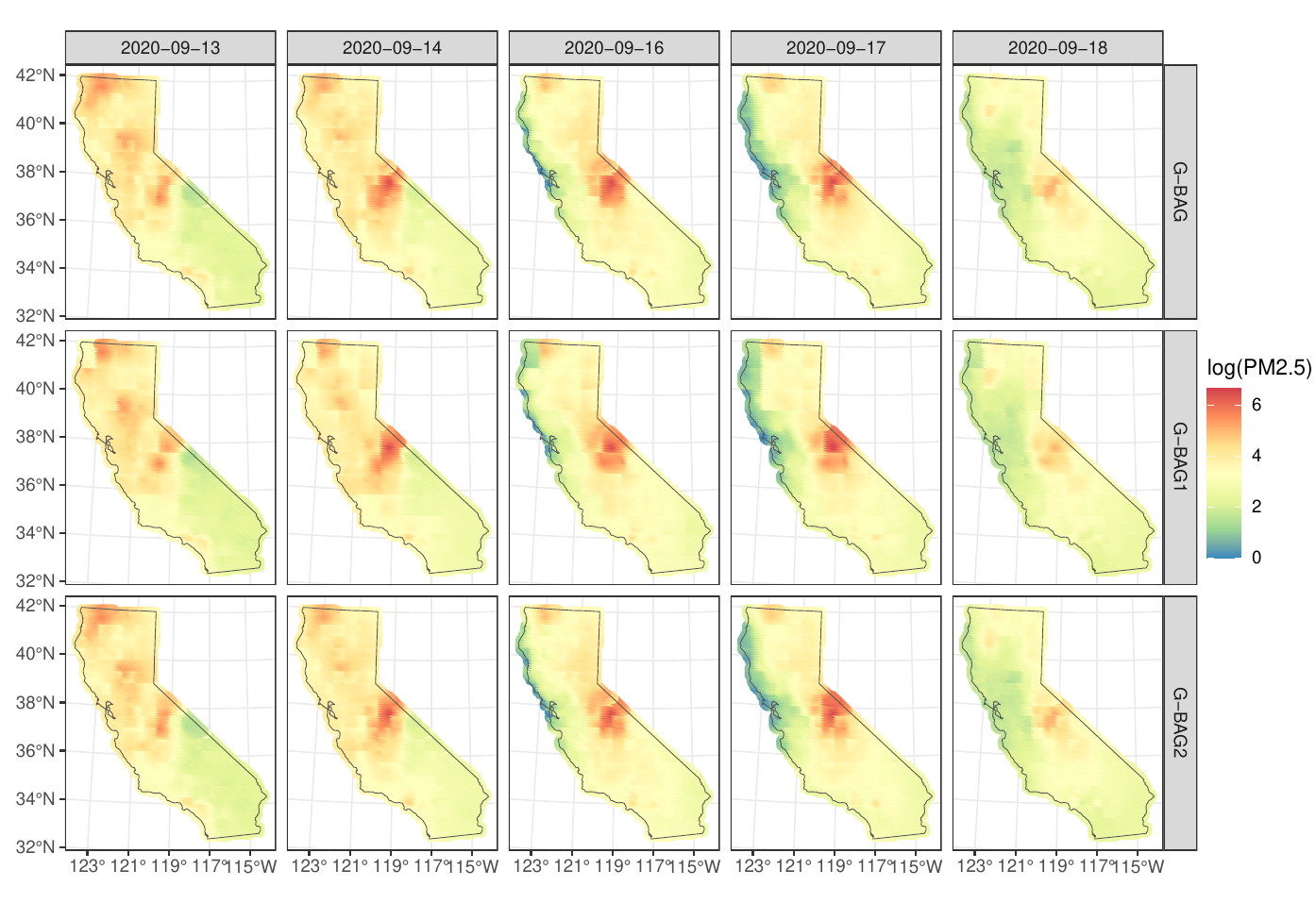}
\caption{Predicted log(PM2.5) by G-BAG models in California on the same dates as Figure \ref{anal:CA_pred_final}.}
\label{anal:CA_sensitivity_pred_final}
\end{figure}  

\bibliographystyle{BJ_apalike}
\bibliography{bibliography}
	
\end{document}